\documentclass[11pt]{article}
\usepackage{graphicx}
\usepackage{graphicx,colordvi}
\usepackage{amssymb}
\usepackage{amsmath,amsfonts}
\usepackage{url}
\usepackage{bm}
\usepackage{longtable}
\usepackage{braket}
\usepackage{subfigure}
\usepackage{color}
\usepackage{multirow}
\usepackage{graphicx}
\usepackage{amsmath}
\usepackage{amssymb}
\usepackage{bm}
\usepackage{dcolumn}

\newcommand{\vn}[1]{{\bf{#1}}}
\newcommand{\vht}[1]{{\boldsymbol{#1}}}

\newcommand{\bege}{\begin{equation}}
\newcommand{\ee}{\end{equation}}
\newcommand{\bal}{\begin{aligned}}

\newcommand{\eal}{\end{aligned}}

\newcommand{\vect}[1]{\ensuremath{\mathbf{#1}}}
\newcommand{\matr}[1]{\ensuremath{\mathcal{#1}}}

\newcommand{\uvect}[1]{\ensuremath{\hat{\mathbf{#1}}}}
\newcommand{\expo}[1]{\ensuremath{\mathrm{e}^{#1}}}

\newcommand{\bsqks}{\ensuremath{b^2_{\vect{k} \uvect{s}}}}
\newcommand{\bsq}{\ensuremath{b^2_{\uvect{s}}}}

\begin{document}
\parskip 1ex

\begin{center}
\section{SCIENTIFIC HIGHLIGHT OF THE MONTH}
\end{center}

\vspace{0.3cm}
\rule{16.0cm}{1mm}
\vspace{2mm}
\setcounter{section}{0}
\setcounter{figure}{0}

\begin{center}

{\Large Anisotropy of spin relaxation and transverse transport in metals}

{\large Yuriy Mokrousov, Hongbin Zhang, Frank Freimuth, Bernd Zimmermann, 
Nguyen H. Long, J\"urgen Weischenberg, Ivo Souza$^*$, Phivos Mavropoulos and Stefan Bl\"ugel}

{\large \it Peter Gr\"unberg Institut and Institute for Advanced Simulation, \\
Forschungszentrum J\"ulich and JARA, 52425 J\"ulich, Germany}

{\large $^*$\it Centro de F\'{\i}sica de Materiales and DIPC, Universidad del
Pa\'\i s Vasco, 20018 San Sebasti\'an, Spain; 
Ikerbasque, Basque Foundation for Science, 48011 Bilbao, Spain}

\end{center}

\def\sigpar{\sigma_m}
\def\sigperp{\sigma_\theta}
\def\sigout{\sigma_z}
\def\sigin{\sigma_x}
\def\alphapt{\alpha^{\rm Pt}}
\def\alphapd{\alpha^{\rm Pd}}
\def\alphafe{\alpha^{\rm Fe}}
\def\uu{\upuparrows}
\def\ud{\uparrow\hspace{-0.055cm}\downarrow}
\def\psins{\psi^{\sigma}_n}
\def\psinsp{\psi_n^{\sigma^{\prime}}}
\def\psims{\psi_m^{\sigma}}
\def\psimsp{\psi_m^{\sigma^{\prime}}}
\def\psinso{\psi_{n,0}^{\sigma}}
\def\psinspo{\psi_{n,0}^{\sigma^{\prime}}}
\def\psimso{\psi_{m,0}^{\sigma}}
\def\psimspo{\psi_{m,0}^{\sigma^{\prime}}}
\def\psipso{\psi_{p,0}^{\sigma^{\prime\prime\prime}}}
\def\psipspo{\psi_{p,0}^{\sigma^{\prime\prime\prime}}}
\def\psilso{\psi_{l,0}^{\sigma^{\prime\prime}}}
\def\psilspo{\psi_{l,0}^{\sigma^{\prime}}}

\begin{abstract}

Using first principles methods we explore the anisotropy of the spin relaxation and
transverse transport properties in bulk metals with respect to the direction of the spin
quantization axis in paramagnets or of the spontaneous magnetization in ferromagnets. 
Owing to the presence of the spin-orbit interaction the orbital and spin character 
of the Bloch states depends sensitively on the orientation of the spins relative to the
crystal axes. This leads to drastic changes in quantities which rely on interband mixing
induced by the spin-orbit interaction.
 The anisotropy is particularly striking for
quantities which exhibit spiky and irregular distribution in the Brillouin zone,
such as the spin-mixing parameter or the Berry curvature of the electronic states. We
demonstrate this for three cases: (i) the Elliott-Yafet spin-relaxation mechanism
in paramagnets with structural inversion symmetry; (ii) the intrinsic anomalous Hall
effect in ferromagnets; and (iii) the spin Hall effect in paramagnets. We discuss the
consequences of the pronounced anisotropic behavior displayed by these properties
for spin-polarized transport applications.

\end{abstract}

\section{Introduction}

 Phenomena belonging to the field of spintronics are associated with the
spin of electrons, which do the job of carrying information accross a device.
In such a situation, the fact that the spin and orbital degrees of freedom of
Bloch electrons in a solid are fundamentally related due to the presence of
spin-orbit interaction (SOI) becomes of great importance. Normally, the 
spin-orbit interaction can be considered as a small perturbation compared to
the other relevant energy scales for electrons in a crystal (such as band gaps,
band widths, or exchange splittings). Its influence is to mix the spin and
orbital character of the Bloch states for each Bloch momentum in a non-trivial
fashion. If we consider now a non-equilibrium situation of an electron moving 
in one of the Bloch bands across the crystal under the influence of an external
electric field, the spin-orbit mediated interaction with other Bloch states 
will determine its spin and orbital dynamics. Consider the case of a paramagnetic crystal.
If we manage to make our
initial incoming electron spin-polarized $-$ a typical situation in a spin-injection
experiment $-$ this spin polarization will decrease in time due to random scattering 
off impurities 
or phonons~\cite{Elliott.1954}, until it completely vanishes. Its exponential decay 
in time is characterized by the
{\it spin relaxation time}, which serves as one of the most basic material 
parameters in spintronics~\cite{Zutic.2004.rmp,mavropoulos:2009}. 
The presence of the SOI modifies in a subtle way the dynamics of Bloch electrons
under an applied electric field, by adding a spin-dependent transverse component
to the velocity. This will result in an {\it
anomalous Hall effect} (AHE) in ferromagnets~\cite{Karplus:1954}, and {\it spin
Hall effect} (SHE) in paramagnets~\cite{Dyakonov:1971}. Conceptually,
understanding of these two phenomena over the past 12$-$15 years generated a
cascade of novel paradigms in modern spintronics and solid state physics
in general. While practically the AHE and SHE entered an everyday toolkit in
experimental spintronics, further exciting research in this field is still ahead 
of us. In particular, a lot remains to be done concerning the microscopic understanding and first principles 
description of spin-relaxation phenomena and transverse transport properties in 
real materials. 

The crystal field in a solid is manifestly anisotropic and it lifts the degeneracy 
between the states with different magnetic quantum numbers. This results in 
a strong dependence of the spin and orbital character of the Bloch states 
on the choice of the spin quantization
axis (SQA) or the direction of the magnetization in the crystal, since the matrix 
elements of the orbital angular momentum operator are strongly anisotropic themselves.
In ferromagnets, the crystal field splitting combined with the anisotropy of the 
orbital angular momentum operator results in a dependence of the eigenvalues 
on the direction of the magnetization, and leads to the
magneto-crystalline anisotropy energy 
(MAE)~\cite{Freeman:1993,Laan:1998,Bruno:1989} $-$ one of the most 
fundamental characteristics of magnetic materials. In the field of transport
phenomena in metals, the anisotropy of the electronic structure with respect 
to the magnetization direction leads to such  prominent phenomena as 
anisotropic magnetoresistance (AMR)~\cite{McGuire:1975}, tunneling 
anisotropic magnetoresistance (TAMR)~\cite{Bode:2002,Molenkamp:2004} 
and ballistic anisotropic magnetoresistance (BAMR)~\cite{BAMR:2005}. As in the case of the
MAE, these effects can be already captured in
many cases by considering only the changes in the band topology in the Brillouin
zone, as the orientation of the magnetization is 
varied~\cite{Seemann:2011,BAMR:2005,Shick:2008,Bode:2002}. This situation 
is in contrast to the case of the AHE and SHE, which are often governed by band
degeneracies at the Fermi level~\cite{AHE-RMP}. In this case the dependence 
of the eigenenergies on the SQA/magnetization is either absent or can often
be neglected, while the anisotropy of the spin and orbital resolution of the 
wavefunctions becomes of primary
importance, and could lead to very large values of the anisotropy of the Hall
conductivities, as speculated already by Fivaz in 1969 for the anomalous Hall
effect~\cite{Fivaz:1969}.

The significant anisotropy of the spin-relaxation and Hall effects is
a valuable tool for tuning the transport properties of spintronics devices. Since 
such anistropy is an intrinsic property of the mono-crystalline solid, it
should be properly averaged when using polycrystalline samples, 
as well as when considering the effect of temperature and magnetization
dynamics on the measured spin-polarization or transverse 
current~\cite{Roman.2009.prl}. Experimentally, only the anisotropy of the 
anomalous Hall effect has been researched in the 
past and in many cases a very large anisotropy was found~\cite{Weissman:1973,Volkenshtein:1961,Hiraoka:1968,Lee:1967,
Ohgushi:2006,Sales:2008,Skokov:2008,Stankiewicz:2011}, while 
evidence of anisotropy in the spin Hall effect~\cite{Sih:2005} 
and spin-relaxation has been presented only recently~\cite{Tombros.2008.prl,
Averkiev.2008}. 

Here, we review the present theoretical understanding
of anisotropy in three phenomena occuring in perfect crystals: (i) spin relaxation, 
(ii) intrinsic anomalous Hall effect, and (iii) intrinsic spin Hall effect, Fig.~1.
We will focus on the developments which took place over the past years~\cite{Roman.2009.prl,
Freimuth.2010.prl,Zhang.2011.prl,Chudnovsky:2009,Zhang.prb.2011,Weischenberg.2011.prl,
Zimmermann:2012}. In particular, 
we present agruments and show from first-principles calculations that due to the
sensitivity of spin-relaxation and Hall effects to the SOI-mediated coupling 
between (nearly) degenerate states in the vicinity of the Fermi level, the
anisotropy of these effects can be gigantic, and has in principal no theoretical
limit. Manifestly, for some directions of the SQA and magnetization in the crystal
the spin relaxation rates and Hall currents can be suppressed by orders of 
magnitude, or even display a change of sign in corresponding conductivity components.  
We discuss possible applications of such large anisotropies, encourage further
experimental studies in this area, and emphasize that a wide range of materials
exhibit anisotropic transverse transport and spin relaxation,
from bulk solids to surfaces and interfaces with essentially lowered
lattice symmetry.  

\subsection{Spin relaxation in paramagnets} 
As a first example of a situation in which the importance of the anisotropy of
the wavefunctions with respect to the choice of the spin quantization axis is very
pronounced we consider the phenomenon of spin relaxation. To be concrete, here we
concentrate on the Elliott-Yafet spin-relaxation mechanism, dominant in materials with structural bulk 
inversion symmetry~\cite{Elliott.1954,mavropoulos:2009}, 
which is due to scattering of electrons off phonons or impurities. 
Owing to the presence of spin-orbit coupling (SOC) in the system such scattering events will flip 
the spin of the electron with a certain probability, which depends on both the wavefunctions of 
the ideal crystal and the scattering potential. However, according to the Elliott approximation~\cite{Elliott.1954},
an estimate of the ratio between
momentum- and spin-relaxation times, $\tau_\mathrm{p}$ and $T_1$, can be given
in a first approximation by neglecting the form of the scattering
potential as follows: $\tau_\mathrm{p} / T_1 \approx 4
b^2$, where $b^2$ is the Elliott-Yafet parameter (EYP) defined below, which is
an intrinsic property of the ideal crystal~\cite{Fabian.1998.prl}.

The coexistence of time-reversal and space-inversion symmetries implies that the
eigenenergies of the crystal at any given Bloch momentum $\vect{k}$ are at
least two-fold degenerate. Following Elliott, we write the corresponding states as
\begin{eqnarray}
{\psi}_{\vect{k} \uvect{s}}^{\uparrow} (\vect{r}) &=& \left[ a_{\vect{k}  \uvect{s}} (\vect{r}) ~ \lvert \uparrow \rangle_{\uvect{s}} + b_{\vect{k}  \uvect{s}} (\vect{r}) ~ \lvert \downarrow \rangle_{\uvect{s}} \right] ~ \expo{i \vect{k} \cdot \vect{r}} ~, \\
{\psi}_{\vect{k} \uvect{s}}^{\downarrow} (\vect{r}) &=& \left[ a_{-\vect{k} \uvect{s}}^{*} (\vect{r}) ~ \lvert \downarrow \rangle_{\uvect{s}} - b_{-\vect{k} \uvect{s}}^{*} (\vect{r}) ~ \lvert \uparrow \rangle_{\uvect{s}} \right] ~ \expo{i \vect{k} \cdot \vect{r}}~.
\label{Elliotts}
\end{eqnarray}
The two spin states $\lvert \uparrow \rangle_{\uvect{s}}$ and $\lvert \downarrow\rangle_{\uvect{s}}$
are eigenstates of $\uvect{s} \cdot \vect{S}$, where $\uvect{s}$ is the unit vector along the chosen SQA,
$\vect{S}=\frac{\hbar}{2}\boldsymbol{\sigma}^p$ is the spin angular momentum operator, and 
$\boldsymbol{\sigma}^p$ are the Pauli matrices. So, for example, $\lvert \uparrow \rangle_{z}$ and 
$\lvert \downarrow\rangle_{z}$ are the eigenstates of the $S_z$ operator. More generally, the
reference frame is specified by the SQA direction $\uvect{s}$, which is chosen to coincide with the
polarization direction of the initial/injected spin population.
 The functions $a_{\vect{k}
  \uvect{s}}(\vect{r})$ and $b_{\vect{k} \uvect{s}}(\vect{r})$ exhibit
the periodicity of the crystal lattice, and we define $\bsqks$ as the unit
cell integral $\int_{\mathrm{u.c.}}{ d^3 r \, \lvert b_{\vect{k}
    \uvect{s}} (\vect{r}) \rvert^2 }$ (similarly for
$a^2_{\vect{k}\uvect{s}}$, which satisfies $a^2_{\vect{k}\uvect{s}}=1-b^2_{\vect{k}\uvect{s}}$ ).

For fixed $\uvect{s}$, the degenerate ${\psi}_{\vect{k}
  \uvect{s}}^{\uparrow}$ and ${\psi}_{\vect{k} \uvect{s}}^{\downarrow}$ states (and
the corresponding $ a_{\vect{k} \uvect{s}} (\vect{r})$ and
$b_{\vect{k} \uvect{s}} (\vect{r})$) can be chosen, via a linear
combination, such that the spin-expectation value
$S_{\vect{k}\uvect{s}} = \langle {\psi}^{\uparrow} \rvert S_{\uvect{s}}
\lvert {\psi}^{\uparrow} \rangle$ is maximal. The spin mixing parameter is
then given by $\bsqks = 1/2 - S_{\vect{k}\uvect{s}}/\hbar$, and is
usually small, due to the weakness of the SOC. In this case the Bloch
states are of nearly pure spin character. 
(Thus, $a_{\vect{k} \uvect{s}}$ represents the "large" component of the spinor, while
$b_{\vect{k} \uvect{s}}$ the "small" component. The relation between the large or small components
of ${\psi}^{\uparrow}_{\vect{k} \uvect{s}}$ and ${\psi}^{\downarrow}_{\vect{k} \uvect{s}}$, $a_{\vect{k} \uvect{s}}=a^*_{-\vect{k} \uvect{s}}$ and $b_{\vect{k} \uvect{s}}=b^*_{-\vect{k} \uvect{s}}$, follows from space and time inversion symmetry).
However, at special spin-flip
hot-spot points in the band structure,~e.g.~accidental degeneracies,
Brillouin zone boundaries or other high symmetry
points~\cite{Fabian.1998.prl}, $\bsqks$ may increase significantly up to
$\frac{1}{2}$, which corresponds to the case of fully spin-mixed
states. Generally, the distribution of the spin-mixing parameter for
a metal with a complicated Fermi surface can be very non-trivial.
The Fermi-surface averaged Elliott-Yafet parameter is given by
\begin{equation}
  \bsq = \frac{1}{n(E_F)} ~ \frac{1}{\hbar} \, \int_\mathrm{FS}{  \frac{  |b_{\mathbf{k}\hat{\mathbf{s}}}|^2   }{ \lvert\vect{v}_F(\vect{k})\rvert } ~ d^2 k }~, \label{integral}
\end{equation}
where $\vect{v}_F(\vect{k})$ is the Fermi velocity. The normalization by the density of states at the Fermi level, $n(E_F) = 1/\hbar \, \int_\mathrm{FS}{ \lvert \vect{v}_F(\vect{k}) \rvert^{-1} ~ d^2 k  }$, ensures that $ 0\leq \bsq \leq \frac{1}{2}$.

For the ensuing discussion it will be useful to divide the spin-orbit operator into spin-conserving and 
spin-flip parts, $\xi(LS^{\uparrow \uparrow})$ and $\xi(LS^{\uparrow \downarrow})$, given respectively by the first
and second terms of the following expression~\cite{Zhang.2011.prl}:
\begin{equation}
  \xi \vect{L} \cdot \vect{S} = \xi L_{\uvect{s}} S_{\uvect{s}} + 
  \xi \left( L_{\uvect{s}}^{+} S_{\uvect{s}}^{-} + 
  L_{\uvect{s}}^{-} S_{\uvect{s}}^{+} \right)/2\,. \label{Eq:spinorbit}
\end{equation}
Here $\xi$ is the spin-orbit coupling strength, $\vect{L}$ is the operator
of the orbital angular momentum, $L_{\uvect{s}} = \vect{L}
\cdot \uvect{s}$, $S_{\uvect{s}} = \vect{S} \cdot \uvect{s}$, and
$L_{\uvect{s}}^{\pm}$ and $S_{\uvect{s}}^{\pm}$ are the 
raising and lowering operators for orbital and spin angular momenta. 
Acting on a state of the crystal obtained 
without SOC, the spin-flip part of the SOI can flip its spin, while the spin-conserving
part will keep it intact.  It is clear that the dot
product $\vect{L} \cdot \vect{S}$ is independent of $\uvect{s}$,
leaving the eigenenergies of the Hamiltonian invariant. However, the
spin-conserving and spin-flip parts, separately, depend on the
choice of the SQA. In ferromagnets, the spin-conserving part of SOI is the one
which is largely responsible for the values of the magneto-crystalline anisotropy energy
and orbital moments~\cite{Bruno:1989,Laan:1998}. 
In paramagnets, the Elliott-Yafet spin-relaxation mechanism
is driven by the spin-flip part of the SOI.

In an experiment, the spin polarization of the electrons subject to
spin relaxation is defined by the direction of the external
magnetic field (e.g.~in conduction-electron spin resonance
experiments) or by the polarization of ferromagnetic leads (e.g.~in
spin-injection experiments).  In a paramagnet, the choice of the
spin-quantization axis, determined by the direction of the spin
polarization, does not influence the band energies, and its most
important manifestation is in the changes of the orbital and spin
character of the Bloch states. Experimentally, the dependence
of the spin-mixing parameter on the SQA was observed in
supported graphene layers~\cite{Tombros.2008.prl} and in
semiconductors~\cite{Averkiev.2008}. However, no microscopic
theory of anisotropic spin-relaxation which explicitly refers to the 
anisotropy of the Bloch states has been given. In bulk metals with
inversion symmetry, 
which are at the focus of this article, the Elliott-Yafet mechanism is dominant. 
Below we will demonstrate that indeed the anisotropy of the EYP in metals can
be gigantic.

\begin{figure}[t!]
\begin{center}
\includegraphics[width=16cm]{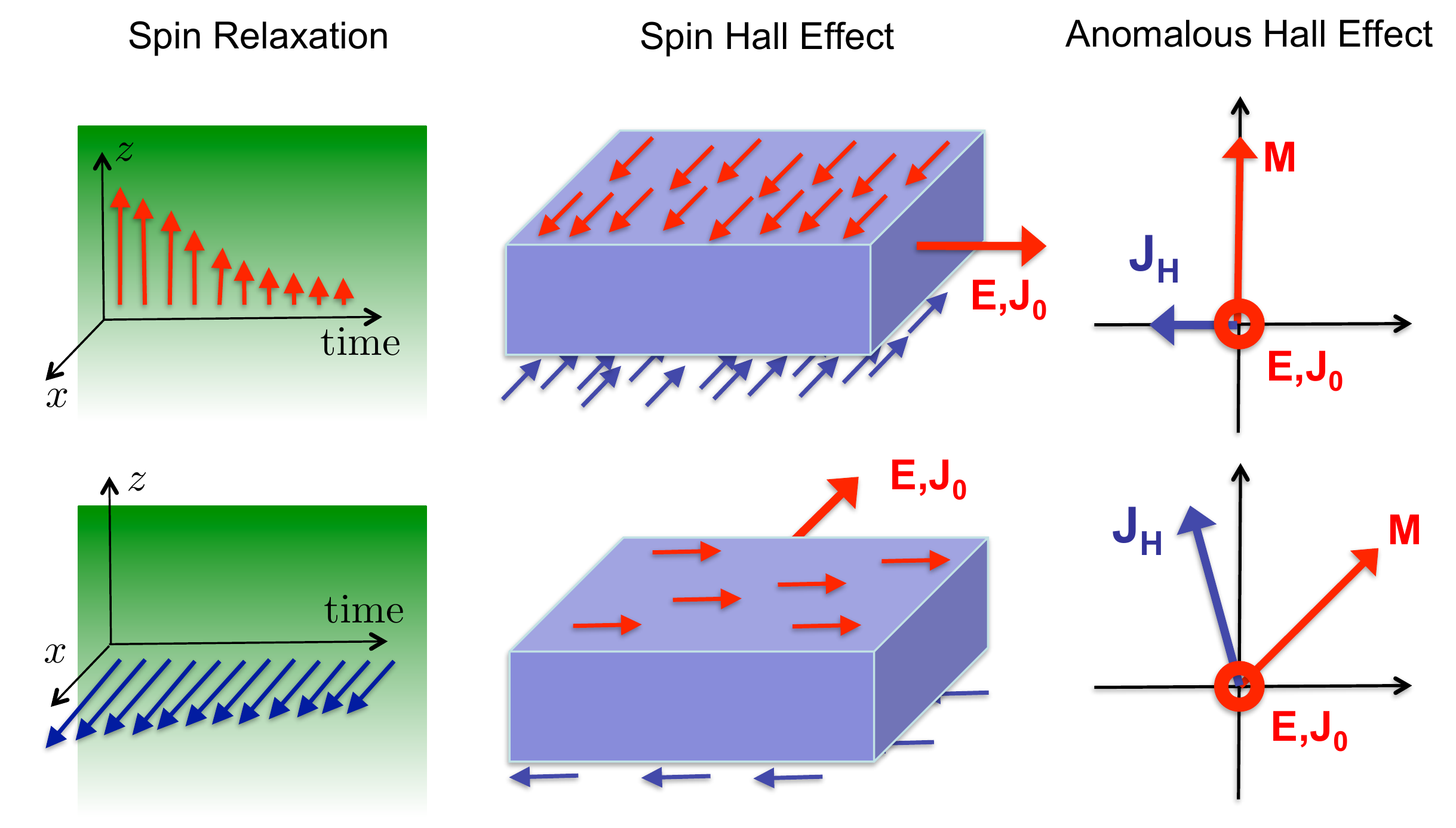}
\caption{Setup of the anisotropic spin-relaxation and transverse spin and anomalous Hall effects
in metals. {\it Anisotropic spin-relaxation} (left): an electron with a certain direction of spin, injected
into a solid which exhibits anisotropic spin-relaxation will lose the memory of its initial
spin polarization over a period of time, which depends on the direction of the spin-polarization
in real space ($z$ or $x$, red and blue arrows correspondingly). {\it Anisotropic spin Hal effect} (middle): the 
magnitude of the spin current and the direction of its spin polarization (related to the sign and magnitude 
of the spin accumulation on the surfaces of the sample) measured along a certain direction 
in a crystal depends on the direction of applied electric field $\mathbf{E}$ which generates
the longitudinal electric current $\mathbf{J}_0$. {\it Anisotropic anomalous Hall effect} (right): for a 
fixed direction of applied electric field $\mathbf{E}$ and corresponding longitudinal electric current 
$\mathbf{J}_0$ (out of the plane of the manuscript), the direction and magnitude of the Hall current
$\mathbf{J}_H$ depend on the direction of magnetization $\mathbf{M}$ in the sample. Note that 
for a general direction of $\mathbf{M}$ away from high-symmetry axes in the crystal, $\mathbf{J}_H$
can be non-orthogonal to $\mathbf{M}$.}  
\end{center}
\end{figure}

\subsection{Anomalous Hall Effect in ferromagnets}  
A second phenomenon, for which multiple 
experimental studies of anisotropy exist~\cite{Weissman:1973,Volkenshtein:1961,Hiraoka:1968,Lee:1967,Ohgushi:2006,Sales:2008,Skokov:2008,Stankiewicz:2011}, but
no quantitative theoretical argumentation for its emergence had been presented until recently~\cite{Roman.2009.prl}, 
is the anomalous Hall effect~\cite{nagaosa:2006,sinitsyn:2008,mokrousov:2009}. The essence of the AHE in a ferromagnet lies in the generation 
of a charge current $\mathbf{J}_H$ transverse to the electric field $\mathbf{E}$ (and corresponding "diagonal" current
$\mathbf{J}_0$), without any applied magnetic field~\cite{AHE-RMP}. Phenomenologically,  
the relation between the $i$th component of the Hall current and the $j$th 
component of the electric field is the following:
\begin{equation}
J_{H,i}=\sigma_{ij}E_j,
\end{equation}
where $\sigma_{ij}$ are the components of the anomalous Hall conductivity (AHC) tensor. Since
in a ferromagnet the anomalous Hall conductivity tensor is second rank antisymmetric, the AHC tensor can be also
seen as the AHC vector $\boldsymbol{\sigma}$, whose components are related to the components
of the AHC tensor as $\sigma_i=\frac{1}{2}\sum_{jk}\epsilon_{ijk}\sigma_{jk}$, through the Levi-Civita tensor 
$\epsilon_{ijk}$:
\begin{equation}
\mathbf{J}_H=\mathbf{E}\times\boldsymbol{\sigma}.
\label{sigma-times-E}
\end{equation}

In general, there can be different contributions to the AHC in a ferromagnet. In a real material 
contributions to the anomalous Hall current which originate from scattering of electrons off impurities or due to
disorder are always present  $-$ this is the so-called {\it extrinsic} AHE~\cite{Gradhand.2010.prl,
Lowitzer.2010.prl,Weischenberg.2011.prl}.
The second important part of the AHE signal $-$ which comes solely 
from the electronic structure of the pristine crystal $-$ is the so-called {\it intrinsic} AHE~\cite{Yao:2004}. 
Namely,
the SOC in a perfect crystal gives rise to a transverse spin-dependent "anomalous velocity" of electrons propagating 
along the direction of the applied electric field $-$ thus leading to the intrinsic anomalous Hall 
current~\cite{Karplus:1954,Berry-RMP}. In the present work we will focus exclusively on the intrinsic contribution to the AHE.

The intrinsic anomalous Hall conductivity is determined by the electronic structure of the pristine crystal, 
which can be accurately calculated using modern first principles methods, see for example a recent
review by Gradhand {\it et al.}~\cite{Gradhand:2012}. Several investigations of intrinsic AHC using the first principles methods have been done, for instance, in SrRuO$_3$~\cite{Fang:2003,Mathieu:2004}, 
Fe~\cite{Yao:2004,Wang:2006}, Ni~\cite{Wang:2007}, Co~\cite{Wang:2007,Roman.2009.prl} and
other ferromagnets. For these materials, the calculated 
intrinsic AHC agrees well with the experimental values, except for the case of fcc Ni~\cite{Wang:2007}, 
which is most probably due to effect of electronic correlations~\cite{Weischenberg.2011.prl,Guo:2011}. 
It is therefore a common 
belief that the AHE in moderately resistive samples of itinerant ferromagnets is often dominated by 
the  intrinsic contribution.

The intrinsic AHC considered in this work can be obtained via the linear response Kubo formula
for the off-diagonal components of the conductivity tensor $\sigma$:
\begin{equation}
\label{eq:kubo}
\begin{split}
\sigma_{ij}\ &=-\ e^2\hbar\int_{\text{BZ}}\frac{d^3k}{8\pi^3}\,\Omega_{ij}(\mathbf{k}), \\
\Omega_{ij}(\mathbf{k})\ &=-\ 2{\rm Im} \sum_{n,m}^{o,e}\frac{\Braket{\psi_{n\mathbf{k}}|v_i|\psi_{m\mathbf{k}}}\Braket{\psi_{m\mathbf{k}}|v_j|\psi_{n\mathbf{k}}}}{(\varepsilon_{n\mathbf{k}}-\varepsilon_{m\mathbf{k}})^2},
\end{split}
\end{equation}
which relates the conductivity tensor to the Brillouin zone (BZ) integral of the $k$-dependent
Berry curvature tensor $\Omega$. In the latter expression $\psi_{n\mathbf{k}}$ and
$\psi_{m\mathbf{k}}$ are respectively the occupied ($o$) and empty ($e$) one-electron
spinor Bloch eigenstates of the crystal, $\varepsilon_{n\mathbf{k}}$ and $\varepsilon_{m\mathbf{k}}$
are their eigenenergies, and $v_i$ and $v_j$ are the Cartesian components of the velocity operator
$\mathbf{v}$. The Berry curvature $\Omega$ appearing in the equation above is the very same quantity which 
arises when the adiabatic dynamics of electrons  in the reciprocal space is considered~\cite{Berry-RMP}. 
In particular, the Berry phase acquired by a Bloch electron as it traverses a closed path in the 
Brillouin zone can be calculated as an integral of the Berry curvature over the enclosed area~\cite{Berry-RMP,Mikitik:1999}. Mathematically, the Berry curvature is the curvature of the fibre
bundle of occupied electronic states in an insulator, and its integral over the whole torus of
allowed Bloch vectors provides the value of the quantized transverse charge conductivity,
as first demonstrated for the case of the quantum Hall effect by Thouless {\it et al.}~\cite{Thouless:1982}. 
The Berry curvature is a key quantity in the field of Chern and topological insulators~\cite{Hasan-RMP,Zhang-RMP}.
Overall, Eq.~(\ref{eq:kubo}) manifests the topological nature of the intrinsic anomalous Hall 
effect in metals. 

Let us briefly outline the concept of anisotropy as it applies to the AHE. The definition is slightly more
complicated than for the Elliott-Yafet parameter discussed previously, owing to the vector nature of the anomalous 
Hall conductivity. The parameter with respect to which the anisotropy of the AHE is studied is the
direction of the magnetization $\mathbf{M}$ in the crystal. 
The anisotropy of the AHC with respect to $\mathbf{M}$ is two-fold: not only the magnitude of
$\boldsymbol{\sigma}$ depends on $\mathbf{M}$, but also the direction of $\boldsymbol{\sigma}$
displays a non-trivial dependence on the direction of magnetization. For a high symmetry direction
of $\mathbf{M}$ the AHC vector is aligned with the magnetization so that the Hall current is perpendicular
to it. For a general direction of $\mathbf{M}$
away from high symmetry axes in the crystal the AHC vector can deviate from the direction of $\mathbf{M}$,
in which case~\cite{Hiraoka:1968,Roman.2009.prl}:
\begin{equation}
\boldsymbol{\sigma}(\mathbf{M}) = \boldsymbol{\sigma}_{\Vert}(\mathbf{M}) +
\boldsymbol{\sigma}_{\bot}(\mathbf{M}),
\label{Vert-Bot}
\end{equation}
where $\boldsymbol{\sigma}_{\Vert}(\mathbf{M})$ is aligned with $\mathbf{M}$ while
$\boldsymbol{\sigma}_{\bot}(\mathbf{M})$ is perpendicular to it~\cite{Roman.2009.prl} (see also Fig.~1). 
The microscopic
origin of the AHE anisotropy is clear from the expression~(\ref{eq:kubo}) for the Berry curvature,
according to which both dependence of wavefunctions as well as eigenenergies on the magnetization
direction leads to the anisotropy of the AHC. It is important to realize that in contrast to the case
of paramagnets with inversion symmetry (considered in the following with respect to the anisotropy
of the SHE and EYP), for which also the eigenspectrum does not change with the SQA, the dependence
of the wavefunctions on the magnetization direction in a ferromagnet is far more complex, owing
to broken time-reversal symmetry. Also, the anisotropy of the velocity matrix elements has to be
taken into account in uniaxial crystals. The largest part of this work is dedicated to analyzing the anisotropy
of the AHE in uniaxial crystals.

\subsection{Spin Hall Effect in paramagnets}
 
The spin Hall effect in paramagnets consists in generation of a {\it spin current} orthogonal
to the direction of an applied electric field $\mathbf{E}$. In a simple picture, the spin 
current in the SHE can be seen as two anomalous Hall currents, propagating in opposite
directions for spin-up and spin-down electrons.
In contrast to the AHE, where the direction of the Hall current is uniquely determined by the 
directions of $\mathbf{E}$ and $\mathbf{M}$, the
spin Hall current propagates in all directions othogonal to $\mathbf{E}$. For each of the directions
of the spin current, the "physical" spin-quantization axis is determined by the direction of the
current's spin polarization. 
First proposed theoretically in 1971~\cite{Dyakonov:1971}, the SHE was 
"re-discovered" in 1999~\cite{Hirsch:1999}, and eventually
experimentally observed in 2004~\cite{Kato:2004}, triggering development of new directions in spintronics~\cite{Uchida:2008,Buhrman:2012} and further research in the direction of quantum spin Hall insulators~\cite{Murakami:2003,Murakami:2004,Bernevig:2006,Konig:2007}. 
In analogy to the anomalous Hall effect, 
the observed SHE in metals contains two types of contributions: one extrinsic (driven by disorder), and
the other intrinsic (diorder-independent)~\cite{Gradhand.2010.prl,Gradhand.2011.01,Freimuth.2010.prl,Lowitzer:2011}.
And while very often the spin Hall effect is associated with the resulting spin accumulation at the boundaries
of the sample, employing inverse SHE it is possible to measure directly the spin Hall conductivities
(SHCs), which are much easier to treat theoretically with {\it ab initio} methods. As in the case of the AHE, 
for transition metals the experimental SHC values agree very often with the values obtained from 
first principles calculations for the intrinsic SHE. 

In the first principles calculations presented below
we consider only the intrinsic~\cite{Murakami:2003,Sinova:2004,Guo:2008,Freimuth.2010.prl,Gradhand.2011.01}
contribution to the SHC, which results from the virtual
interband transitions in the
presence of an external electric field. It may be expressed using a linear response Kubo formula analogous to 
Eq.~(\ref{eq:kubo}) for the AHC:
\begin{equation}
\label{eq:kubo:shc}
\begin{split}
\sigma_{ij}^s\ &=-e\hbar\int_{\text{BZ}}\frac{d^3k}{8\pi^3}\,\Omega_{ij}^s(\mathbf{k}), \\
\Omega_{ij}^s(\mathbf{k})\ &=-\ 2{\rm Im} \sum_{n,m}^{o,e}\frac{\Braket{\psi_{n\mathbf{k}}|Q_i^s|\psi_{m\mathbf{k}}}\Braket{\psi_{m\mathbf{k}}|v_j|\psi_{n\mathbf{k}}}}{(\varepsilon_{n\mathbf{k}}-\varepsilon_{m\mathbf{k}})^2},
\end{split}
\end{equation}
where $Q_{i}^{s}$ is the spatial $i$- and spin $s$-component of the
spin velocity operator, and tensor $\Omega$ is sometimes referred to as the "spin Berry curvature".
If only the spin-conserving part of the SOI is taken into account, the spin-projection along the direction
of the spin-polarization of the current $\hat{\mathbf{s}}$ is
a good quantum number, and the
spin velocity operator may be written as
$Q_{i}^{s}=\frac{\hbar}{2}\,v_{i}\sigma^p_{s}$. In this case the SHC equals twice the value
of (scaled) anomalous Hall conductivity for spin-up electrons only.
Here, $\sigma^p_{s}$ is a Pauli matrix used to express the $s$-component
of the spin operator.
In order to treat the spin-nonconserving part of the SOI correctly,
we used the
definition of the spin current density operator
given in Ref.~\cite{Shi:2006}.

To our knowledge, experimentally, the anisotropy of the spin Hall effect has been discussed
only once for AlGaAs quantum wells~\cite{Sih:2005}. In metals the anisotropy of the SHE was investigated
recently by Freimuth {\it et al.} from first principles~\cite{Freimuth.2010.prl}. 
In many aspects, the SHC anisotropy
is analogous to that of the anomalous Hall conductivity. It is remarkable, however, that due to the
higher symmetry of the problem the anisotropy of the SHC in transition metals is what we call 
purely {\it geometrical}. By this term we mean that it is exactly absent in case of a cubic crystal, while generally 
the dependence of the magnitude and spin polarization of the spin current on its direction can 
be reconstructed {\it exactly} from corresponding values for high-symmetry
directions in the crystal.  This is in sharp contrast to the behavior of the AHC or EYP, which 
exhibit anisotropy already in cubic crystals, and for which the dependence of the magnitude
of the EYP (anomalous Hall current) on the direction of the SQA (magnetization) cannot be 
reconstructed  from the respective "high-symmetry" values. The middle part of this review is 
dedicated to the anisotropy of spin Hall effect in transition
metals.

\section{Computational methods}
\label{comp_details}

For calculations of the Elliott-Yafet parameter and corresponding Fermi surfaces 
we used density functional theory in the local density approximation
\cite{Vosko.1980.CanJPhys} to calculate the underlying electronic structure.  
For the self-consistent calculations we employed the Korringa-Kohn-Rostoker 
(KKR) Green-function method~\cite{KKR-code} in the atomic sphere 
approximation and solve the Dirac equation with angular-momentum expansion 
up to $\ell_\mathrm{max}=4$. The Fermi surface is determined by the KKR
secular equation, $\mathrm{det}(\matr{M}(\vect{k},E_F)) = 0$, which is
equivalent to the condition that at least one eigenvalue of the KKR
matrix $\matr{M}$ vanishes. We search for the $\vect{k}$-vectors
fulfilling this condition with a tetrahedron method using linear
interpolation of the complex eigenvalues of $\mathcal{M}$. We choose a
grid of 200 $\vect{k}$-points for each direction in the full Brillouin
zone, resulting in about $10^7$ Fermi-surface points. We followed the
procedure described in Ref.~\cite{Heers.PhD} to maximize the
spin component $S_{\vect{k} \uvect{s}}$ at the Fermi-surface points. The
integration (Eq.~\eqref{integral}) is done by evaluating the integrand
at the Fermi-surface points and interpolating linearly within the
connecting triangles (the details of the method will be published elsewhere).

For calculations of the intrinsic anomalous Hall and spin Hall conductivities we
employed the full-potential linearized augmented plane-wave (FLAPW) method, as implemented
in the J\"ulich code \texttt{FLEUR}~\cite{fleur}. We used the generalized gradient approximation to 
the DFT and experimental lattice constants of the transition-metals. The self-consistent 
calculations with SOC were done in second variation with k$_{\rm max}$ between 3.7 and 
4.0~a.u.$^{-1}$ and about 8000$-$16000 $k$-points in the full Brillouin zone. For intermediate 
alloys, for instance, (Fe$_{0.5}$Co$_{0.5}$)Pt, the virtual crystal approximation (VCA) was applied 
on the $3d$ atomic sites, where the composition-averaged core potential was used instead of that 
of pure $3d$ elements, together with corresponding number of valence electrons, and interpolated 
lattice constants from the neighboring compounds~\cite{Zhang.prb.2011}. 
For the calculations of the conductivities
 we applied the Wannier interpolation technique of Wang {\it et al.}~\cite{Wang:2006}. We followed
 the method introduced in Refs.~\cite{Souza:2002} and~\cite{Freimuth:2008} to construct the 
 maximally-localized Wannier functions (MLWFs) from the FLAPW Bloch states $\psi_{\mathbf{k}m}$:
\begin{equation}
W^{\phantom{k}}_{\mathbf{R}n}(\mathbf{r})=
\frac{1}{N}\sum_{\mathbf{k}}e^{-i\mathbf{k}\cdot\mathbf{R}}
\sum_{m}U_{mn}^{(\mathbf{k})}\psi^{\phantom{k}}_{\mathbf{k}m}(\mathbf{r}),
\label{WF-expression}
\end{equation}
where  
$W_{\mathbf{R}n}$ denotes the $n$-th WF centered at lattice site $\mathbf{R}$,
$U^{(\mathbf{k})}_{mn}$ refers to the unitary transformation among the Bloch states at $\mathbf{k}$
which minimizes the spread of the Wannier functions.
Using the self-consistent charge density with SOC included, 18 spinor MLWFs per transition-metal atom,
corresponding to $s,p,d$-type orbitals, were generated using {\tt wannier90} code~\cite{wannier90}.
 
Working in the basis of the maximally-localized WFs allows us to construct a real-space 
tight-binding Hamiltonian of the crystal, which can reproduce the electronic bands with any 
given accuracy at any $k$-point in the Brillouin zone, given that a necessary number of $k$-points was
used for the generation of the WFs~\cite{Wang:2006}. 
The real-space tight-binding hopping parameters can be calculated as:
\begin{equation}
\label{eq_hamilt_rs}
H_{mm'}(\mathbf{R})= \frac{1}{N}\sum_{\mathbf{k}n}
\epsilon^{\phantom{k}}_{\mathbf{k}n}
e^{-i\mathbf{k}\cdot\mathbf{R}}
\left(
U_{nm}^{(\mathbf{k})}
\right)^{*}
U_{nm'}^{(\mathbf{k})},
\end{equation}
where $H_{mm'}(\mathbf{R})$ denotes the hopping parameter between 
Wannier orbitals $W_{\mathbf{R}m'}(\mathbf{r})$ and 
$W_{\mathbf{0}m}(\mathbf{r})$. 
Based on those parameters, the Hamiltonian $H(\mathbf{k})$, matrix elements of the velocity 
operator as well as charge and spin  Berry curvature in reciprocal space can be efficiently 
evaluated using the Wannier interpolation technique~\cite{Wang:2006}.

In the section on the anomalous Hall effect, we evaluate the derived perturbation 
theory expressions for the AHC in L1$_0$ FePt. 
In order to apply the perturbation theory in the basis of Wannier functions, we use the basis 
of Wannier functions constructed without SOI to calculate the 
matrix elements of the spin-orbit interaction. To do this, the scalar-relativistic Hamiltonian without SOI
is set up for the majority and minority states, and diagonalized in order to obtain 
the Bloch functions $\psi_{\mathbf{k}n}^{\sigma}(\mathbf{r})$, with 
$\sigma$ = $\uparrow$ or $\downarrow$.
The matrix elements of SOC in the basis of Bloch states can be then calculated:
\begin{equation}
\label{eq:socdef}
V^{(\mathbf{k})}_{n\sigma,n'\sigma'}=\sum_{\mu}
\frac{ \mu^{\phantom{k}}_{ \text{B}} }{\hbar m_{\text{e}} e c^{2} }\Braket{\psi_{\mathbf{k}n}^{\sigma}|
\frac{1}{r}\frac{dV^{\mu}(r)}{dr}\mathbf{L}^{\mu}\cdot\mathbf{S}|\psi_{\mathbf{k}n'}^{\sigma'}},
\end{equation}
where $\mathbf{L}^{\mu}$ is the atomic orbital 
momentum operator associated with atom $\mu$ (with the potential $V^{\mu}$).
In the scalar-relativistic approximation, the Hamiltonian $\tilde{H}_{mm'}^{\sigma}(\mathbf{R})$
can be obtained as:
\begin{equation}
\label{eq_hamilt_rs}
\tilde{H}^{\sigma}_{mm'}(\mathbf{R})
=
\frac{1}{N}\sum_{\mathbf{k}n}
\epsilon^{\phantom{k}}_{\mathbf{k}n\sigma}
e^{-i\mathbf{k}\cdot\mathbf{R}}
\left(
U_{nm\sigma}^{(\mathbf{k})}
\right)^{*}
U_{nm'\sigma}^{(\mathbf{k})}.
\end{equation}
Likewise, the matrix elements $V^{(\mathbf{k})}_{n\sigma,n'\sigma'}$ are
transformed into the basis set of Wannier functions:   
\begin{equation}\label{eq_socmat_rs}
V_{mm'}^{\sigma\sigma'}(\mathbf{R})
= \frac{1}{N}\sum_{\mathbf{k}nn'}
V_{n\sigma,n'\sigma'}^{(\mathbf{k})}
e^{-i\mathbf{k}\cdot\mathbf{R}}
\left(
U_{nm\sigma}^{(\mathbf{k})}
\right)^{*}
U_{n'm'\sigma'}^{(\mathbf{k})}.
\end{equation}
The complete Hamiltonian with SOC in the WF-basis is then given by
\begin{equation}
\label{eq:ham}
H_{mm'}^{\sigma\sigma'}(\mathbf{R})
=
\tilde{H}^{\sigma}_{mm'}(\mathbf{R})\delta_{\sigma\sigma'}+
V_{mm'}^{\sigma\sigma'}(\mathbf{R}).
\end{equation}
Such a separation enables us to perform the perturbation treatment of SOC.
By calculating from first principles an atomic shell 
averaged SOC parameter $\xi\overset{def}{=}\Braket{\frac{1}{r}\frac{dV(r)}{dr}}$, it is possible 
to write the SOC operator approximately as $\xi \mathbf{L}\cdot\mathbf{S}$, where
$\mathbf{L}\ (\mathbf{S})$ is the total orbital (spin) angular momentum
operator. Calculated in such a way the SOC strength $\xi$ for Pt (about 0.6 eV) is one order of magnitude larger
than that of $3d$ elements, for instance, $\xi=0.06$~eV for Fe.

\section{Anisotropy of spin relaxation in metals}

First of all, let us work out the perturbation theory expression for the spin-mixing parameter of a
certain state $\psi_n$ (we omit the explicit $k$-dependence for the moment). Let us assume that the 
spin-conserving part of the spin-orbit interaction has
been included in the Hamiltonian, which has $\psi_n$ as an eigenstate. In this case, 
since the spin-conserving SOC keeps the spin a good quantum number, in a 
paramagnet with structural inversion symmetry, the state $\psi_n$ can be characterized by
a certain value of spin, say, $\psi_n=\psi_n^{\uparrow}$. This has an exact replica but of 
opposite spin $\psi_n^{\downarrow}$. Upon including into the picture
the spin-flip SOI, $\psi_n^{\uparrow}$ will acquire an admixture of the down spin which we will
denote as $(\psi_n^{\uparrow})^{\downarrow}$, and which corresponds to the part that includes 
$b_{\mathbf{k}\hat{\mathbf{s}}}$ in Eq.~(\ref{Elliotts}). It is clear that since spin-flip SOI is off-diagonal
in orbital character,
the down-spin admixture of $\psi_n^{\uparrow}$ does not come from interaction with 
$\psi_n^{\downarrow}$ at the same energy, but comes from the interaction with the other states
in the system. Within  first order non-degenerate perturbation theory $(\psi_n^{\uparrow})^{\downarrow}$ can be calculated as:
\begin{equation}
(\psi_n^{\uparrow})^{\downarrow}=\xi\sum_{m\neq n}\frac{\langle \psi_m^{\uparrow}|LS^{\uparrow \downarrow}|\psi_n^{\downarrow}\rangle}{\varepsilon_n-\varepsilon_m}\psi_m^{\downarrow}.
\label{EY-WF-PT}
\end{equation}
Since the spin-mixing parameter $b_n^2$ is equal to $|(\psi_n^{\uparrow})^{\downarrow}|^2$,
we readily obtain from the latter expression that
\begin{equation}
b_n^2 \approx \xi^2\sum_{m\neq n}\frac{|\langle \psi_n^{\uparrow}|LS^{\uparrow \downarrow}|\psi_m^{\downarrow}\rangle|^2}{(\varepsilon_n-\varepsilon_m)^2}.
\label{EY-PT}
\end{equation}
Therefore, the spin-mixing parameter of a certain state is just a sum of amplitudes
for  SOC-mediated spin-flip transitions from this state to other states and back. This picture 
of the Elliott-Yafet
parameter in solids had been suggested by Elliott already in 1954~\cite{Elliott.1954}. 
Later on in this work, we will 
apply a similar approach in order to perform a perturbation theory analysis of the Hall effects. 

\begin{figure}[t!]
\begin{center}
\includegraphics[width=15cm]{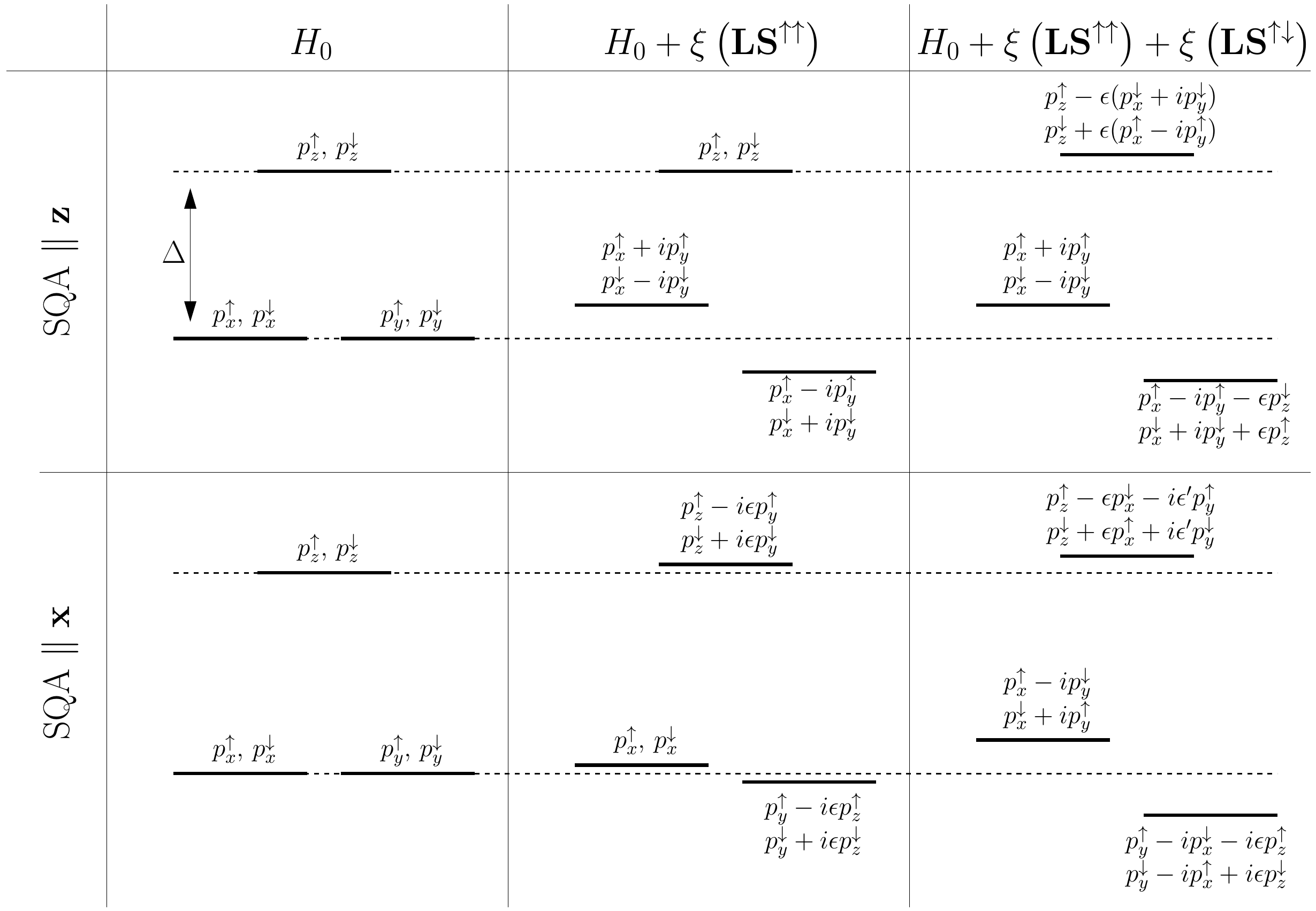}
\caption{Eigenvalues of the spin-degenerate $p$-states described by the Hamiltonian $\mathrm{diag} (\varepsilon, \varepsilon, \varepsilon + \Delta) \otimes \mathtt{1}_{2 \times 2}$ (left column), including the
spin-conserving part of SOC for two different quantization axes (middle column), and including
both spin-conserving and spin-flip parts of SOC for two different quantization axes (right column).
For each doubly degenerate state the orbital and spin decomposition is given with a label. Note that while
the energy spectrum is identical for both SQA's in the right column, the spin character of the two
low-lying states is different, which gives rise to the anisotropy of the spin-mixing parameter. In 
the labels, "$\epsilon$" and "$\epsilon'$" denote a small admixture of the corresponding state
of the order of $\xi/\Delta$. 
\label{EYP-Model-Levels}}
\end{center}
\end{figure}

Before proceeding with {\it ab initio} calculations, we consider a simple model which is able 
to capture the origin and essential properties of the anisotropy of the spin-mixing parameter in a
solid.
 Namely, let us consider
six $p$-orbitals, $p_x^{\sigma}$, $p_y^{\sigma}$ and $p_z^{\sigma}$ with $\sigma = (\uparrow, \downarrow$) standing for the spin of the orbitals. 
In order to consider the spin-mixing separately, we explicitly separate the two SOC terms. We have:
\begin{equation}
  H = H_0 + \xi (LS)^{\uu} + \xi (LS)^{\ud}= \mathrm{diag} (\varepsilon, \varepsilon+\delta, \varepsilon + \Delta) 
  \otimes 
  \mathtt{1}_{2 \times 2} + \xi (LS)^{\uu} + \xi (LS)^{\ud}, 
  \label{eq:TBmodel}
\end{equation}
where in the on-site part the four states $p_x^{\sigma}$ and $p_y^{\sigma}$ are chosen to be almost
 degenerate at energy $\varepsilon$ (separated by energy $\delta$: $\delta/\Delta << 1$), and the 
 $p_z^{\sigma}$ orbitals are shifted to higher energy $\varepsilon + \Delta$ in order to mimic the 
 uniaxiality of the lattice. The SOC strength is given by $\xi$, with $\xi/\Delta << 1$. The 
 energetic levels and their orbital character without spin-orbit are shown in the left column of 
 Fig.~\ref{EYP-Model-Levels}, in which $\delta$ was put to zero.

Let us first consider the case when the SQA points along the $z$-axis and $\delta=0$. 
When only spin-conserving SOC is added to $H_0$, the 
eigenstates are $(p_x^{\sigma} \pm i p_y^{\sigma})/\sqrt{2}$ and $p_z^{\sigma}$ ($\sigma = \uparrow, \downarrow$), with energies as sketched in Fig.~\ref{EYP-Model-Levels}. The only non-vanishing 
matrix elements of the spin-flip SOC are $\langle p_x^{\uparrow} - i p_y^{\uparrow} \rvert LS^{\uparrow \downarrow} \lvert p_z^{\downarrow} \rangle = 2$ and $\langle p_x^{\downarrow} + i p_y^{\downarrow} \rvert LS^{\uparrow \downarrow} \lvert p_z^{\uparrow} \rangle = -2$ which come from the states
that are well-separated in energy. According to Eqs.~(\ref{EY-WF-PT}) and (\ref{EY-PT}) this leads to
a small admixture of the $p_z$-state of opposite spin in the lowest lying eigenstate, and  corresponding
spin-mixing parameter of the order of $(\xi/\Delta)^2$ when the spin-flip SOC is included. On the other hand, when the SQA is chosen along the $x$-axis, the spin-conserving part of 
SOC mixes small $\xi / \Delta$-portions of $p_y^{\sigma}$ with $p_z^{\sigma}$ orbitals, see Fig.~\ref{EYP-Model-Levels}. Now, there are four non-vanishing matrix elements of spin-flip SOI, all of order 1, 
among which two transitions are very close in energy with a separation of $\sim \xi^2/\Delta$, giving rise to a much larger spin-mixing of the order of $\Delta/\xi$. This results in a very strong spin-mixing between
the two low-lying orbitals when the spin-flip SOC is added, and leads to a very large 
spin-mixing parameter of the lowest-lying state of the order of $(\Delta/\xi)^2$, meaning that higher-order perturbation 
theory is needed since the spin-mixing parameter cannot exceed $\frac{1}{2}$. The resulting orbital and spin
character of the states when the Hamiltonian with complete SOC is diagonalized, is presented in the
right column of Fig.~\ref{EYP-Model-Levels}. Note, that the final eigenenergies are the same, while the character of the states is different among the two directions of the SQA. Obviously, the resulting anisotropy of the spin-mixing parameter of the lowest-lying state with respect to the choice of the SQA 
is very large if $\xi/\Delta << 1$.

\begin{figure}[t!]
\begin{center}
\includegraphics[width=8.5cm]{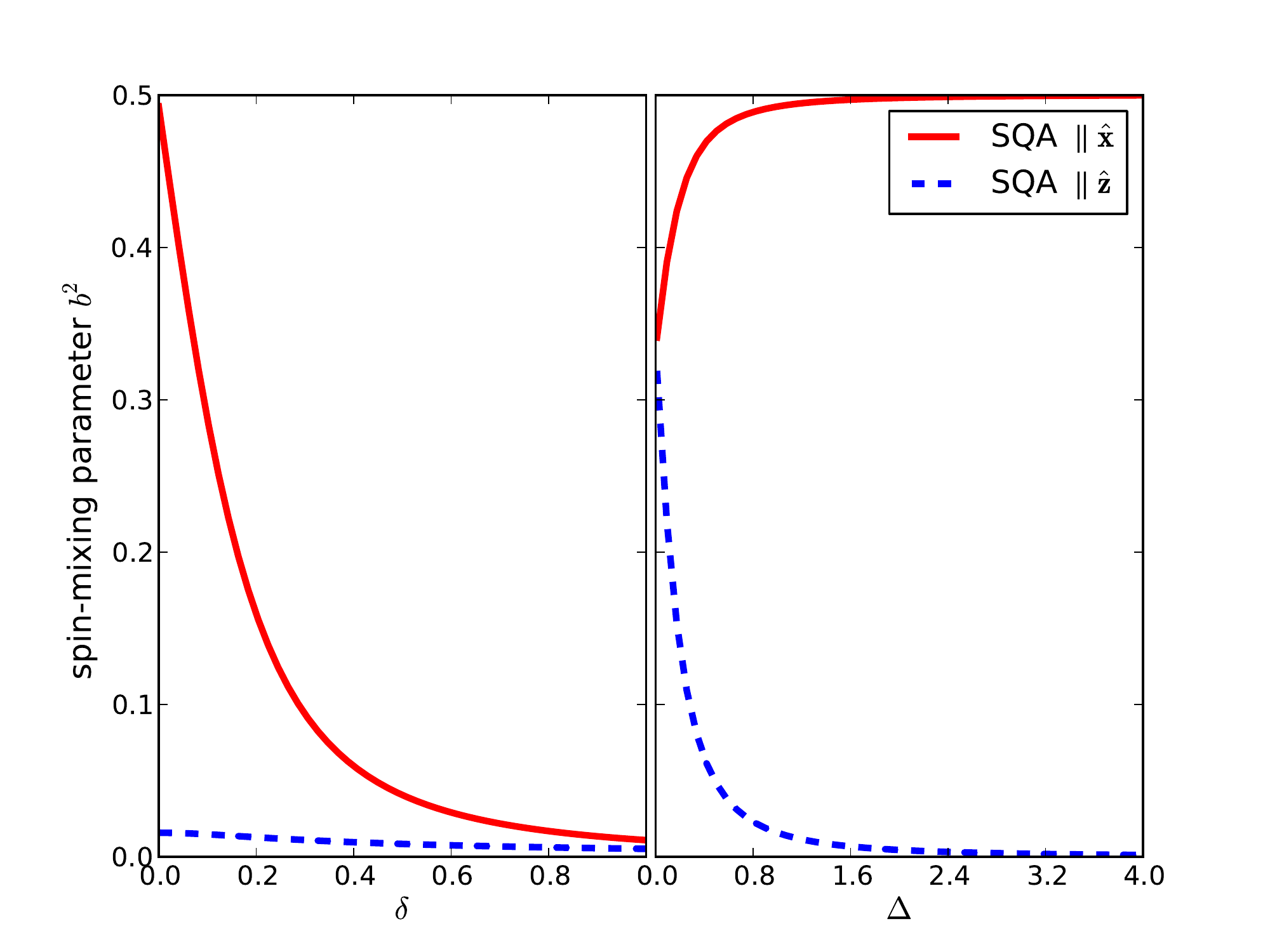}
\includegraphics[width=7cm]{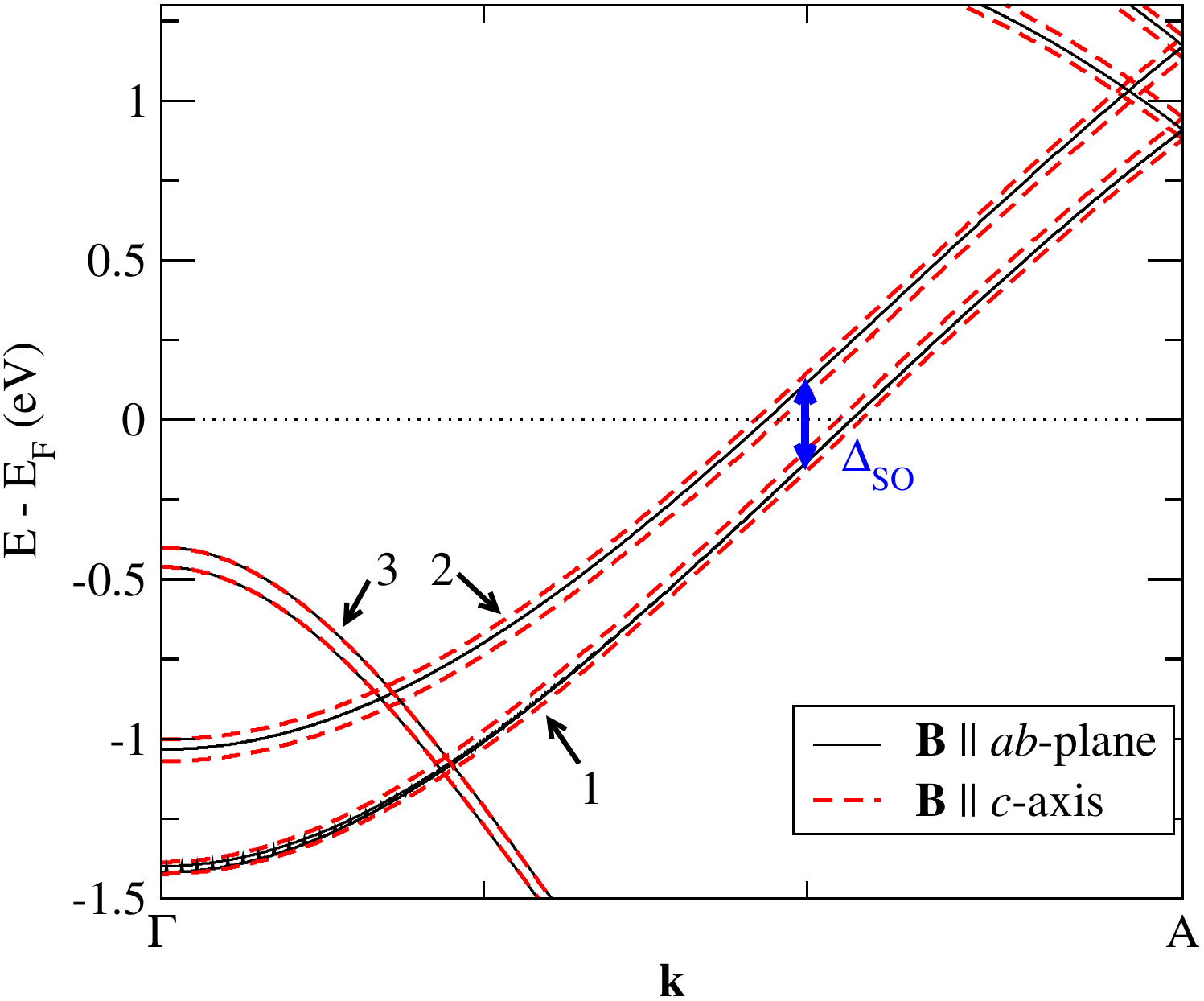}
\caption{Left: Dependence of the spin-mixing parameter of the lowest lying state in the 
model from Fig.~\ref{EYP-Model-Levels} on the separation $\delta$ (in eV) between  
nearly degenerate 
$p_x$ and $p_y$ states at constant separation with the $p_z$ orbital $\Delta=1$~eV (left), and on the 
separation of degenerate $p_x$ and $p_y$ states ($\delta=0$~eV) with the $p_z$ orbital $\Delta$ 
(in eV, right). The SOC strength of 0.1~eV was taken for these calculations. 
Clearly, the largest anisotropy of the spin-mixing parameter is acquired when the states $p_x$ and $p_y$
are perfectly degenerate and are lying far away from the $p_z$ orbital. Right: Band structure of hcp 
osmium around the Fermi level, in the direction $\Gamma-A$ of the BZ with applied $\vect{B}$-field 
of $40$~meV. The exchange splitting of the two bands crossing the Fermi energy (1 and 2)
depends on the direction of $\vect{B}$, reflecting the anisotropy of the spin-mixing at the hot-spot 
``H'' in Fig.~\ref{EY-Os-Fermi}.  
\label{EYP-Model-Anisotropy}}
\end{center}
\end{figure}

The model presented above allows us to make some  statements concerning the general 
conditions under which a large anisotropy of the spin-mixing parameter in a metal can be expected. 
First of all, crucial is the presence of a degeneracy or near-degeneracy at $E_F$, of Bloch states 
originating from the atomic orbitals $\phi_m$ and $\phi_{m'}$, with the orbital characters $|m-m'|\neq 1$, 
which are the eigenstates of the $L_{\hat{\mathbf{s}}}$ operator for some direction of the SQA (say, $z$).
In this case ($|m-m'|\neq 1$) no direct coupling is allowed between them by the spin-flip part of the SOC Hamiltonian, and the system is "protected'' against large-amplitude spin-flip transitions, since the
spin-mixing occurs due to interaction with other, energetically different, states. Correspondingly, the further away 
these other states
are from the Fermi energy, the smaller the spin-mixing parameter will be, as exemplified for our model
in Fig.~\ref{EYP-Model-Anisotropy} where at $\delta=0$ (we remind that $\delta$ stands for the 
splitting between $p_x$ and $p_y$ orbitals), the parameter $\Delta$ is varied. In this case
also relative position of the states $\phi_m$ and $\phi_{m'}$ with respect to each other is not that
important for the spin-mixing parameter, see dotted line in Fig.~\ref{EYP-Model-Anisotropy}, in which at 
constant $\Delta$, parameter $\delta$ is varied. On the other hand, for the SQA along the $x$-axis,
the spin-mixing between the nearly degenerate states $\phi_m$ and $\phi_{m'}$ is favored and reaches 
very large values, decaying as a function of the separation
$\delta$, see full line in Fig.~\ref{EYP-Model-Anisotropy}. The spin-mixing of the nearly degenerate states
with the other states is minimal on the other hand, but it reduces the overall value of the spin-mixing
parameter. The suppression of the spin-mixing parameter of the states at the Fermi energy due to 
interaction with the higher lying states is reduced, the further the latter are from the Fermi energy,
as clearly visible in Fig.~\ref{EYP-Model-Anisotropy}, in which $\Delta$ is varied at constant $\delta=0$.
Overall, by examining Fig.~\ref{EYP-Model-Anisotropy}, we conclude that the largest anisotropy
of the spin-mixing parameter in a metal will be favored when the states with $|m-m'|\neq 1$ at the 
Fermi are perfectly degenerate, and are positioned very far away from other states.

Let us now turn to hcp osmium, which we choose as an example of a typical transition metal with
a uniaxial crystal structure and the properties of which we calcualted from first principles. First, we 
take a look at the bandstructure of Os along the 
$\Gamma-A$ path  from the Brillouin zone center along the $z$-axis, presented in Fig.~\ref{EYP-Model-Anisotropy} (right).
The splitting of the two bands (each band is doubly degenerate) which cross the Fermi level here 
("1" and "2", full lines), is due to the spin-orbit interaction, as can be verified from the fact that they fall on top 
of each other when scaling down the spin-orbit coupling strength. Also, without SOC, it is straightforward 
to determine the orbital character of the bands: in this case the bands have a $d_{+1}$ and $d_{-1}$
character, which are superimposed to form the $d_{xz}$ and $d_{yz}$ states.
Overall, we have all prerequisites for a large anisotropy of the Elliott-Yafet parameter at this point
of the Fermi surface, according to the arguments presented above, since bands "1" and "2" are 
well-separated from other bands at the crossing with the Fermi level. Before proceeding with
an explicit calculation of the Elliott-Yafet parameter distribution over the whole Fermi surface
of Os, we perform a numerical experiment in order to examine the anisotropy of the response of the bands
in Fig.~\ref{EYP-Model-Anisotropy} to a small Zeeman-like field $\vect{B}$ with the magntiude of 
40~meV. The small Zeeman field which we apply by hand lifts the remanent degeneracy 
owing to the coupling to the Bloch states of the form $\vect{B} \cdot \boldsymbol{\sigma}^p$, 
which breaks the time-reversal symmetry and defines a spin-quantization axis in the direction of 
$\vect{B}$.  In Fig.~\ref{EYP-Model-Anisotropy} we clearly observe a splitting of bands
``1'' and ``2'' for $\vect{B}$ along the $c$-axis in the crystal (dashed lines). However, for 
$\vect{B}$ in the $ab$-plane, the degenerate pairs ``1'' and ``2'' do not split (solid
lines), which marks a very anisotropic response to a Zeeman magnetic field. 
We can understand this result by employing the perturbation theory arguments from before: 
in first order, the energy shift of a state due to the presence of a small Zeeman field is 
proportional to its spin polarization, which is similar to that depicted in Fig.~\ref{EYP-Model-Levels}. Correspondingly,
while in the latter case the states are fully spin mixed and the bands do not split, in the former situation
a large Zeeman splitting is achieved.  
 
\begin{figure}[t!]
\includegraphics[width=16cm]{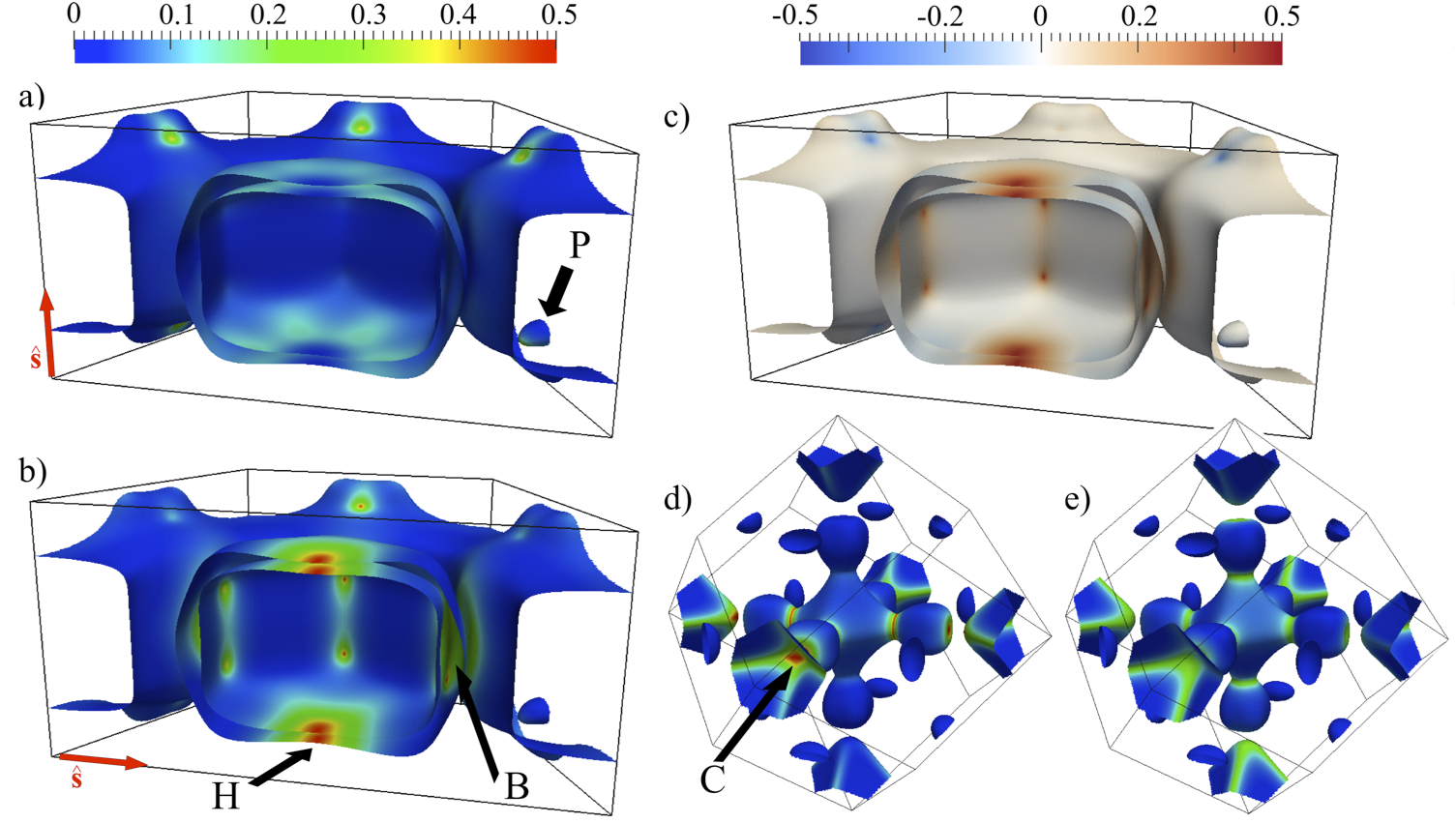}
\caption{Fermi surfaces of Os (a-c) and W (d-e). For
    an illustration of the nested sheets only half of the Fermi
    surface of Os is shown. The Elliott-Yafet parameter $\bsqks$ is
    shown with a color code on the Fermi surface with the SQA along
    the $c$-axis (a) and in the $ab$-plane (b). Red arrows at the
    left-lower corner of (a) and (b) indicate the direction of the
    SQA. The difference of $\bsqks$ between the two directions of
    $\uvect{s}$ is shown in (c). Analogously, $\bsqks$ for $\uvect{s}$
    along $[001]$ and $[111]$ in W is shown in (d) and (e), respectively.
    The averaged values of $b_{\uvect{s}}^2$ over all directions of $\uvect{s}$,
    corresponding to polycrystalline samples, are 0.0666 for Os and 
    0.0627 for W. Note that the color scale on the right refers only to (c). 
    Taken from~\cite{Zimmermann:2012}.
    \label{EY-Os-Fermi}}
\end{figure}

The calculated Fermi surface of Os presented in Fig.~\ref{EY-Os-Fermi}(a-c) consists 
of two nested sheets, a surrounding surface crossing the Brillouin zone boundary and little hole
pockets (denoted by ``P''). The latter ones are ellipsoids in an
extended zone scheme centered around a point on the  Brillouin zone
  boundary.  Analyzing the distribution of the spin mixing parameter
$\bsqks$ on the Fermi surface, we observe a strong dependence on the
SQA, evident from comparing Figs.~\ref{EY-Os-Fermi}(a) and (b). For $\uvect{s}$ along
the $c$-axis of the crystal, Fig.~\ref{EY-Os-Fermi}(a), the
spin mixing is relatively uniform ($\bsqks \approx 0.05$) for large
areas of the Fermi surface, reaching higher values near the
pockets. However, this picture changes drastically when $\uvect{s}$ is
in the $ab$-plane (Fig.~\ref{EY-Os-Fermi}(b)). In this case, the
areas with full spin mixing (red, $\bsqks \approx 0.5$) are prominent,
most clearly at the caps of the two nested Fermi-surface sheets, indicated by ``H'',
which are formed by bands "1" and "2" crossing the Fermi level in 
Fig.~\ref{EYP-Model-Anisotropy}. Additionally, large areas with smaller, but
still strong spin mixing ($\bsqks \approx 0.3$) are visible,~e.g.~in
the region denoted by ``B''. Overall, for the two considered cases
there is a strong qualitative difference in the $\vect{k}$-dependent
spin-mixing parameter $\bsqks$. 

\begin{figure}[t!]
\includegraphics[width=16cm]{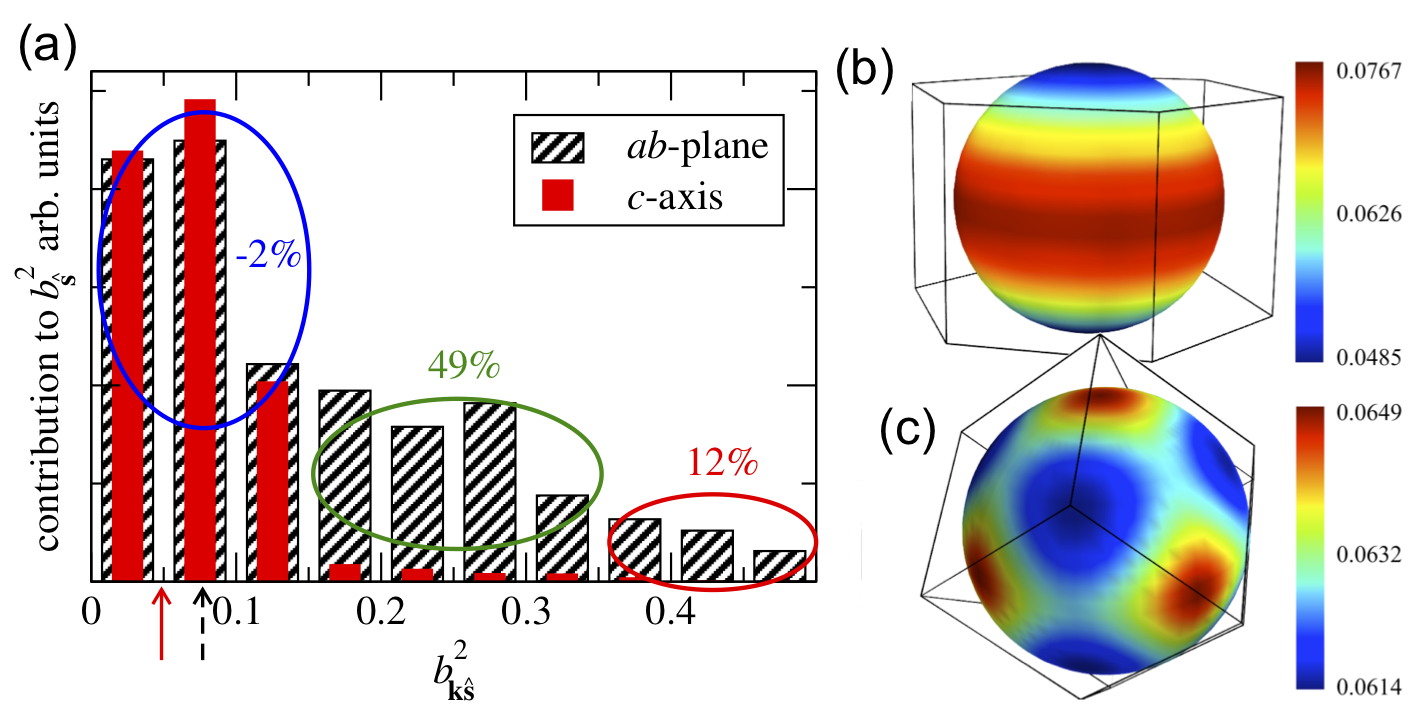}
\caption{(a) Contribution to the Fermi-surface average, $\bsq$, as function of $\bsqks$ (see text). The solid (dashed) arrow denotes $\bsq$ for the SQA along the $c$-axis (in the $ab$-plane). The numbers correspond to the respective contribution of each region to the anisotropy $\tilde{\mathcal{A}}$, leading to a total anisotropy $\mathcal{A}$ of 59\%. 
Note, that while the anisotropy of the EYP is clearly dominated by intermediate region of $\bsqks$, 
the major contribution to the EYP for both directions of the SQA comes from the region with
 $\bsqks<0.15$ (96 and 60\% for the SQA along the $c$-axis and in the $ab$-plane, respectively).
In (b) and (c),
    the integrated Elliott-Yafet parameter $\bsq$ is shown as function
    of the SQA direction for Os and W, respectively (different color
    scale). Taken from~\cite{Zimmermann:2012}.
    \label{EY-Os-Fermi-2}}
\end{figure}

 As for the Fermi-surface averaged values $\bsq$, we
find
$\bsq$ of $4.85 \times 10^{-2}$ and $7.69 \times 10^{-2}$ for
$\uvect{s}$ along the $c$-axis and the $ab$-plane, respectively,
yielding thus a gigantic anisotropy of the Elliott-Yafet parameter,
defined as $\mathcal{A} = \left[ \mathrm{max}_{\hat{\mathbf{s}}}(\bsq) -
  \mathrm{min}_{\hat{\mathbf{s}}}(\bsq) \right]/\mathrm{min}_{\hat{\mathbf{s}}}(\bsq)$, of 59\%. On the other hand, the
anisotropy $\mathcal{A}$ with respect to rotations of the SQA within
 the $ab$-plane is negligible. These two limiting
cases are depicted in Fig.~\ref{EY-Os-Fermi-2}(b), in which the value of $\bsq$ is
shown as a function of $\uvect{s}$ for all possible directions of
$\uvect{s}$. The absent (or very small) anisotropy in the $ab$-plane
is reflected in the rotationally invariant color-scale around the
$c$-axis, as opposed to the large difference between the $ab$-plane
and the $c$-axis.  The difference of $b^2_{\vect{k}}$ for the two
limiting cases of SQA for each point at the Fermi surface is shown in
Fig.~\ref{EY-Os-Fermi}(c). Large areas of the Fermi surface show
small orientational dependence of $b_{\vect{k}}^2$ (white areas). The
anisotropy at the hot spots is very large, but the sign is different
between the hot spot ``H'' and the hot spots near the pockets.
The magnitude of the effect is strongly enhanced by the large
extension of the two near-degenerate, parallel sheets of the Fermi
surface, resulting in a spin-flip "hot area'' around "H" instead of
a single "hot spot''. In addition, the reduced symmetry helps:
if the crystal had cubic symmetry, then upon change of the SQA from
$z$ to $x$ the effects at rotationally equivalent parts of the Fermi
surface would mutually cancel.

Next, we analyze the hot-spot contribution to the averaged $\bsq$ and
the anisotropy $\mathcal{A}$. We perform integrals similar to
Eq. \eqref{integral}, but restricting the integration to the part of
the Fermi surface where $\bsqks$ lies in certain intervals, $x_i <
\bsqks \leq x_{i+1}$, with $x_i=0,0.05,0.10,..$. This integration
results in values $\tilde{b}^2_{\uvect{s}}$ which form the histogram
presented in Fig.~\ref{EY-Os-Fermi-2}(a). As we can see, for the SQA along
the $c$-axis, $\bsq$ is mainly determined by regions with relatively
low spin-mixing parameter ($\bsqks < 0.15$), leading to the total
value of $4.85\times 10^{-2}$ (denoted by the solid arrow). For
$\uvect{s}$ in the $ab$-plane there is also a considerable contribution from
regions with $\bsqks > 0.15$ increasing the total value to $7.69
\times 10^{-2}$ (dashed arrow). Comparing the two histograms for
different SQA, we can draw conclusions about the respective
contribution of each region to the total anisotropy,
$\tilde{\mathcal{A}} = (\tilde{b}^2_{ab} - \tilde{b}^2_{c}) /
b^2_{c}$. Interestingly, the anisotropy originates not only from the
hot spots with $\bsqks > 0.35$ leading to $\tilde{\mathcal{A}} =
12\%$, but mainly from the areas with smaller spin mixing $0.15 <
\bsqks \leq 0.35$ around the hot spots and regions ``B'', resulting in
$\tilde{\mathcal{A}}= 49\%$. The larger area with low spin-mixing,
$\bsqks \leq 0.15$, does not contribute to the anisotropy
significantly ($\tilde{\mathcal{A}}=-2\%$).

Let us now turn to tungsten, which has a bcc lattice structure, and see whether
higher symmetry of the lattice brings qualitative changes in the anisotropy
of the EYP. When
$\uvect{s} \parallel [001]$, $\bsqks$ exhibits hot-spots in directions
perpendicular to $\uvect{s}$ (denoted by ``C'')
(Fig.~\ref{EY-Os-Fermi}(d)), following a formation scenario
similar to that at the ``$H$''-point in Os.  Additionally, many states
with smaller spin mixing ($0.2 < \bsqks < 0.3$) are present at the
Fermi surface, leading to $\bsq = 6.49 \times 10^{-2}$. For the SQA along
another high symmetry direction of the lattice, $\uvect{s} \parallel
[111]$ in Fig.~\ref{EY-Os-Fermi}(e), the intensity at the point
``C'' is reduced, but a large area with smaller spin mixing is clearly
present, resulting in $\bsq = 6.14 \times 10^{-2}$. For SQA along
$[110]$, we find $\bsq = 6.26 \times 10^{-2}$. This leads to an
anisotropy $\mathcal{A}=6\%$, which is still large but one order of
magnitude smaller than in hcp Os. This observation is similar to the
dependence of the magnetocrystalline anisotropy energy~\cite{Zener:1954} and anisotropy
of the intrinsic anomalous Hall conductivity~\cite{Roman.2009.prl} on the symmetry of
the lattice in ferromagnetic crystals: the cubic W crystal exhibits a
fourfold rotational axis, causing SOC to contribute to $\mathcal{A}$
in fourth order. In the uniaxial hcp structure, an axis perpendicular
to the $c$-axis is only twofold, and SOC enters $\mathcal{A}$ already in
second order. Nevertheless, the comparatively large anisotropy of
spin-relaxation in W is partly a consequence of the $d$-states, which
yield a strong directional anisotropy of the Fermi surface. In
contrast to this, the Fermi surface of gold consists of $s$-like
states and can be regarded as almost spherical. For the Elliott-Yafet
parameter in Au, we find a value of $\bsq \approx 3.25
\times 10^{-2}$ which is comparable in magnitude to that in W and Os, but the anisotropy is one order of
magnitude smaller than in W. Looking at the
symmetry of $\bsqks$ in W, we recognize that it is lower than the
symmetry of the lattice. And although for a spin quantization axis along
$[001]$ the fourfold rotational symmetry of the lattice around this
axis is retained by $\bsqks$, Fig.~\ref{EY-Os-Fermi}(d), further symmetry
breaking will occur for an arbitrary direction of the SQA, leaving only those point-group
symmetry operations of the lattice that map the SQA to itself, plus the inversion symmetry $\mathbf{k}\rightarrow -\mathbf{k}$. In
contrast, for the integrated value $\bsq$,
Fig.~\ref{EY-Os-Fermi-2}(c), the full symmetry of the lattice is
obviously retained.

To conclude, we underline that the spin relaxation in metals can strongly depend on the orientation of 
the injected-electron spin axis due to a corresponding anisotropy of the Elliott-Yafet 
coefficient~\cite{Zimmermann:2012}. 
The anisotropy is expected to be largest in non-cubic crystals, in the presence of extended, 
nested Fermi-surface sheets that are almost degenerate, resulting in delocalized "spin-flip hot areas" 
instead of singular "spin-flip hot spots". Especially critical are cases where the splitting is caused primarily 
by the spin-orbit coupling. Since there is no theoretical limit on the area of the nested sheets in this 
scenario, the anisotropy of the EYP can be in principal {\it colossal}, by far exceeding the values 
calculated and presented here for Os, Au and W.

\section{Anisotropy of intrinsic spin Hall effect in metals}
\label{Spin-Hall-Effect}

\begin{figure}
\includegraphics[width=16cm]{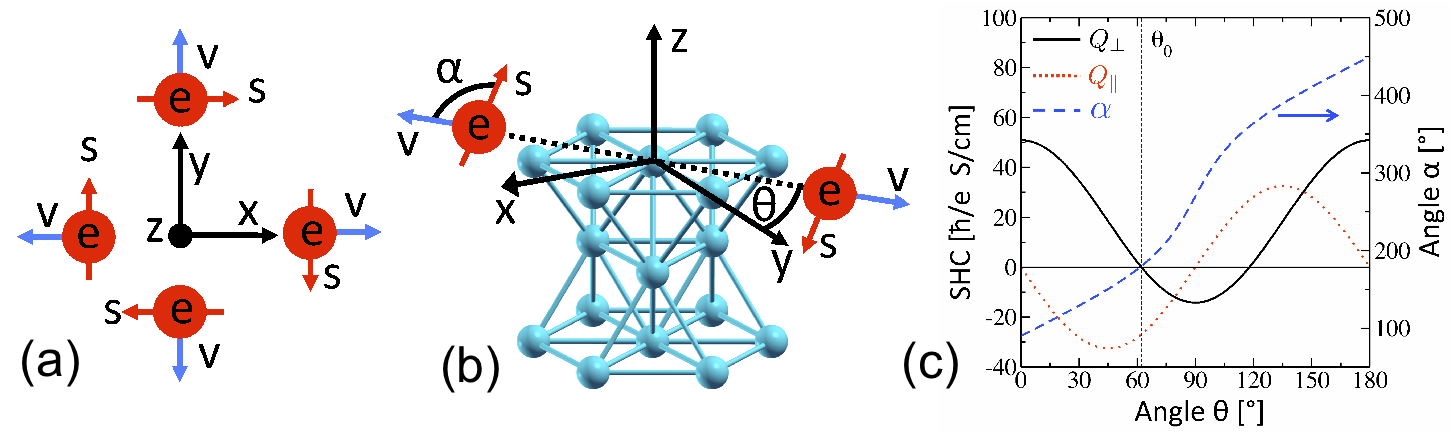}
\caption{\label{fig_she_tensor} (a) Spin currents in cubic systems induced by
an electric field along the $z$-axis. For electrons \textbf{e} with velocity \textbf{v}
in $y$-direction the
spin \textbf{s} points in $x$-direction, while for electrons going in $x$-direction
the spin points in minus $y$-direction.
(b) Hexagonal hcp structure of the transition metal Ti. The spin
current in direction $\hat{\vn{n}}=(0,\cos\theta,\sin\theta)^{T}$ induced
by an electric field in $x$-direction is not perpendicular to the
velocity \textbf{v} for a general angle $\theta$, i.e., \textbf{v} and \textbf{s}
enclose an
angle $\alpha\neq 90^{\circ}$. (c) Decomposition of the SHC of Sc into perpendicular and
parallel components following Eq.~\eqref{eq_q_perp}. The angle
$\alpha$ enclosed by the direction of the spin current and the
direction of the spin polarization is also shown. At the
angle $\theta_{0}$=62.2$^{\circ}$ the component of the spin polarization
perpendicular to the spin current vanishes and $\alpha$=180$^{\circ}$.
Taken from~\cite{Freimuth.2010.prl}.
}
\end{figure}

The spin current is characterized by
velocity and spin polarization.
Hence, the spin current density $\vn{Q}$ is a tensor in
$\mathbb{R}^{3}\otimes\mathbb{R}^{3}$
spanned by the basis vectors $\hat{\vn{e}}_{i}\otimes\hat{\vn{f}}_{k}$.
For clarity we use the symbols $\hat{\vn{f}}_{x},\hat{\vn{f}}_{y}$ and $\hat{\vn{f}}_{z}$
to denote the unit vectors of spin polarization while
$\hat{\vn{e}}_{x}$, $\hat{\vn{e}}_{y}$ and $\hat{\vn{e}}_{z}$ are
the unit vectors of velocity.
In terms of the spin Hall conductivity tensor, $\sigma_{ij}^{k}$ (which has
three indices: $i$ denotes the direction of spin current,
$j$ the direction of applied external electric field, and
$k$ the direction of spin polarization of the spin current), the spin current density 
for a general direction
of electric field is given by
\bege\label{eq_spincurrent_compact}
\vn{Q}=
\sum_{ijk}\sigma_{ij}^{k}\hat{\vn{e}}_{i}\otimes\hat{\vn{f}}_{k}E_{j}.
\ee
While the anisotropy of the AHE manifests itself in the 
dependency of the magnitude of the
conductivity vector on the magnetization direction,
in the case of the SHE in paramagnets there is no
magnetization vector $\vn{M}$ to control, only the direction of the
applied electric field can be varied. However, the spin polarization
of the induced spin current depends on the direction in which the
spin current is measured (see Fig.~\ref{fig_she_tensor}(a)). Hence, for a fixed electric field
a given spin polarization $\hat{\vn{s}}$
is measured only in a certain direction.
Thus, in analogy to
Eq.~\eqref{sigma-times-E} we may write
\bege\label{eq_define_spin_hall_conductivity_vector}
\vn{Q}^{\hat{\vn{s}}}=\vn{E}\times\vht{\sigma}(\hat{\vn{s}}),
\ee
where $\vn{Q}^{\hat{\vn{s}}}$ is the spin current density for spin polarization along $\hat{\vn{s}}$
and $\vht{\sigma}(\hat{\vn{s}})$ is the SHC vector. (We remind that while $\vn{Q}^{\hat{\vn{s}}}$ lives
in the space of basis vectors $\hat{\vn{e}}_{x}$, $\hat{\vn{e}}_{y}$ and $\hat{\vn{e}}_{z}$, the direction of the
spin polarization $\hat{\vn{s}}$ of the spin current is spanned by basis vectors $\hat{\vn{f}}_{x},\hat{\vn{f}}_{y}$ 
and $\hat{\vn{f}}_{z}$).
If the magnitude of the SHC vector depends on the
spin polarization direction $\hat{\vn{s}}$ in a material, the SHE in this material is said to be 
{\it anisotropic}. The spin Hall
conductivity vector and SHC tensor, in analogy to the anomalous Hall effect (see introduction), are related
as follows:
\bege
\sigma_{l}(\hat{\vn{s}})=\frac{1}{2}\sum_{ijk}\epsilon_{ijl}\sigma_{ij}^{k}s_{k},
\ee
where $\sigma_{l}$ is the $l$-th component of the conductivity vector,
$\hat{\vn{s}}=(s_{x},s_{y},s_{z})^{T}$ and $\epsilon_{ijl}$ is
the Levy-Civita symbol.

In cubic systems symmetry requires
that $\sigma_{ij}^{k}=\sigma_{xy}^{z}\epsilon_{ijk}$.
Thus, the SHC may be expressed in terms of
one material parameter,
Eq.~\eqref{eq_define_spin_hall_conductivity_vector} simplifies to
$\vn{Q}^{\hat{\vn{s}}}=\sigma_{xy}^{z}\vn{E}\times\hat{\vn{s}}$, and the conductivity vector is
$\vht{\sigma}(\hat{\vn{s}})=\sigma_{xy}^{z}\hat{\vn{s}}$. Since the magnitude of the
conductivity vector, $\sigma_{xy}^{z}$, is independent of $\hat{\vn{s}}$,
the SHE is isotropic in cubic systems.
The relationship between the direction of spin current and
the direction of spin polarization in cubic systems
is illustrated in Fig.~\ref{fig_she_tensor}(a). 
For the spin-mixing parameter and the anomalous Hall effect
the dependence on the direction of the SQA and magnetization in the sample, respectively, can be
more  complicated even in cubic crystals, see~e.g.~Fig.~\ref{EY-Os-Fermi-2}(c).

Let us consider now rigorously the situation of the SHE in transition metals with hcp structure, 
Fig.~\ref{fig_she_tensor}(b), keeping in mind that the following results remain valid also for
general uniaxial structures. 
If the electric field is applied along the
$x$-direction, the magnitude of the spin current in $y$-direction will generally differ
from the one in $z$-direction since the $x$-axis exhibits only 2-fold rotational symmetry.
The spin current density in direction $\hat{\vn{n}}=(0,\cos\theta,\sin\theta)^{T}$ is
\bege\label{eq_e_field_along_x}
\hat{\vn{n}}\cdot\vn{Q}=
-(
\sigma^{z}_{xy}\hat{\vn{f}}_{z}\cos\theta-
\sigma^{y}_{zx}\hat{\vn{f}}_{y}\sin\theta
)E_{x}.
\ee
Note that according to Eq.~\eqref{eq_spincurrent_compact} $\hat{\vn{n}}\cdot\vn{Q}$ is a
vector pointing in the direction of spin polarization.
We define the anisotropy of the SHE for spin polarization
in the $yz$-plane as $\Delta_{zy}=\sigma_{xy}^{z}-\sigma_{zx}^{y}$. Physically, if $\sigma_{xy}^{z}$ and 
$\sigma_{yz}^{x}$ have the same sign, parameter $\Delta_{zy}$ quantifies the difference in the magnitude 
of the spin current measured along $y$ and along $z$ axes, when the electric field points along $x$.  
For a general angle $\theta$ the components of
the spin current with spin polarization parallel
to $\hat{\vn{n}}$ ($Q_{\Vert}$) and spin polarization perpendicular
to $\hat{\vn{n}}$ ($Q_{\perp}$) are
given by
\bege
\begin{aligned}
\label{eq_q_perp}
Q_{\Vert}&=\hat{\vn{n}}\cdot \vn{Q}\cdot\hat{\vn{n}}=
-\frac{1}{2}\Delta_{zy}\sin(2\theta)E_{x},\\
Q_{\perp}&=(\sigma_{zx}^{y}+\Delta_{zy}\cos^{2}\theta)E_{x}.
\end{aligned}
\ee
If $\Delta_{zy}\neq 0$, the spin polarization is
perpendicular to $\hat{\vn{n}}$ only if $\hat{\vn{n}}$ is along
the $y$ or $z$-direction, otherwise spin polarization and direction of
spin current enclose the
angle $\alpha=\arctan(Q_{\perp}/Q_{\Vert})\neq 90^{\circ}$,
as shown in Fig.~\ref{fig_she_tensor}(b).
It follows from Eq.~\eqref{eq_q_perp} that
$Q_{\perp}$ is zero at the angle
\bege\label{eq_theta_zero}
\theta_{0}=\arccos\sqrt{-\sigma_{zx}^{y}   /\Delta_{zy}  }
\ee
if $\sigma_{xy}^{z}$ and $\sigma_{zx}^{y}$ differ in sign. At this angle $\theta_{0}$ 
the spin polarization and the spin current are collinear. This is an interesting constellation, 
which cannot occur in cubic systems. An analogous situation can also occur  
for the anomalous Hall effect,~i.e.~different sign of the AHC for two different 
high-symmetry directions of the magnetization in the crystal, 
as we show in section~\ref{abinitio}. In the latter case, there exists a direction of the 
magnetization in the crystal for which the Hall current (spin current in SHE) and the 
magnetization (spin-polarization in SHE) are collinear. Motivated by the rotational
sense of the Hall current as the magnetization direction is rotated, we call this effect 
the {\it anti-ordinary Hall effect}, see section~\ref{abinitio}.

The case of spin current in $x$-direction and
electric field $\vn{E}=(0,E\cos\theta,E\sin\theta)^{T}$ in the $yz$-plane is simply related
to the previous one by a minus sign:
The components of the spin current with spin polarization parallel and
perpendicular to
the electric field $\vn{E}$ are given by
$Q_{\Vert}=\frac{1}{2}\Delta_{zy}\sin(2\theta)E$ and
$Q_{\perp}=-(\sigma_{zx}^{y}+\Delta_{zy}\cos^{2}\theta)E_{x}$,
respectively.
At the angle $\theta_{0}$, Eq.~\eqref{eq_theta_zero}, the spin polarization
and the electric field are collinear.
Thus, one can achieve collinearity of spin polarization and electric field,
or collinearity of spin polarization and direction of spin current if
$\sigma_{xy}^{z}$ and $\sigma_{zx}^{y}$ differ in sign. For the
anti-ordinary anomalous Hall effect this means that we can find a 
direction of $\mathbf{E}$ such that the Hall current is collinear to the magnetization,
while transverse Hall current is not zero.

If the electric field is applied along the $z$-axis, the same magnitude of the
spin current will be measured in all directions perpendicular to the $z$-axis,
since the $z$-axis exhibits 3-fold rotational symmetry.
The spin current in direction $\hat{\vn{n}}=(\cos\theta,\sin\theta,0)^{T}$
is in this case
\bege\label{eq_e_field_along_z_direction}
\hat{\vn{n}}\cdot\vn{Q}=
(
\sigma_{yz}^{x}\hat{\vn{f}}_{x}\sin\theta
-
\sigma_{zx}^{y}\hat{\vn{f}}_{y}\cos\theta
)E_{z}.
\ee
Symmetry requires that $\sigma_{zx}^{y}=\sigma_{yz}^{x}$.
Consequently, the magnitude of the spin current is independent of $\theta$
and the spin polarization is perpendicular to both the electric field and
$\hat{\vn{n}}$.

In the case of the hcp structure the conductivity vector
and the spin current density, Eq.~\eqref{eq_define_spin_hall_conductivity_vector}, may be
expressed in terms of the anisotropy as
\bege
\label{eq_conductivity_vector_spin_polarization}
\begin{aligned}
\vht{\sigma}(\hat{\vn{s}})&=\sigma_{yz}^{x}\hat{\vn{s}}+(0,0,\Delta_{zy}s_{z})^{T},\\
\vn{Q}^{\hat{\vn{s}}}&=
\sigma_{yz}^{x}\vn{E}\times\hat{\vn{s}}+\Delta_{zy}s_{z}(E_{y},-E_{x},0)^{T}.
\end{aligned}
\ee
Hence, only two parameters, $\sigma_{yz}^{x}$ and $\Delta_{zy}$, suffice
to describe the SHE in hcp nonmagnetic metals. The fact that one needs only two parameters
to reconstruct the exact analytical dependence of the spin polarization 
on the direction in which the spin current is measured is a manifestation
of the {\it geometrical anisotropy} of the SHE. This is a major difference to the 
anomalous Hall effect and spin relaxation, for which the conductivity vector and the EYP
have to be recalculated anew for each direction of the magnetization and SQA, 
since the EYP and the AHC for a general direction of the SQA/magnetization
cannot be related to the corresponding values for the high-symmetry axes.
For example, in case of the AHE,
already four parameters are needed for an approximate
expansion of the conductivity of hcp crystal up to third order in the directional 
cosines~\cite{Roman.2009.prl}. 

\begin{figure}[t!]
\begin{center}
\includegraphics[width=15cm]{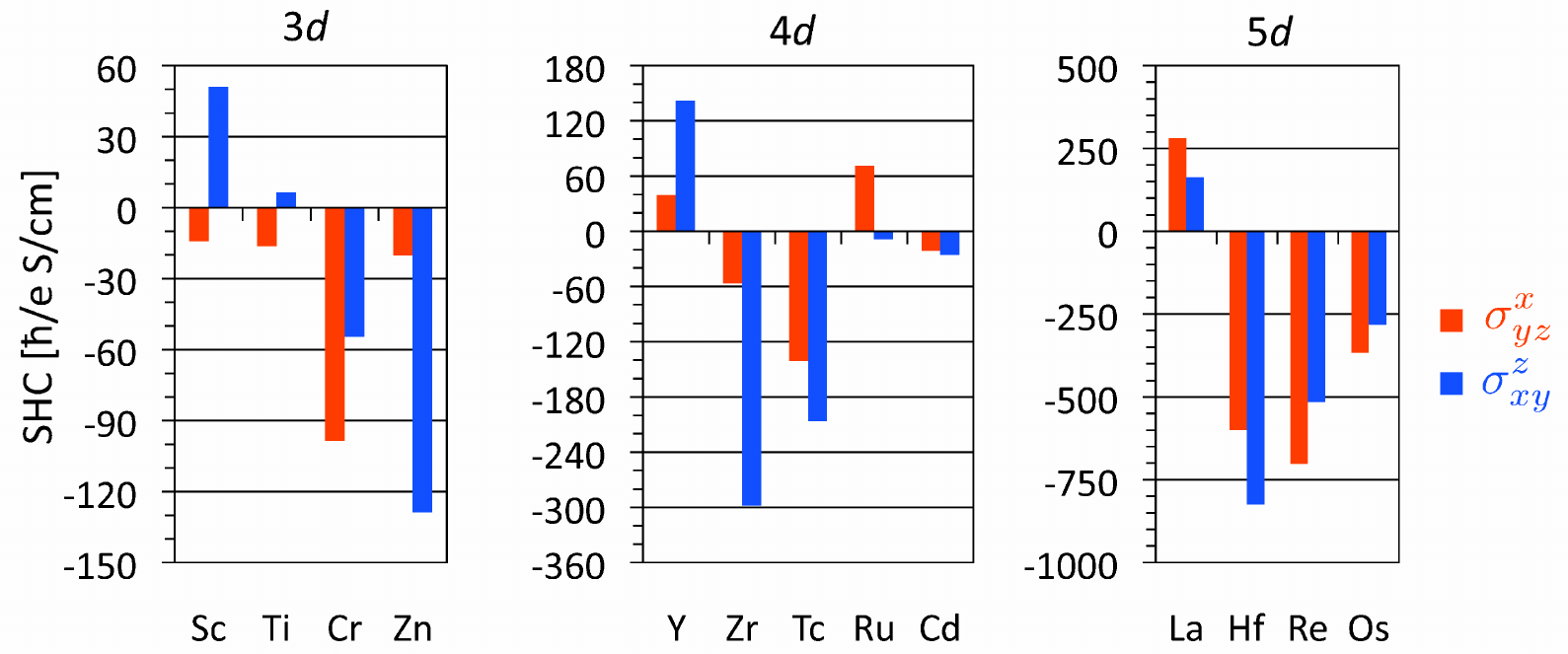}
\end{center}
\caption{\label{conductivies_anisotropic_she} For the hcp metals
Sc, Ti, Zn, Y, Zr, Tc, Ru, Cd, La, Hf, Re and Os and for
antiferromagnetic
Cr the spin Hall conductivities
$\sigma^{x}_{yz}$ and $\sigma^{z}_{xy}$ are shown as light (red) and
dark (blue) bars, respectively. Taken from~\cite{Freimuth.2010.prl}.}
\end{figure}

Next, we present in Fig.~\ref{conductivies_anisotropic_she} the results of  first principles calculations of the intrinsic SHC, 
Eq.~\eqref{eq:kubo:shc}, for the hcp metals Sc, Ti, Zn, Y, Zr, Tc, Ru, Cd, La, Hf, Re and Os 
and for antiferromagnetic Cr (see section 2 for computational details). In the case of Cr we neglected 
the spin-density wave and considered 
the antiferromagnetic structure with two atoms in the unit cell and with the magnetic moments 
parallel and antiparallel to the $z$-axis. 
Except for Cd all metals studied in this work exhibit a large anisotropy of SHE, which we expect 
to be clearly visible in experiments. Of particular interest are the hcp metals Sc, Ti and Ru, where 
the sign of the conductivity changes as the spin polarization is rotated from the $z$-axis into the 
$xy$-plane. As discussed before, collinearity of the spin polarization and the electric field
(or the spin polarization and the spin current) may be achieved if the electric field (the spin current) 
lies in the $yz$-plane at the angle $\theta_{0}$, Eq.~\eqref{eq_theta_zero}, from the $y$-axis.
To illustrate this we plot in Fig.~\ref{fig_she_tensor}(c) the angle $\alpha$ enclosed by the direction 
of the spin 
current and the direction of the spin polarization as well as the SHCs associated with $Q_{\Vert}$ 
and $Q_{\perp}$ (see Eq.~\eqref{eq_q_perp}) as a function of the angle $\theta$ for Sc.
The critical angles at which the perpendicular component of the spin polarization
vanishes are $\theta_{0}$=62.2$^{\circ}$, $\theta_{0}$=32.1$^{\circ}$, and 
$\theta_{0}$=19.1$^{\circ}$ for Sc, Ti, and Ru, respectively. Note that in case of Ru we have the case of a {\it colossal anisotropy} of the SHE: the values of the two 
calculated conductivities differ by an order of magnitude. In the case of Cr the SHE is anisotropic 
as the cubic symmetry is broken by the staggered magnetization: If the spin polarization of the spin 
current is perpendicular to the staggered magnetization the SHC is larger by a factor of 1.8 compared 
to the case of spin polarization parallel to the staggered magnetization. We can thus claim that in
antiferromagnets the direction of the local spins presents an additinal channel for the SHE anisotropy.
Such anisotropy is not anymore geometrical, however, due to the dependence of the electronic structure 
on the direction of local magnetization, similarly to the case of the AHE in ferromagnets.

Generally, a simple analysis of the SHC and its anisotropy in terms of a simple model becomes
very difficult, since (i) the integrand in Eq.~\eqref{eq:kubo:shc} varies very strongly as a function 
of $\vn{k}$ (see for example Fig.~\ref{FePt-berry}) and the entire Brillouin zone has to be
considered in the integration in order to reproduce the SHC
quantitatively correctly; (ii) for the anisotropy of the SHC not only the anisotropy of the
wavefunctions with respect to the SQA, discussed in the previous section, has to be taken
into account, but also the anisotropy of the velocity matrix elements has to be
necessarily accounted for. This makes it hardly possible to interpret the spin Hall conductivity 
in terms of a small number of virtual interband transitions. Even the sign and order of magnitude 
of the SHC are difficult to predict based on simple arguments. 

One aspect we would like to remark
on is the importance of transitions in Eq.~\eqref{eq:kubo:shc} which are driven by spin-flip
SOI, and the difference between the AHE and SHE as far as the anisotropy of the conducitivities
is concerned.
Let us consider a situation of two doubly-degenerate Bloch states at a certain $k$-point, occupied $\psi_n$ and 
unoccupied $\psi_m$. Let us also assume that these states
are well-separated in energy,~i.e., the first-order perturbation 
theory as given by equation~(\ref{EY-WF-PT}), applies. In this case, consider the contribution 
to the spin Berry curvature $\Omega_{xy}^z$ which comes from the products of the type:
\begin{equation}
\sim \langle \psi_m | v_x \sigma^p_z| \psi_n \rangle
\langle \psi_n | v_y                | \psi_m \rangle = 
\langle a_m^{\uparrow} + b_m^{\downarrow} | v_x \sigma^p_z| a_n^{\uparrow} + b_n^{\downarrow} \rangle
\langle a_n^{\uparrow} + b_n^{\downarrow} | v_y| a_m^{\uparrow} + b_m^{\downarrow} \rangle,
\end{equation}
where the SQA is chosen along the $z$-axis, and $a_m^{\uparrow}= a_m|\uparrow\rangle$, etc., 
according to the expansion~(\ref{Elliotts}). If we neglect the relativistic correction to the velocity
operator (as our {\it ab initio} calculations show it is a very good approximation in most of the cases),
the velocity operator does not couple states of different spin  and the spin mixing parameter
enters with the terms of the order of $b_m b_n $, which means that the spin-flip spin-orbit
appears only in contributions to the spin Berry curvature which are proportional to $\xi^2$ and higher
even powers of $\xi$ (we remind that $\xi$ is the SOC strength in the system). 
It is clear that in this case the dominant contribution to the SHC comes from $\sim \xi$
 spin-conserving transitions. We prove a similar result in the next section for the anomalous Hall
 effect.

\begin{figure}[t!]
\begin{center}
\includegraphics[width=15cm]{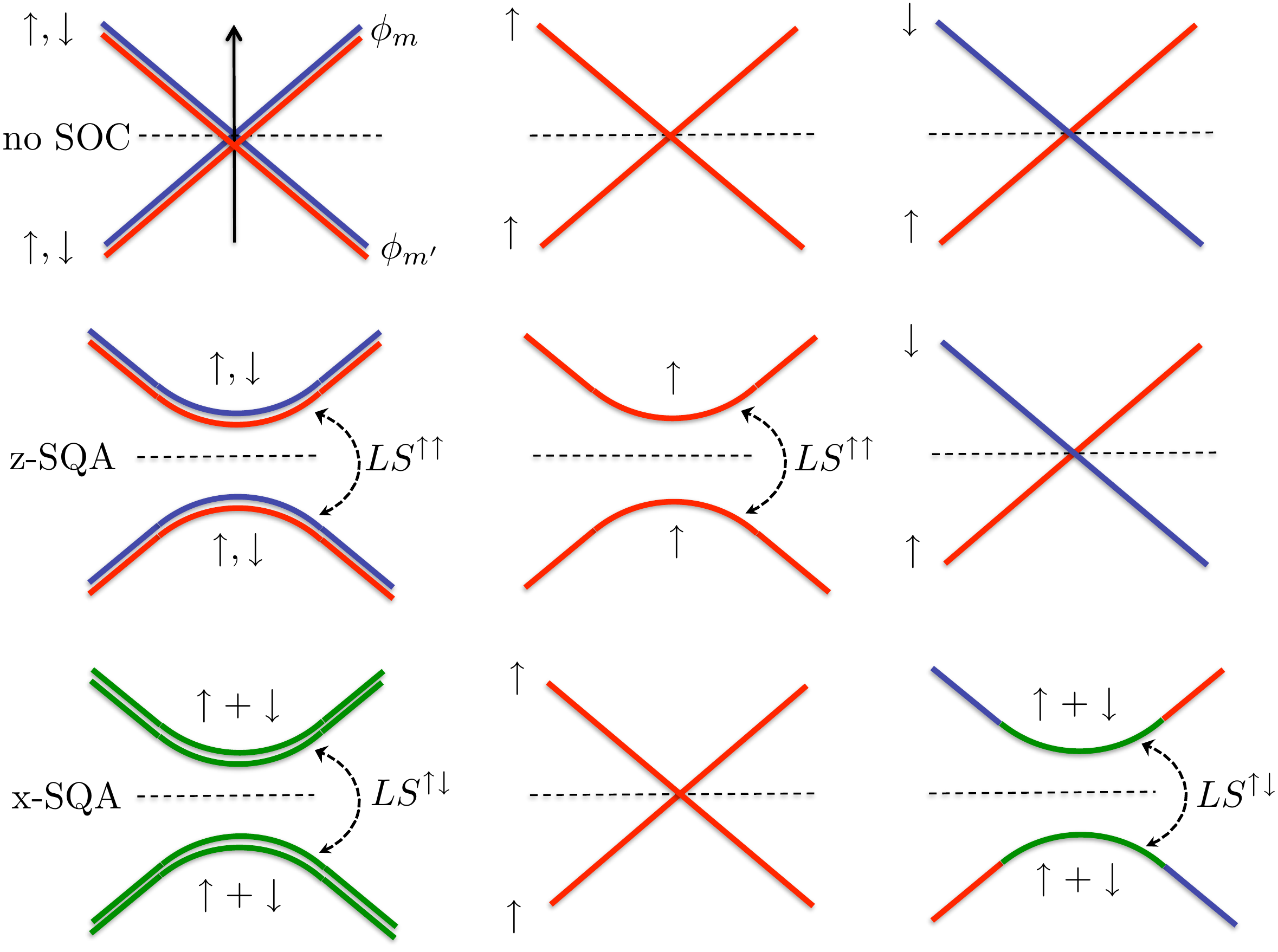}
\end{center}
\caption{\label{Crossings} Left column: a band degeneracy due to bands of $m$ and $m'$ orbital character, $|m-m'|\neq 1$,
in a paramagnet with structural inversion symmetry without SOC, is lifted due to spin-conserving SOC for SQA along $z$,
and by spin-flip SOC for SQA along $x$. In a ferromagnet the band degeneracies between the bands of the same spin character
(middle column), and of the opposite spin character (right column) have to be considered instead. Note that in this case the 
energy shifts due to SOC depend strongly on the direction of the magnetization and the type of the crosssing.
The red (blue) color stands for spin-up (-down) character of the states, while 
green color marks the states which are of essentially mixed spin character.}
\end{figure}

While in the largest part of the Brillouin zone the SHC originates mainly from the spin-conserving
SOC, in the vicinity of a degeneracy (or crossing) point, such as depicted in Fig.~\ref{Crossings} (left, no SOC), both
spin-flip and spin-conserving SOC  can provide very large contributions to the SHC. Generally speaking,
depending on the orbital character of the states which cross, the role of the spin-conserving and
spin-flip SOC for the SHC around such points can be interchanged by changing the SQA. Consider for 
example a situation from the previous section, where the (doubly-degenerate) bands which cross have 
a dominant $m$ and $m'$ orbital character, $|m-m'|\neq 1$, and are well separated from other states,
Fig.~\ref{Crossings} (left column). In this case for the SQA along $z$ the degeneracy between the states
is lifted by $LS^{\uparrow\uparrow}$, and the states keep their almost pure spin character. On the other hand,
by pointing the SQA along $x$, the degeneracy is lifted due to $LS^{\uparrow\downarrow}$, and the
states become strongly mixed in spin. Such anisotropy of the wavefunctions will contribute to the 
anisotropy of  the SHC, but what also has to be taken into account is that while in the first case the $v_x\sigma^p_z$
and $v_y$ velocity operators have to be considered in the expression above, in the second case they have
to be replaced with $v_y\sigma^p_x$ and $v_z$. One has to realize that for a $k$-point away from high symmetry
directions in the Brillouin zone and for the bands which have mixed orbital character the mixture of $LS^{\uparrow\uparrow}$
and $LS^{\uparrow\downarrow}$ in the spin-orbit Hamiltonian  can be complicated and the Berry curvature can vary in a 
non-trivial
fashion with the SQA. Nevertheless, by looking at the situation depicted on the left of Fig.~\ref{Crossings},
it becomes intuitively clear why the anisotropy of the SHE in paramagnets with structural inversion symmetry is
geometric: upon rotation of the SQA the spin-conserving part of SOI is continuously rotated into the 
spin-flip part, while the energy spectrum remains unchanged and the crystal basically remains "the same"
system. Indeed, none of the gound state properties of such a crystal are sensitive to the direction of the SQA,
and it is the non-equilibrium nature of the spin-relaxation and transport phenomena which makes them sensitive to it. 
The condition for such a continuous transformation between $LS^{\uparrow\uparrow}$ and 
$LS^{\uparrow\downarrow}$ obviously lies in the availability of both spin-up and spin-down states for 
each $k$-point and energy,~i.e.,~the spin degeneracy.

In a ferromagnet, the spin degeneracy is lifted due to the presence of the magnetization. In this case, since the
spin-up and spin-down subspaces are separated in energy, the spin-conserving SOC does not transform continuously
into the spin-flip SOC at a given $k$-point and energy when the magnetization is rotated, and two types of band 
crossings (or degeneracies) without SOC should be considered: between bands of the same, and of the opposite spin,
see Fig.~\ref{Crossings}. Depending on the orbital character of the states, the degeneracy between them will be lifted in
the first case by $LS^{\uparrow\uparrow}$ for one direction of the magnetization only (say, $z$), while in the second case 
it will be  lifted by $LS^{\uparrow\downarrow}$ only for another (say, $x$), see Fig.~\ref{Crossings}. Simply speaking,
since the position of the two types of degeneracies in energy and in the Brillouin zone, as well as their number, is different 
in a ferromagnet, effectively, for the two different directions of the magnetization we have two different systems with 
different energy spectrum and different eigenstates. In particular,  such asymmetry is the reason for 
the anisotropy of the orbital moments and total energy in ferromagnets. Since the difference in,~e.g., eigenspectrum for the 
two different magnetization directions can be hardly reconstructed analytically due to the complexity of the Halmiltonian in
transition metals, this leads to a complicated behavior of the AHC as a function of the direction of the magnetization in the crystal. 
The same holds true for the Elliott-Yafet parameter, considered in the preceeding section, in which case the time-reversal symmetry 
in the system is effectively broken by a certain spin direction of the injected electron, which is able to "probe" the spin-mixing
parameter of the states. 

Finally, we would like to make two remarks. The first one concerns the topology of the degeneracy points in the 
Brillouin zone. The type of degeneracy shown in the upper row of Fig.~\ref{Crossings} provides the so-called
"monopole" contribution to the Berry curvature~\cite{AHE-RMP}, intensively discussed in the literature, especially 
with respect to topological insulators~\cite{Zhang-RMP}. Such degeneracies arise at single (often high-symmetry) 
points in the Brillouin zone, as well as along so-called "hot loops"~\cite{Zhang.2011.prl,Zhang:2012:Bi}. 
Another important contribution to the Berry curvature can be also given by transitions 
between the pairs of parallel bands degenerate along whole (often high-symmetry) lines or even areas in the Brillouin 
zone $-$ these are the so-called "ladder transitions"~\cite{Zhang.2011.prl}. In the language of spin-relaxation, 
such transitions would occur at the spin-flip hot areas in the vicinity of the Fermi surface. The conclusions of the 
discussion above hold true for both cases. Secondly, it is important to underline that although, referring to the 
perturbation theory arguments, the same matrix elements of SOI enter into the expressions for energy shifts, 
Hall conductitivities and spin-mixing parameter, these expressions are fundamentally different. This means that, for 
example, even though the band degeneracy in Fig.~\ref{Crossings} would be lifted by the spin-conserving SOC, it can 
happen that the major contribution to the Berry curvature is provided by the spin-flip SOI, and the other way around.
Concerning this issue, see also the discussion around second-order perturbation theory expression for the AHC
in section~\ref{Pert-Theo}.

\section{Anisotropy of intrinsic anomalous Hall effect in metallic ferromagnets}

\subsection{Anisotropic AHE in uniaxial ferromagnets: first principles studies}
\label{abinitio}

Experimentally, the $-$ sometimes strong $-$ anisotropy of the anomalous Hall effect in metals is a rather well-known
phenomenon:  see,~e.g., experimental data for bcc Fe~\cite{Weissman:1973}, fcc Ni~\cite{Volkenshtein:1961,Hiraoka:1968}, hcp Gd~\cite{Lee:1967}, as well as FeCr$_2$S$_4$~\cite{Ohgushi:2006},
Yb$_{14}$MnSb$_{11}$~\cite{Sales:2008}, Y$_2$Fe$_{17-x}$Co$_x$~\cite{Skokov:2008}
and $R_2$Fe$_{17}$ ($R$ = Y, Tb, Gd)~\cite{Stankiewicz:2011}. Theoretically, the first argumentation
for a strongly anisotropic behavior of the AHC in transition metals was provided by Roman {\it et al.}~\cite{Roman.2009.prl}.
In that work it was argued that the main reason for the observed anisotropy of the intrinsic
AHC in uniaxial hcp cobalt, which reaches as much as a factor of four between the conductivities
for the magnetization in-plane and out-of-plane (see Table~\ref{AHE-TMs}), in agreement to experiment, 
lies in the irregular 
and spiky behavior of the Berry curvature in the reciprocal space. The main
conclusions and analysis presented in the following three subsections holds true also for the case
of the spin Hall effect.

\begin{figure}
\begin{center}
\includegraphics[width=13cm]{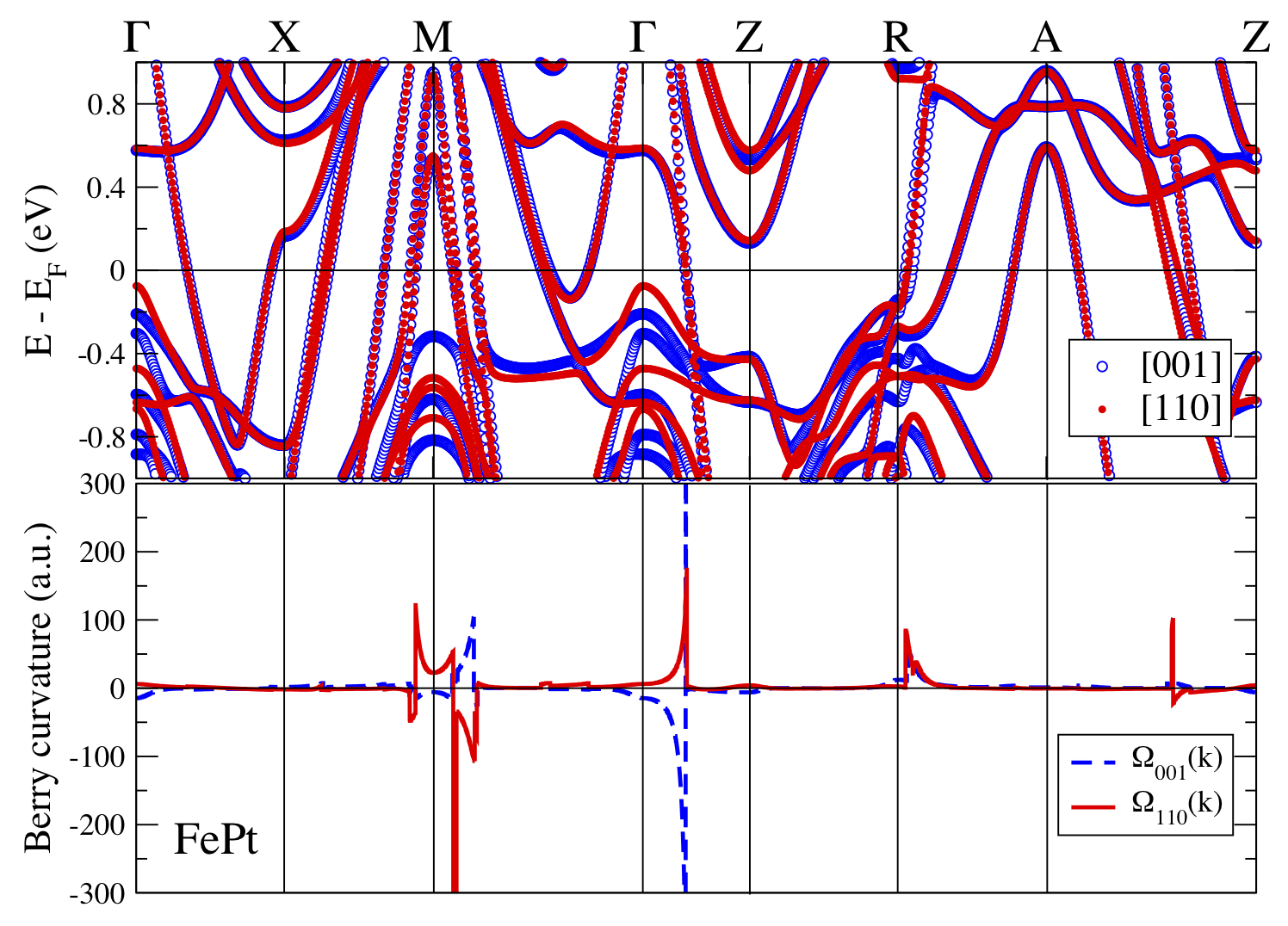}\vspace{0.5cm}
\includegraphics[width=12.9cm]{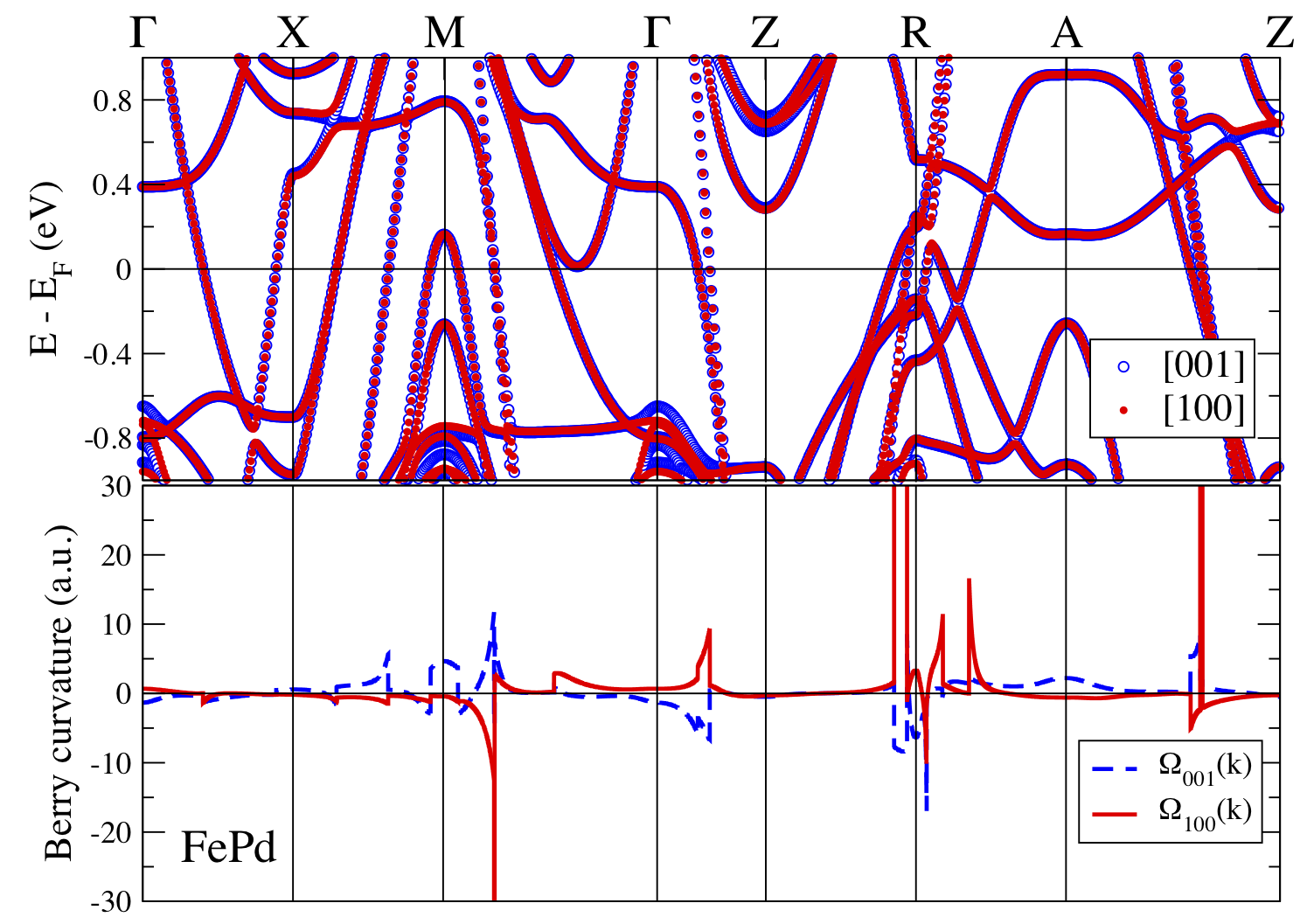}
\caption{Bandstructures and distribution of the Berry curvature along high symmetry 
directions in the Brillouin zone for the magnetization along [001] and [110] crystallographic 
axes in L1$_0$ FePt and FePd. Here $\Omega_{[ijk]}$ marks the (only non-vanishing) component 
of the Berry cruvature vector along the direction of the magnetization $[ijk]$.
\label{FePt-berry}}
\end{center}
\end{figure}

\begin{figure}[t!]
\begin{center}
\includegraphics[width=16cm]{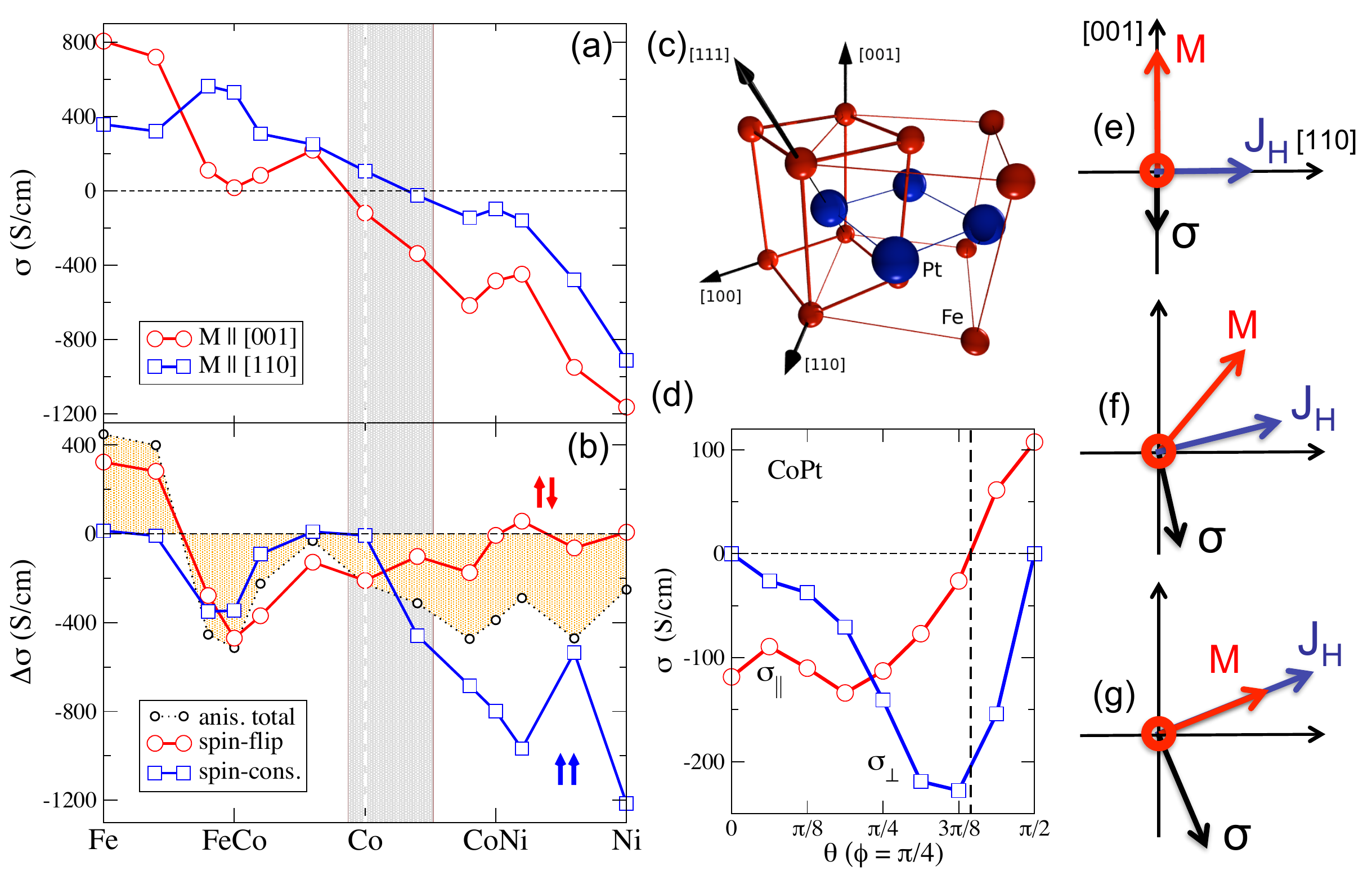}
\caption{Anomalous Hall conductivity of $3d$Pt alloys, (a),
for $\mathbf{M}$ along [001] ($\sigout$, open circles) and [110] ($\sigin$,
open squares), and anisotropy, (b) ($\Delta\sigma^{tot}=\sigout - \sigin$, small open 
circles), with respect to the band filling mimicked within the virtual crystal approximation. 
(c) Crystal structure of L1$_0$ FePt alloy. 
Small (red) spheres stand for the Fe atoms, while large (blue) spheres mark the Pt ions. 
The primitive unit cell used in the calculations is enclosed with thicker lines. In the text, 
$z$ stands for the [001] axis, while $x$ stands for the [110] direction in the crystal.
(d) Anti-ordinary Hall effect in CoPt. Red circles (blue squares) denote the 
$\sigma_{\parallel}$ ($\sigma_{\perp}$) component of AHC, as a function of the 
angle $\theta$ of the magnetization $\mathbf{M}$ with [001]-axis upon
rotating it into the [110] direction. (e)-(g) depict the relative orientation of the Hall current 
$\mathbf{J}_H$, AHC $\boldsymbol{\sigma}$ and magnetization $\mathbf{M}$ in the situation 
of the anti-ordinary AHE. In (c)-(e) the magnetization is confined to the ($\bar{1}$10)-plane.
We note, that due to symmetry considerations, the conductivity vector stays together with
the magnetization in the ($\bar{1}$10)-plane.
Taken from~\cite{Zhang.prb.2011}.
\label{Anti-Ordinary}}
\end{center}
\end{figure}

Let us take a look at the distribution of the Berry curvature along the high symmetry lines in 
the Brillouin zone for another uniaxial ferromagnet, L1$_0$ FePt (see Fig.~\ref{Anti-Ordinary}(c) for
a sketch of the structure and definition of the crystallographic directions), presented in 
Fig.~\ref{FePt-berry}. The characteristic spikes 
in the vicinity of points of near-degeneracy across the Fermi energy can be seen,~e.g., around 
the $M$-point or in the middle of the $\Gamma Z$-path. As we shall see in the next section,
similarly to the case of the Elliott-Yafet parameter, the large values of the Berry curvature in the
vicinity of such points can be inevitably related to the matrix elements of the spin-orbit interaction
between the occupied and unoccupied states, scaled by the energy difference between them. As 
discussed for the case of the EYP and SHE, those matrix elements are 
strongly anisotropic with respect to the SQA, or, direction of $\mathbf{M}$, resulting in  the
remarkable anisotropy of the Berry curvature in Fig.~\ref{FePt-berry}, both in magnitude (in the
middle of the $\Gamma Z$-path) and sign (close to $M$-point). When integrated over the whole
Brillouin zone, the anisotropy of the Berry curvature leads to a factor-of-two reduction in the
AHC in FePt as the magnetization is changed from out-of-plane to in-plane (see Table~\ref{AHE-TMs}). 
In general, in cubic
crystals, the anisotropy of the AHE with respect to the directional cosines of the magnetization
appears in all odd (owing to the anti-symmetricity of $\boldsymbol{\sigma}$ with respect to 
$\mathbf{M}$) orders starting from the third, and it is normally much weaker than that for the 
uniaxial crystals, for which the anisotropy is present already in the first order~\cite{Roman.2009.prl}, 
compare~e.g.~values for bcc Fe to the ones for the uniaxial ferromagnets in Table~\ref{AHE-TMs}.

\begin{table}[t!]
\begin{center}
\begin{tabular}{l|rrrrrr}
                            & bcc Fe & hcp Co & FePd& FePt & CoPt &  NiPt \\ \hline
 $\mathbf{M}\Vert [001]$ &  767     &  477         &  135     &  818    &    $-$119 &  $-$1165\\ 
  $\mathbf{M}\Vert [100]$ &       &     100     &       &             &      &  \\ 
$\mathbf{M}\Vert [110]$ &  810     &            &  276    &  409     &   107  &  $-$914 \\
$\mathbf{M}\Vert [111]$  &  842         &            &        &             & \\
\end{tabular}
\caption{Calculated from first principles anomalous Hall conductivity as a function of direction
of magnetization in the crystal. The data are taken from Ref.~\cite{Weischenberg.2011.prl} for bcc Fe and hcp Co
(similar values for hcp Co were obtained in Ref.~\cite{Roman.2009.prl}),
from Refs.~\cite{Seemann:2010,Zhang.2011.prl} for L1$_0$ FePt and FePd, and from Ref.~\cite{Zhang.prb.2011} for
L1$_0$ CoPt and NiPt. 
\label{AHE-TMs}}
\end{center}
\end{table}

When we compare the anisotropy of the AHE to the anisotropy of the Elliott-Yafet parameter in metals, 
several things come to mind. Firstly, the anisotropy of the AHC is a more complex quantity, which hinders
analysis in terms of a simple line of arguments, as it can be done for the anisotropy of the EYP. This is
due to the fact that while for the emergence of the EYP only spin-flip part of SOC plays a role and the
transitions between the spin-degenerate bands can be ignored, for the AHC both spin-conserving and spin-flip
parts of SOI have to be taken into account when transitions between occupied and unoccupied bands
of various orbital and spin character varying in the Brillouin zone have to be considered according 
to Eq.~(\ref{eq:kubo}) for the AHC (see also considerations at the end of section~\ref{Spin-Hall-Effect}). 
We analyze this in more detail in the following two sections. Secondly,
in addition to the anisotropy of the SOI matrix elements, which leads to the anisotropy of the wavefunctions and eigenenergies, similarly to the SHE also anisotropy of the {\it velocity} matrix elements
matters for the total value of the AHC anisotropy. Finally, the Berry curvature is not 
confined to the Fermi surface, but has a finite spread in energy. While it is already clear 
from Eq.~(\ref{eq:kubo}), in order to further clarify this point we refer to the distribution of the Berry curvature 
for L1$_0$ FePd alloy in Fig.~\ref{FePt-berry}. In this plot, the presence of wide regions in $k$-space
is evident for which the Berry curvature arises due to transitions between bands well-separated in energy. 
Since in such regions the Berry curvature also displays a very anisotropic behavior (see also
Table~\ref{AHE-TMs}) it seems reasonable to ask on whether there is a certain threshold 
in energy beyond which the transitions between bands can be neglected for the anisotropy of the AHC. 

In order to answer this question, following Roman {\it et al.}~\cite{Roman.2009.prl}, we introduce the cumulative
anomalous Hall conductivity $A(\omega)$, which accumulates all transitions in Eq.~(\ref{eq:kubo})
for which $\varepsilon_{n\mathbf{k}}-\varepsilon_{m\mathbf{k}}$ is larger than $\hbar\omega$. In the
limit of $\omega\rightarrow 0$ all transitions in Eq.~(\ref{eq:kubo}) are accounted for, and 
$A(\omega=0)$ equals the full AHC. A closer inspection of the cumulative AHC presented as a function
of energy and magnetization direction in Fig.~\ref{fig:mcd} (left) for FePt, CoPt and NiPt (the latter two also 
exhibit a very large anisotropy of the AHC, see Table~\ref{AHE-TMs}), as well as for hcp Co in 
Ref.~\cite{Roman.2009.prl}, shows that the energetical distribution of transitions which provide the anisotropy 
of the AHC is concentrated in a narrow 1~eV window around the Fermi energy. This is in contrast 
to the distribution of $A(\omega)$ for each of the magnetization directions separately, which decays slowly
over a much larger energy scale of several eV. This means that for the anisotropy of the AHC the interband
transitions in the close vicinity of the Fermi level, which give very large contributions to the AHC, are more 
important than for the values of the AHC themselves. In principle, this is in agreement to the results for the EYP 
we obtained in Os, Fig.~\ref{EY-Os-Fermi-2}: there, the contributions from the areas where the spin-mixing parameter (Berry curvature in the case of the AHE) was enhanced, are dominant,
while for the values of the EYP themselves the regions with smaller spin-mixing parameter are more important,
for each of the directions of the SQA, see caption to Fig.~\ref{EY-Os-Fermi-2}.

Generally, the AHC, as well as its anisotropy,  displays a strong dependence on the exact position of 
band (near-) degeneracies with respect to the position 
of the Fermi energy. We therefore expect a non-trivial behavior of the AHC anisotropy for a 
ferromagnet with a complex electronic structure when the Fermi energy, or other parameters, such
as the lattice constant or exchange splitting, are smoothly varied. An example of this phenomenon
can be seen in Fig.~\ref{Anti-Ordinary}(a), in which the AHC and its anisotropy are plotted as a function
of the "band filling" in L1$_0$ $3d$Pt alloys (see section~\ref{comp_details} for details of the calculations).
In this plot we observe that when going from FePt to NiPt the AHC anisotropy undergoes a change in
sign and large changes in magnitude.
With grey shaded area in Fig.~\ref{Anti-Ordinary}(a) the region around the
CoPt alloy is highlighted, where both AHC for $\mathbf{M}\Vert [001]$ ($\sigout$) and $\mathbf{M}\Vert [110]$ ($\sigin$) change their sign. This sign change leads
to the occurrence of two key phenomena with respect to the anisotropic AHE.
The first one $-$ the {\it colossal anisotropy} of the AHE $-$ according to
calculations in Fig.~\ref{Anti-Ordinary}, occurs for Fe$_x$Co$_{1-x}$Pt alloy with $x\approx 0.1$ 
and for  Co$_x$Ni$_{1-x}$Pt alloy with $x\approx 0.85$. For these two compounds one of the 
conductivities crosses zero, which marks the complete disappearance of the intrinsic anomalous Hall
current $\mathbf{J}_H$ for one of the magnetization directions in the crystal. This is reminiscent
of the situation for the spin Hall effect in Ru, see Fig.~\ref{conductivies_anisotropic_she}. In terms of the longitudinal transport within the setup of~e.g.~anisotropic magnetoresistance (AMR) experiment~\cite{Seemann:2011}, 
the occurrence of the colossal anisotropy of the diagonal conductivity would results in a metal-insulator transition in the crystal $-$ in case of the
colossal AHE anisotropy observed in $3d$Pt alloys all compounds remain metallic for all
magnetization directions, however, and retain their complicated electronic
structure around the Fermi energy.

For CoPt alloy the situation, depicted in Fig.~\ref{Anti-Ordinary}(d-g), is completely different. 
Remarkably, $\sigma_{\Vert}$  turns to zero at $\theta_0=70^{\circ}$, which manifests the occurrence 
of the  {\it anti-ordinary} Hall effect in the crystal of CoPt, discussed already within the framework of 
the spin Hall effect in the preceeding section. At this "magic" angle, the magnitude of the anomalous 
Hall current $\mathbf{J}_H$ is almost twice larger than it is for $\mathbf{M}\Vert z$, however,
due to non-vanishing $\sigma_{\perp}$ component of the AHC vector, $\mathbf{J}_H$ is aligned 
{\it along} the direction of the magnetization. By analyzing Fig.~\ref{Anti-Ordinary}(d-g) we observe that the 
rotational sense of the anomalous Hall current is opposite to that observed in the ordinary Hall effect 
(OHE) of free electron gas. For OHE, Lorentz forces $\sim[\mathbf{H}\times\mathbf{v}]$ are acting 
on electrons with velocity $\mathbf{v}$ in the presence of magnetic field $\mathbf{H}$. The resulted 
ordinary Hall current of free electrons is always perpendicular to $\mathbf{H}$ irrespective of its
direction, opposite to the situation of the anti-ordinary anomalous Hall effect, observed in CoPt. 
Here, turning the magnetization clockwise in the $(\bar{1}10)$-plane results in an anti-clockwise 
rotation of $\mathbf{J}_H$, with its value staying rather large all the time. On the other hand, in
analogy to the spin Hall effect, for the anti-ordinary Hall effect, it is possible to find a direction of
the electric field $\mathbf{E}$ such that the Hall current is perpendicular to $\mathbf{E}$ and 
$\mathbf{M}$, which are, in turn, collinear to each other. 
Again, such situation obviously cannot occur for the ordinary Hall effect of free electrons.

In the region of $3d$Pt alloys in the vicinity of L1$_0$ CoPt the anisotropy of the AHE manifests 
itself in crucial ways suggesting new functionalities of the AHE-based devices. In this region, large 
changes in the magnitude of the anomalous Hall current as well as relative orientation of the Hall 
current with respect to the magnetization can be easily achieved by simple reorientation of the 
sample's magnetization. While the former could be used in order to~e.g.~tune the relative magnitudes 
of the extrinsic and intrinsic anomalous Hall signal~\cite{Seemann:2010, Lowitzer.2010.prl}, among 
most straightforward applications of the latter could be a realization of the planar Hall effect 
(PHE)~\cite{Yau:1971,Seemann:2011} which is related to the Hall effect in ferromagnetic materials  
with electric field, magnetization and the Hall current sharing same 
sample plane. So far, it is believed that in most of the cases the PHE originates from anisotropic magnetoresistance in metallic ferromagnets, although a PHE mechanism stemming from the anomalous 
Hall effect due to non-collinearity of the magnetization in semiconductor-based materials has 
been also suggested~\cite{Bowen:2005}. Within the scope of the anti-ordinary Hall effect, described 
in this work, it would be possible to observe the PHE coming solely from the anisotropic nature of 
the collinear ferromagnetic materials.

\subsection{Perturbation theory treatment: FePt}
\label{Pert-Theo}

In this section we present the perturbation treatment of the intrinsic AHC given by the linear 
response Kubo formula Eq.~(\ref{eq:kubo}). According to first order non-degenerate perturbation theory, 
the perturbed wavefunction $\psims$ originating from the unperturbed wave function $\psimso$
with spin $\sigma$ upon including the spin-orbit interaction is given by
\begin{equation}
\label{eq:pert1}
\Ket{\psims} =\Ket{\psimso} +\sum^{\sigma^{\prime\prime\prime}}_{p\neq m}
\frac{\Braket{\psipspo|\xi\mathbf{L\cdot S}|\psimso}}{\varepsilon_{m,0}-\varepsilon_{p,0}}\ket{\psipspo},
\end{equation}
where $\varepsilon_{m,0}$ and $\varepsilon_{p,0}$ denote the
unperturbed eigenenergies, and $k$-point indices have been omitted for simplicity. 
Compared to Eq.~(\ref{EY-WF-PT}), here
we consider the complete SOC Hamiltonian, and not only its spin-flip part.
Following Eq.~(\ref{eq:kubo}), in order to obtain an expression
for $\sigma_z$, the key is to evaluate the imaginary part of the following product:
\begin{equation}
\label{eq:pert2}
\frac{1}{(\varepsilon^{(1)}_{n}-\varepsilon^{(1)}_{m})^2}\Braket{\psinsp|v_x|\psims}\Braket{\psims| v_y|\psinsp}.
\end{equation}
In this expression $v_y$ and $v_z$ operators have to be considered instead of $v_x$
and $v_y$, respectively, if $\sigin$ is to be evaluated. The energies $\varepsilon^{(1)}_{n}$ and 
$\varepsilon^{(1)}_{m}$ stand for the first-order perturbed eigenvalues. 
Substituting Eq.~(\ref{eq:pert1}) into 
Eq.~(\ref{eq:pert2}), we can sort out the terms which appear at different orders with respect 
to the SOI strength $\xi$. The purpose of this is the 
general analysis of simplified expressions, and discussion of the orders with respect to $\xi$ and 
their energy scales, which remain valid also when the (degenerate) perturbation theory is applied rigorously
in higher orders. In the following we assume that the velocity operator does not contain the
relativistic correction due to spin-orbit coupling, which we generally find to be a very good
approximation.

A typical first-order in $\xi$ contribution to Eq.~(\ref{eq:pert2}) involves a 
sum over additional transitions via auxiliary states $\psilso$, and looks like:
\begin{equation}
\label{eq:1storder}
\frac{\xi}{(\varepsilon^{(1)}_{n}-\varepsilon^{(1)}_{m})^2}\Braket{\psinspo| v_x|\psimso}\sum^{\sigma^{\prime\prime}}_{l\neq n}
\frac{\Braket{\psilso|\mathbf{L\cdot S} |\psinspo}}
{\varepsilon_{n,0}-\varepsilon_{l,0}}\Braket{\psimso|v_y|\psilso},
\end{equation}
while the second-order contribution to the product of the matrix elements of the velocity 
operators involves already two sums of additional transitions via auxiliary states $\psipspo$ 
and $\psilso$, and consists of terms with the following structure:
\begin{equation}
\label{eq:2ndorder}
\frac{\xi^2}{(\varepsilon^{(1)}_{n}-\varepsilon^{(1)}_{m})^2}\sum^{\sigma^{\prime\prime\prime}}_{p\neq m}\frac{\Braket{\psipspo|\mathbf{L\cdot S}|\psimso}}
{\varepsilon_{m,0}-\varepsilon_{p,0}}\Braket{\psinspo| v_x|\psipspo}\sum^{\sigma^{\prime\prime}}_{l\neq n}
\frac{\Braket{\psilso| \mathbf{L\cdot S}|\psinspo}}
{\varepsilon_{n,0}-\varepsilon_{l,0}}\Braket{\psimso| v_y|\psilso}.
\end{equation}
For the first order terms, Eq.~(\ref{eq:1storder}), the initial state $\psimso$ and final state $\psinspo$ 
must have the same spin, since the velocity operator does not act on the spin part of the wavefunction
(if we neglect the relativistic correction to the velocity operator). 
This means that the state $\psilso$ has to be of the same spin as states $n$ and $m$. This can happen
only due to the spin-conserving part of the spin-orbit interaction ${LS}^{\uu}$, as was also found by 
Cooper~\cite{Cooper:1965}, meaning that only spin-conserving SOI contributes to the AHC in the first 
order with respect to $\xi$. Thus, within the non-degenerate perturbation theory, we would expect the 
largest contribution to the AHC from the spin-conserving part of the spin-orbit interaction. It should be 
kept in mind, however, that in materials containing heavy atoms the SOI cannot be treated as a small 
perturbation. Moreover, as follows from our previous discussion, the important role for the AHC of near 
degeneracies across the Fermi level cannot be denied, for which the above arguments, based on non-degenerate perturbation  theory, do not apply (see also discussion at the end of section~\ref{Spin-Hall-Effect}).

\begin{figure}
\begin{center}
\includegraphics[width=13cm]{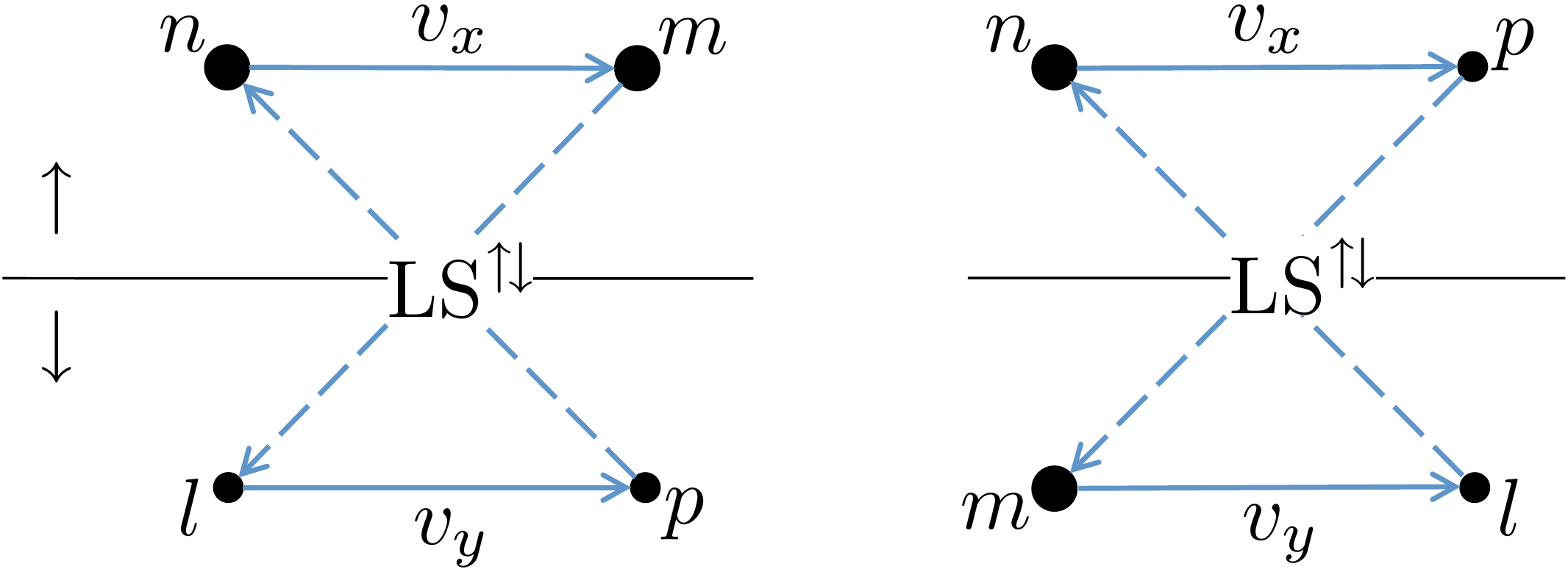}
\caption{Diagrams depict the second-order in SOI strength $\xi$ contributions
to the AHC upon expanding the wavefunctions according to the first-order non-degenerate
perturbation theory, see Eq.~(\ref{eq:2ndorder}). 
The solid (dashed) lines stand for the matrix elements of the velocity
operator (${LS}^{\ud}$-operator) between the bra- and ket-states, marked with 
indeces next to the dots, with the direction of the lines from the bra- to the ket-state. The 
horisontal thin lines separate the $\uparrow$-states from the $\downarrow$-states. 
\label{fig:diagrams}}
\end{center}
\end{figure}

On the other hand, the spin-flip processes contribute only in second- and 
higher-order terms. We have also come to this conclusion for the spin Hall effect in paramagnets,
following a somewhat different argumentation. Analysing the terms second-order in $\xi$ which come
from the first-order perturbed wavefunctions, we find four types of contributions
to the AHC,  analogous to the one, given by expression~(\ref{eq:2ndorder}). All four 
terms include two summations over auxiliary states $\psipspo$ and $\psilso$, and include 
two products of the matrix elements of the SOI and components of the velocity operator. 
Assuming for simplicity that the occupied state $\psinspo$ has $\sigma^{\prime}$ = $\uparrow$, 
the four types of contributions can be related (omitting the energy denominators for simplicity) 
to the products of the velocity and SOI matrix elements arranged in the way, presented in 
Fig.~\ref{fig:diagrams}. In this figure, the diagram on the left stands for the product of 
$\langle\psi_{n,0}^{\uparrow}|v_x|\psi_{m,0}^{\uparrow}\rangle$,
$\langle\psi_{m,0}^{\uparrow}|{LS}^{\ud}|\psi_{l,0}^{\downarrow}\rangle$,
$\langle\psi_{l,0}^{\downarrow}|v_y|\psi_{p,0}^{\downarrow}\rangle$ and 
$\langle\psi_{p,0}^{\downarrow}|{LS}^{\ud}|\psi_{n,0}^{\uparrow}\rangle$, while
the diagram on the right side stands for the product of 
$\langle\psi_{n,0}^{\uparrow}|v_x|\psi_{p,0}^{\uparrow}\rangle$,
$\langle\psi_{p,0}^{\uparrow}|{LS}^{\ud}|\psi_{m,0}^{\downarrow}\rangle$,
$\langle\psi_{m,0}^{\downarrow}|v_y|\psi_{l,0}^{\downarrow}\rangle$ and
$\langle\psi_{l,0}^{\downarrow}|{LS}^{\ud}|\psi_{n,0}^{\uparrow}\rangle$.
The other two contributions to the second order AHC come from the diagrams in
Fig.~\ref{fig:diagrams}, in which the $v_x$ and $v_y$ operators are interchanged, 
while all directions of the arrows are reversed. 

As it is evident from Fig.~\ref{fig:diagrams}, all diagrams contributing to the second-order AHC
include the matrix elements of the spin-non-conserving part of 
the spin-orbit interaction. Interestingly, although a single act of ${LS}^{\ud}$
on a wavefunction is to flip its spin, in addition to the contribution to the AHC from the 
occupied $n$ and unoccupied $m$ states of different spin character (right diagram), 
there can also be a non-vanishing contribution from the second-order transition between
the $n$ and $m$ states of the same spin. Finally, in addition to the contributions to 
the second-order AHC depicted in Fig.~\ref{fig:diagrams}, one could in principle consider 
a situation when all the $n,p,l,m$-states have the same spin character and the SOI matrix
elements are coming from its spin-conserving part. Assuming 
the elements of the velocity operator (at a given $k$-point) to be real, however, it is easy 
to see then that the product of the SOI matrix elements would give a real number, since the 
${LS}^{\uu}$-operator is purely complex, and thus the imaginary part of the
second-order AHC contribution would vanish in this case. As confirmed by our first
principles calculations, this assumption is valid.

By calculating the spin-orbit matrix elements in the basis of states unperturbed
by SOI as described in section 2, we applied the non-degenerate first-order 
perturbation theory in the  
wavefunctions with respect to SOC and computed the corresponding orders of
contributions to the AHC in FePt. The results of these calculations are shown in 
Table~\ref{table:pert} for $\mathbf{M}\Vert x$ and $\mathbf{M}\Vert z$. 
Here, we considered separately the spin-conserving and spin-flip
parts of our first principles Hamiltonian, converged the system, and applied the
first order non-degenerate perturbation theory in wavefunctions, as described above.
If we keep only ${LS}^{\uu}$ spin-orbit in our calculations, we arrive at the value
of the AHC which we denote as  $\sigma^{\uu}$, while keeping exclusively the  
${LS}^{\ud}$ SOC leads to the value of the conductivity $\sigma^{\ud}$. If  
$\sigma^{\uu}$ and $\sigma^{\ud}$ are analogously calculated non-perturbatively 
from first principles, see right column of Table~\ref{table:pert}, then, as our calculations show
\begin{equation}
\sigma \approx \sigma^{\uu} + \sigma^{\ud},
\end{equation}  
where $\sigma$ stands for the total AHC calculated with the complete SOC 
Hamiltonian. 
Moreover, if the first order perturbation theory in wavefunctions is applied to evaluate
the corresponding conductivities, the above decomposition is exact.
To obtain the perturbation theory values
in Table~\ref{table:pert} we used a tolerance parameter $\Delta$ of 50~meV:
that is, when for a considered unperturbed state $|m\rangle$ the difference in
energy $|\varepsilon_{m,0}-\varepsilon_{p,0}|$ was less than $\Delta$ in 
Eq.~(\ref{eq:pert1}), the
projection on state $|p\rangle$ was considered
to be zero and the corresponding term in Eq.~(\ref{eq:pert1}) was neglected.

\begin{table}
\begin{center}
\begin{tabular}{l|rrrrrc}
                            & 1$^{\rm st}$ order  & 2$^{\rm nd}$ order & 3$^{\rm rd}$ order & 4$^{\rm th}$ order & $\Sigma$ & First principles \\ \hline
 $\sigout^{\uu}$ &  581     & 0       &  84     &  0    &          665        &577 \\ 
  $\sigout^{\ud}$   &  0      &  84     &   0     & $-$34       & 50  &133  \\ 
$\sigin^{\uu}$ &  557     &  0      &  184    &   0    &           741        &585   \\
$\sigin^{\ud}$    &    0      & $-$238     &  0      &  $-$106 &  $-$344   &$-$184  \\
\end{tabular}
\caption{Decomposition of the AHC of FePt into contributions of different orders and their sum ($\Sigma$)
based on a perturbative treatment of the spin-orbit interaction, in 
comparison to first-principles non-perturbative values. All values are in S/cm. See main text for details.
\label{table:pert}}
\end{center}
\end{table}

Overall, by inspecting Table~\ref{table:pert} we can conclude that the
agreement of the perturbation theory results with first-principles results presented in 
Table~\ref{AHE-TMs} and Table~\ref{table:pert} for FePt is reasonable. 
The major 
contribution to $\sigma^{\uu}$ comes in the first order with respect to $\xi$
and constitutes around 550~S/cm for both magnetization directions, while
within the next contributing order these values are corrected by at most
30\%. Thus, up to considered orders, the $\sigma^{\uu}$ calculated from
the perturbation theory agrees qualitatively with the correspoding values computed
from the Berry curvature, and also displays a small anisotropy of the AHC
when the magnetization direction is varied, in agreement with the non-perturbative calculations.
As far as the spin-flip AHC is concerned, the perturbation theory predicts its much
smaller values as compared to the spin-conserving AHC. Moreover, as we can see
from Table~\ref{table:pert}, the sign of $\sigma^{\ud}$ is positive for $\mathbf{M}\Vert z$
and negative for $\mathbf{M}\Vert x$, and the flip-AHC anisotropy reaches a large 
value of around 400~S/cm, as compared to the first-principles value of 320~S/cm. 
All of these observations are in agreement to the results
for $\sigma^{\ud}$ calculated from the non-perturbative Berry curvature. 
A reasonable agreement of perturbation theory values with the first principles ones is certainly
coincidencial for FePt, in which the AHC seems not to be dominated by transitions between the bands
separated by less than 50~meV in energy. Nevertheless, while 
within our approach the singular contributions to the Berry
curvature (arising from (near) degeneracies between the states), which can play an important
role for the AHC, are not taken into account, certain general features of the spin-flip and
spin-conserving AHC apparent from this analysis are universal, as discussed in the next section.

\subsection{Spin-flip and spin-conserving transitions}

As discussed in the previous section, and confirmed by explicit calculations within the 
first order non-degenerate perturbation theory, the 
spin-conserving part of SOI contributes in the first and higher odd orders with respect to
the spin-orbit strength, while the spin-non-conserving part of SOI leads to non-vanishing
AHC in second and higher even orders. Such oddness and evenness of $\sigma^{\uu}$
and $\sigma^{\ud}$ with respect to $\xi$ can be also demonstrated in higher orders of
(degenerate) perturbation theory, although we do not provide explicit expressions here.
As follows from our calculations, this remains true even when the anomalous Hall conductivity
is treated non-perturbatively within the first principles methods, outlined in section 2.
Here, we present explicit calcualtions of the $\sigma^{\uu}$ and $\sigma^{\ud}$ in L1$_0$
FePt and NiPt as a function of the spin-orbit strength in the system, $\xi$. The results of the
calculations, in which the SOC strength was scaled uniformly on $3d$ and Pt atoms with
respect to the unscaled values $\xi_0$, are presented in Fig.~\ref{fig:scaling1}. Indeed, 
from this plot we observe that, within the accuracy of the calculations, in NiPt the conductivity 
$\sigma^{\uu}(\xi)=-\sigma^{\uu}(-\xi)$, while in FePt  
$\sigma^{\ud}(\xi)=\sigma^{\ud}(-\xi)$, allowing thus for an expansion of $\sigma^{\uu}(\xi)$
($\sigma^{\ud}(\xi)$) in odd (even) powers of $\xi$. Note also that for larger values of $\xi$ 
the behavior of $\sigma^{\uu}$ and $\sigma^{\ud}$ is manifestly different from linear and quadratic,
respectively, marking thus the importance of higher order terms.

\begin{figure}[t!]
\begin{center}
\includegraphics[width=7.3cm]{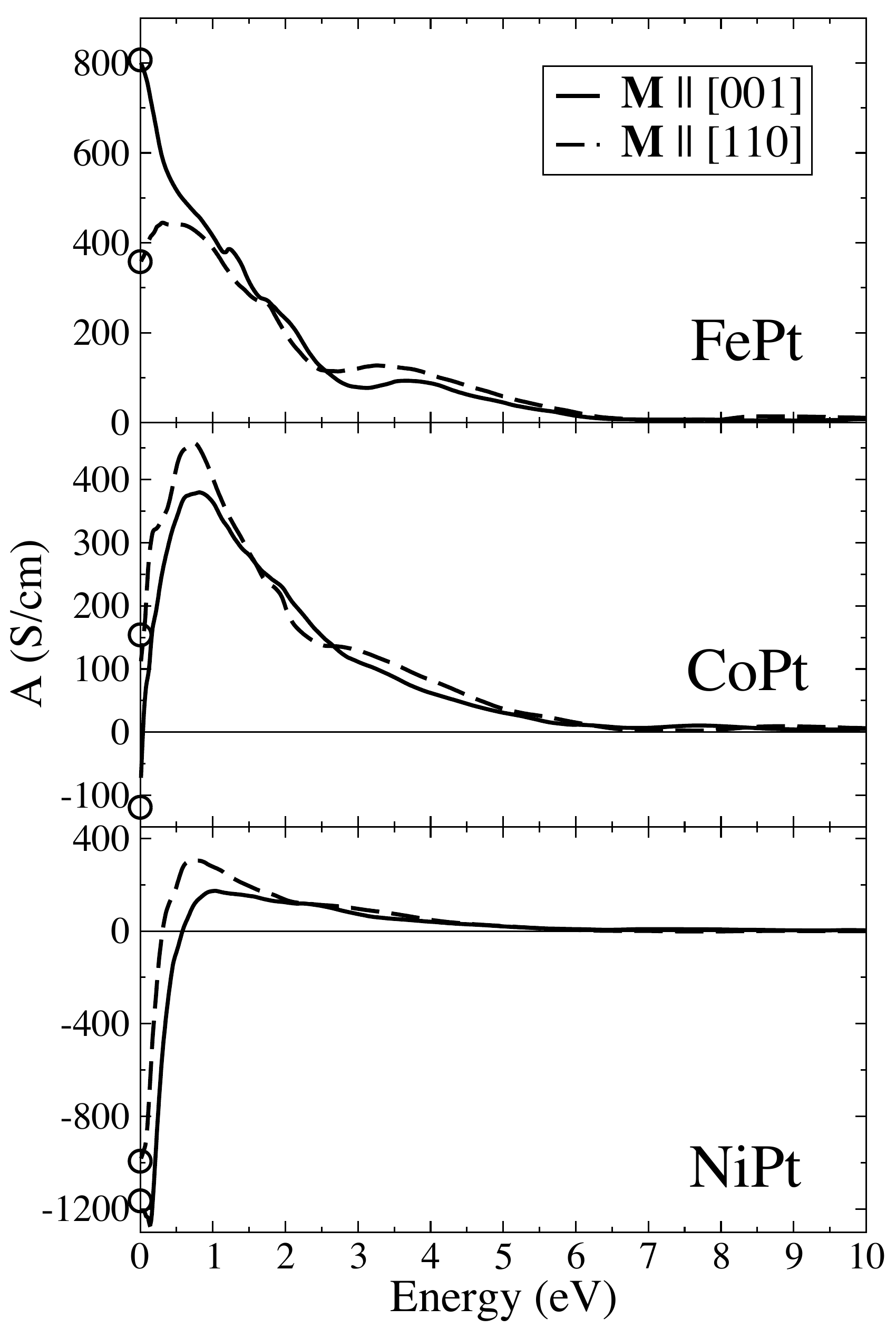}
\includegraphics[width=7.3cm]{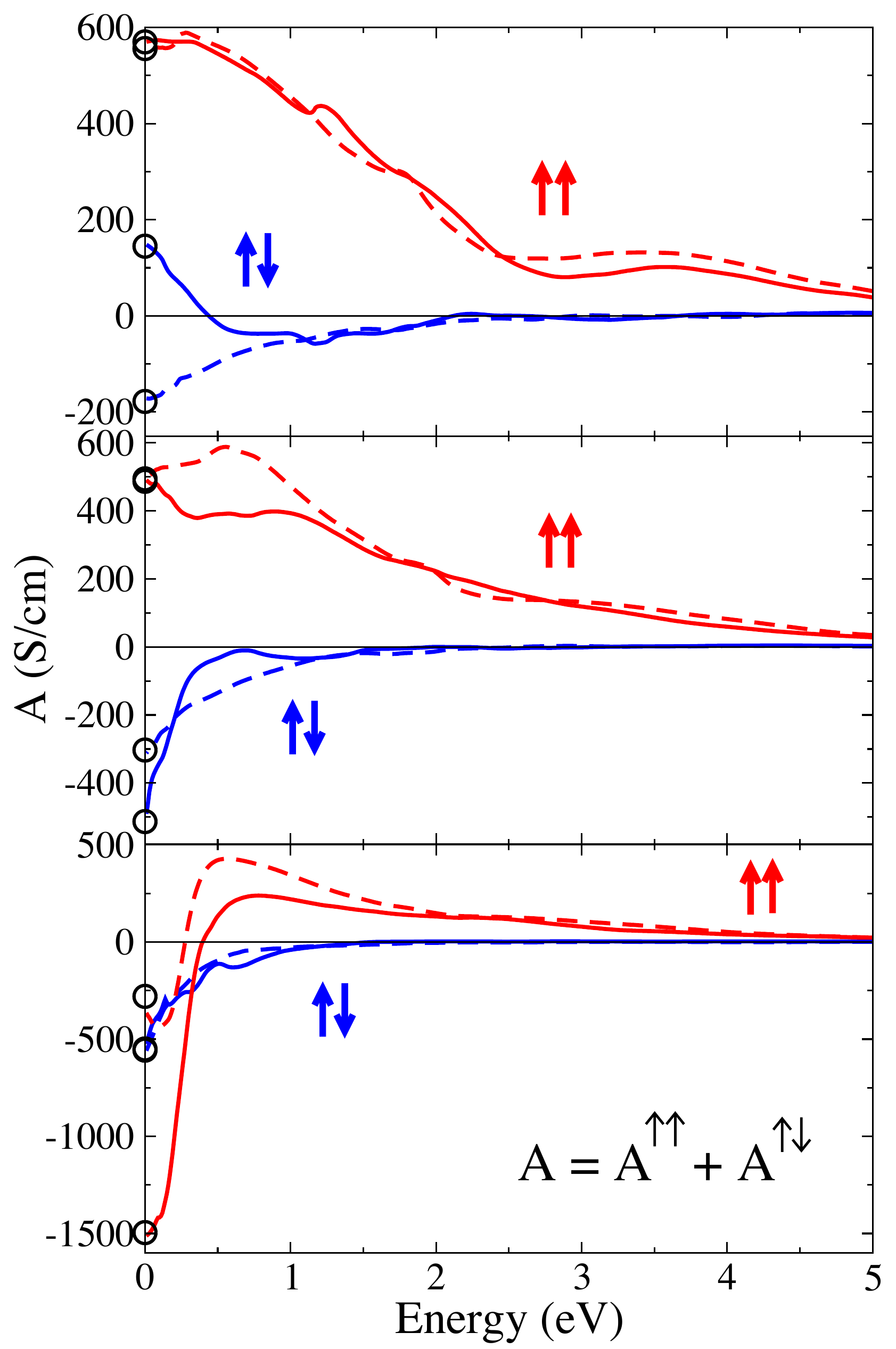}
\end{center}
\caption{\label{fig:mcd}
  Total (left, black), spin-conserving (right, red) and spin-flip (right, blue) cumulative AHC as a function
of energy in FePt, CoPt and NiPt for two different magnetization directions (solid and dashed lines). 
}
\end{figure}

\begin{figure}[t!]
\centering
\includegraphics[width=7.6cm]{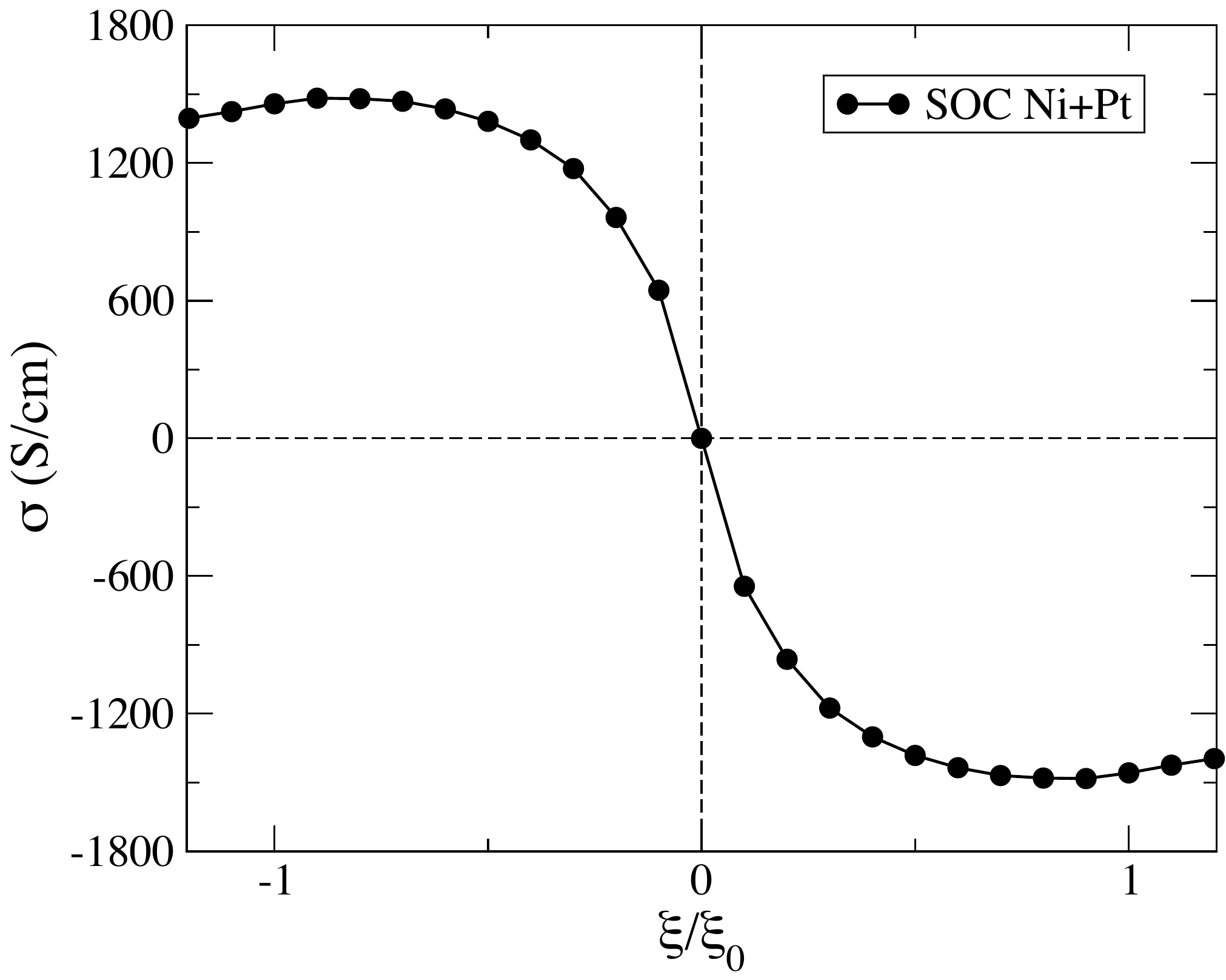}
\includegraphics[width=7.4cm]{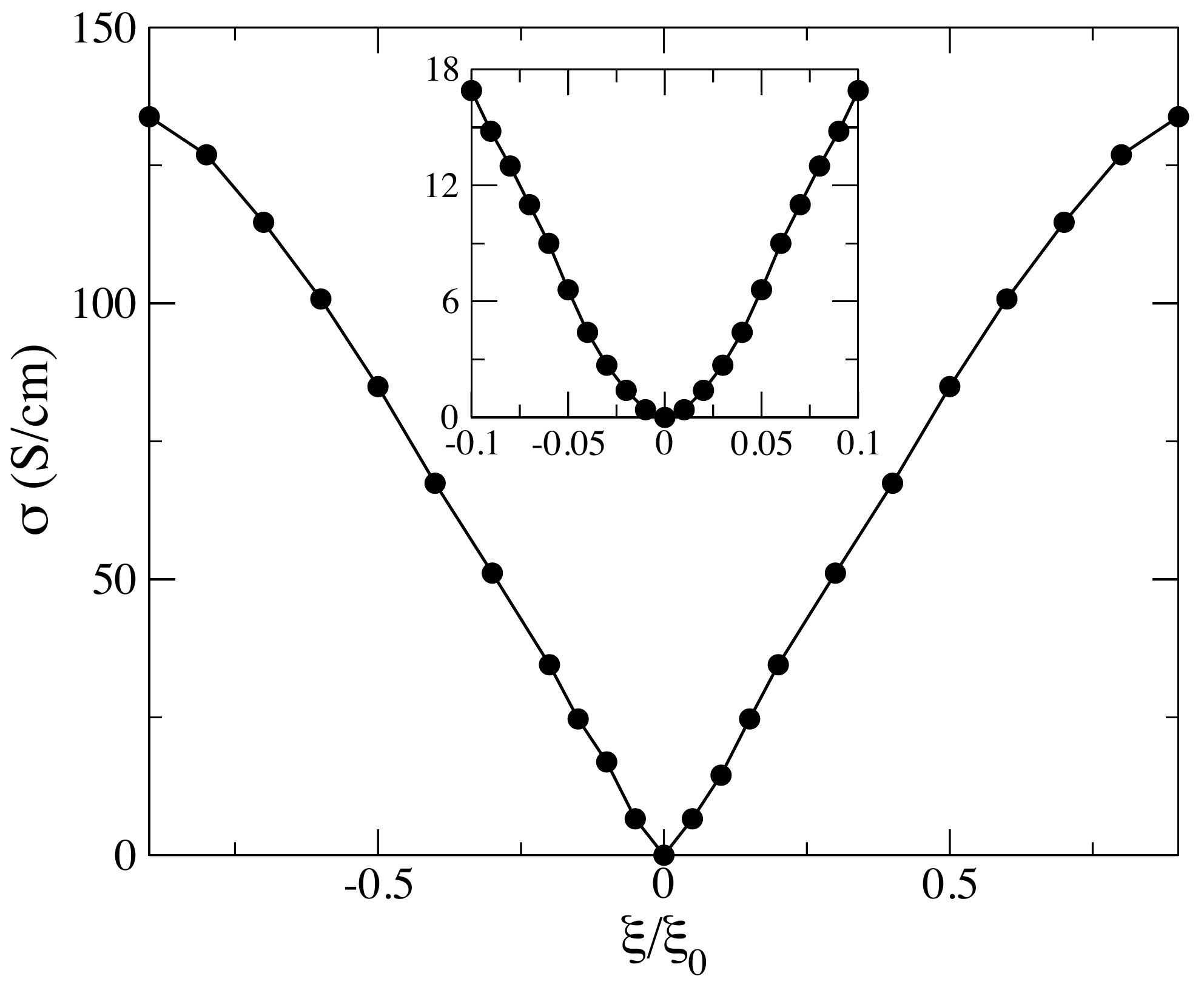}
\caption{Left: Scaling of the spin-conserving AHC in NiPt with respect to the strength 
of spin-conserving SOC on both Ni and Pt sites. Right: Scaling of the spin-flip 
AHC in FePt with respect to the strength of the spin-flip SOC. SOC on both Fe and 
Pt sites was scaled uniformly. Inset displays the zoom-in into the region of small SOI 
strength. Note, that for the sake of a numerical experiment we also consider 
the regime of negative values of the SOI strength in our first principles calculations.
\label{fig:scaling1}}
\end{figure}

In the latter Fig.~\ref{fig:scaling1}, by zooming into the region of very small $\xi$,  we observe that
the behavior of $\sigma^{\uu}$ in NiPt (not shown) and of $\sigma^{\ud}$ in FePt
(inset of Fig.~\ref{fig:scaling1}) is clearly dominated by the linear and quadratic in $\xi$ terms,
respectively. In general, the fact that the spin-conserving transitions appear 
already in the first order with respect to $\xi$, while the spin-flip transitions in second
and higher, has two essential consequences. Firstly, it means that the energetic spread
of the spin-flip conductivity will be much narrower than that of the spin-conserving AHC, due
to the higher power of the energy denominator in Eq.~(\ref{eq:2ndorder}) of the order of 
$(\varepsilon_n-\varepsilon_m)^4$, as compared to that of the order of 
$(\varepsilon_n-\varepsilon_m)^3$ in Eq.~(\ref{eq:1storder}). This can be clearly
seen in Fig.~\ref{fig:mcd} (right), in which the cumulative AHC for FePt, CoPt and NiPt is 
decomposed, analogously to the total AHC, into spin-conserving and spin-flip contributions:
\begin{equation}
A(\omega)\approx A^{\uu}(\omega) + A^{\ud}(\omega).
\end{equation}
Noticably, while $A^{\uu}(\omega)$ decays on the scale of the bandwidth of several eV,
the spin-flip cumulative AHC is localized in a much narrower energy region of the order
of 1~eV. It is important to mention that the anisotropy of the total AHC can present a 
competition between the anisotropy of the spin-conserving and spin-flip parts, depending
on the exact details of the electronic structure, see for example Fig.~\ref{fig:mcd} (right) and
Fig.~\ref{Anti-Ordinary}(b) $-$ we refer here also to the discussion at the end of section~\ref{Spin-Hall-Effect}.

Remarkably, the energetic scale of the spin-flip transitions in Fig.~\ref{fig:mcd} roughly corresponds
to the energy scale of the spin-orbit interaction of Pt atoms. This observation brings us to the second
conclusion that  we can make out from the perturbation theory analysis: the contribution of spin-conserving 
transitions to the AHC is normally dominant over the spin-flip transitions (this can be clearly seen 
in Fig.~\ref{fig:mcd}), since the latter appear only starting from the second order in SOC strength. 
Correspondingly, in order to promote the spin-flip contribution to the AHC, the spin-orbit strength
in the material has to be enhanced. Let us consider this point in detail, and prove that
the spin-flip processes in FePt are induced mostly by the strong SOI on the Pt atoms~\cite{Zhang.2011.prl}. To do this, 
we selectively turn off the SOI on each atomic species inside the crystal. The atom-resolved
spin-orbit Hamiltonian reads
\begin{equation}
H_{\rm SO} =  \xi_{\rm Fe}\mathbf{L}^{\rm Fe}\cdot\mathbf{S}+
 \xi_{\rm Pt}\mathbf{L}^{\rm Pt}\cdot\mathbf{S},
 \label{atomic-SOC}
\end{equation}
where $\mathbf{L}^{\mu}$ is the orbital angular momentum operator
associated with atomic species $\mu$ (Fe or Pt), and $\xi_{\mu}$ is the spin-orbit
coupling strength averaged over valence $d$-orbitals. In FePt we find
$\xi_{\rm Fe}^0=0.06$~eV and $\xi_{\rm Pt}^0=0.54$~eV, where
$\xi^0_\mu$ denotes the value calculated from first-principles.

\begin{figure}[t!]
\begin{center}
\includegraphics[width=10.0cm]{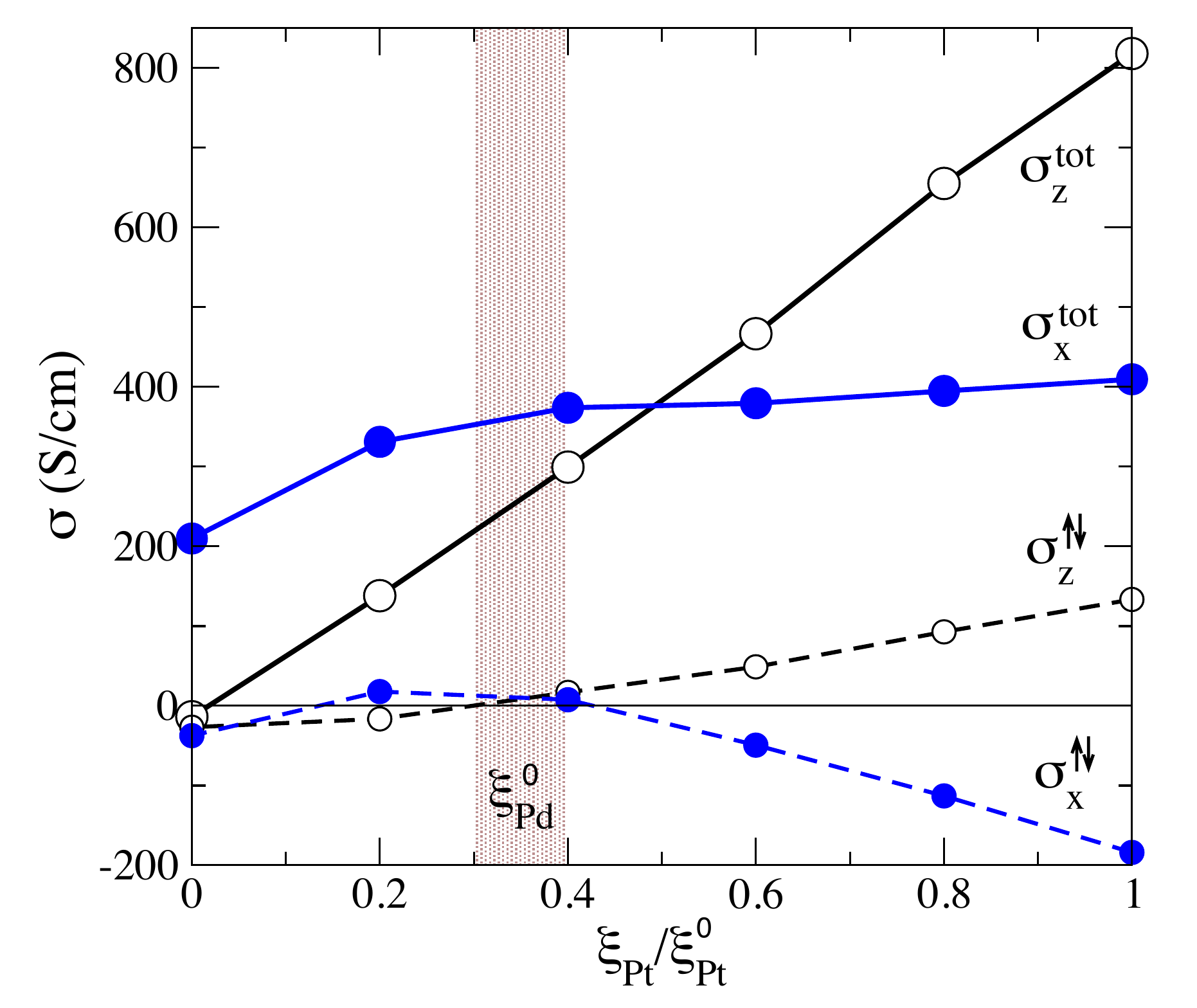}
\end{center}
\caption{\label{fig:soc-scaling}
  Dependence of the total ($\sigout^{\rm tot}$ and $\sigin^{\rm tot}$)
  and spin-flip ($\sigout^{\ud}$ and $\sigin^{\ud}$) AHC in FePt alloy
  on the strength $\xi_{\rm Pt}$ of the SOI inside the Pt atoms. Taken from~\cite{Zhang.2011.prl}.}
\end{figure}

\begin{table}[t!]
\begin{center}
\begin{tabular}{l|rrr|rrrrrr}
                            & $\sigma^{\rm tot}$ & $\sigma^{\uu}$ & $\sigma^{\ud}$ & Fe$^{\rm tot}$    & Fe$^{\uu}$ & Fe$^{\ud}$ & Pt$^{\rm tot}$ & Pt$^{\uu}$ &  Pt$^{\ud}$ \\ \hline
 $\mathbf{M}\Vert [001]$ & 818 & 577 &  133  &   14        &  18         &  $-$27    &  848            &    541    &  282         \\ 
  $\mathbf{M}\Vert [100]$ & 409 & 585 &  $-$184   & 210      &   254      &   $-$38   &     65           &    426    &  $-$361   \\
\end{tabular}
\caption{AHC in FePt for two magnetization directions, resolved into spin-flip
  and spin-conserving contributions from the SOI on each atomic
  species (left part), compared to the corresponding total values (right part). 
  All values are in S/cm. Taken from~\cite{Zhang.2011.prl}.
\label{TableX}}
\end{center}
\end{table}

Let us recalculate now the AHC after setting to zero either $\xi_{\rm
  Fe}$ or $\xi_{\rm Pt}$ in Eq.~(\ref{atomic-SOC}), and then further
  decompose the conductivity into the spin-flip and spin-conserving parts.
   The results are presented in Table~\ref{TableX}.
Although such a decomposition is not exact, it reproduces the total values rather well. 
Namely, the sum of the total conductivities driven by
  SOI on Fe (Fe$^{\rm tot}$ in Table~\ref{TableX}) and on Pt (Pt$^{\rm tot}$ in
  Table~\ref{TableX}) is in reasonable agreement to the  values of $\sigma^{\rm tot}$
  for both magnetization directions. Moreover, the decomposition
  of the total atom-resolved AHCs into spin-conserving and spin-flip parts
  is almost exact, as can be seen from Table~\ref{TableX}.
Consider first the AHC driven by $\xi_{\rm Fe}$. For both magnetization
  directions the spin-flip contribution is very small, while the
  spin-conserving part is small along [001] but large along [100]. As
  for the AHC induced by $\xi_{\rm Pt}$, the spin-conserving part is large but fairly isotropic,
  while the spin-flip part is highly anisotropic, changing from a
  large positive value along [001] to a large negative value along
  [100]. This confirms that the large and strongly anisotropic
  $\sigma^{\ud}$ is governed by the SOI inside the Pt atoms.

 Let us confirm the conclusion we draw from the perturbation theory description via
 nonperturbative calculations where we tune by hand the SOI strength $\xi_{\rm Pt}$ 
 on the Pt atoms. The results for the total and spin-flip AHC are shown in Fig.~\ref{fig:soc-scaling} as a 
 function of $\xi_{\rm Pt}/\xi_{\rm Pt}^0$. It can be seen that for $\xi_{\rm Pt}$ less 
 than $\xi_{\rm Pt}^0/2$, the absolute value of the spin-flip AHC does not exceed a 
 modest value of 50~S/cm. In this regime $\sigout^{\rm tot}$ and $\sigin^{\rm tot}$
  are dominated by spin-conserving processes.  Moreover, we note that
  while the decrease in $\sigout^{\rm tot}$ is almost perfectly linear,
  $\sigin^{\rm tot}$ stays fairly constant over a wide region of
  $\xi_{\rm Pt}$ values. This can be understood from the fact that for
  $\mathbf{M}\Vert x$ the spin-conserving and spin-flip
  contributions arising from Pt largely cancel one another
  (see Table~\ref{TableX}), so that the total AHC is mostly driven by the SOI on
  the Fe atoms. In contrast, for $\mathbf{M}\Vert z $ it is the
  SOI on the Pt atoms which dictates the AHC.
The artificial tuning of $\xi_{\rm Pt}$ performed above describes rather
  well what happens if the Pt atoms are replaced with Pd, to form the
  isoelectronic FePd alloy~\cite{Seemann:2010}.  This can be seen
  by comparing the values of $\sigma^{\rm tot}$ and $\sigma^{\ud}$
  of 135 (276) and 24 (62)~S/cm for $\mathbf{M}\Vert [001]$ ($\mathbf{M}\Vert [100]$),
  respectively,  
  in FePd with the values taken from the shaded area 
  in Fig.~\ref{fig:soc-scaling}, where $\xi_{\rm Pt}\approx \xi_{\rm Pd}^0=0.19$~eV.  In particular,
  the sign of the AHC anisotropy in FePd, which is opposite from that in FePt,
  is correctly reproduced by the scaled calculations on FePt.

\section{Outlook}

In this review, we briefly outlined the recent progress in understanding and predicting
the anisotropy of the spin relaxation and intrinsic anomalous and spin Hall effect in metals
from first principles. In case of the spin-relaxation this anisotropy is the consequence of
the anisotropy in the wavefunctions upon changing the spin-quantization axis in the crystal,
which can be probed via a non-equilibrium process such as an injection of an electron
with a certain direction of spin polarization into a material which exhibits the anisotropy
of the Elliott-Yafet coefficient. In case of the spin Hall effect, in addition to the anisotropy
of the wavefunctions with respect to the SQA, the anisotropy of the velocity matrix elements
comes into play in non-cubic crystals, which leads to an anisotropic correlation between the
direction of an applied electric field, direction of the spin current and its spin polarization. 
For ferromagnets exhibiting the anomalous Hall effect, in addition, eigenvalues and wavefunctions
display a very non-trivial dependence on the direction of magnetization in the crystal, which results
in a complicated relation between the orientation of magnetization and direction of the Hall
current, as well as its magnitude. The anisotropy of the spin relaxation and Hall currents in
perfect crystals can be so strong that it can reach colossal values. 
For spin and anomalous Hall effects, the magnitude of the Hall current can be even
completely suppressed via a suitable choice of the direction of the electric field and/or 
magnetization. Such strong anisotropy should manifest itself clearly in an experiment, and 
one of the purposes of the current review is to stimulate further experimental  studies with the aim 
of extending the functionalities of future spintronic devices.

\begin{figure}[t!]
\begin{center}
\includegraphics[width=6.6cm]{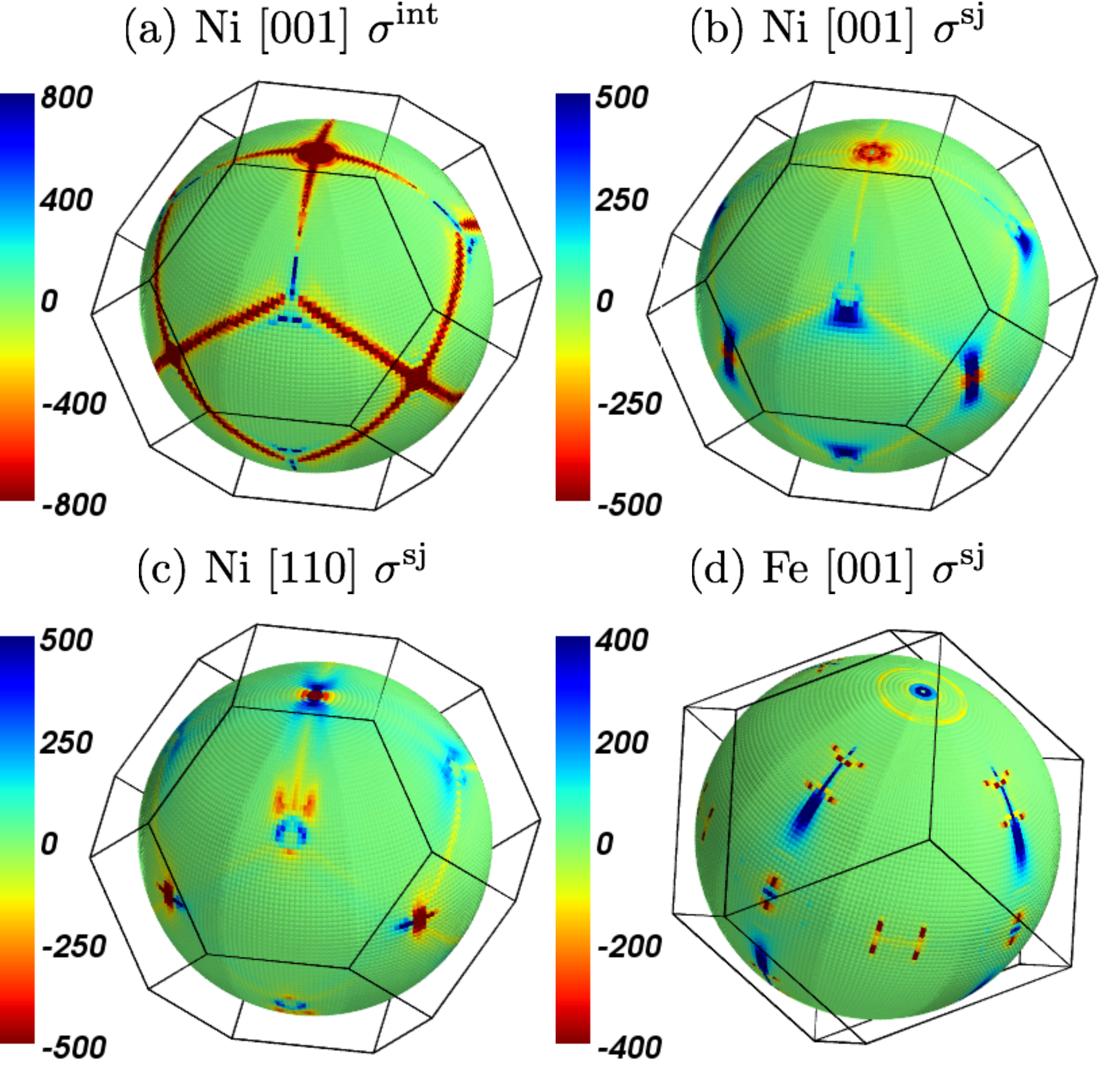}\hspace{0.8cm}
\includegraphics[width=6.5cm]{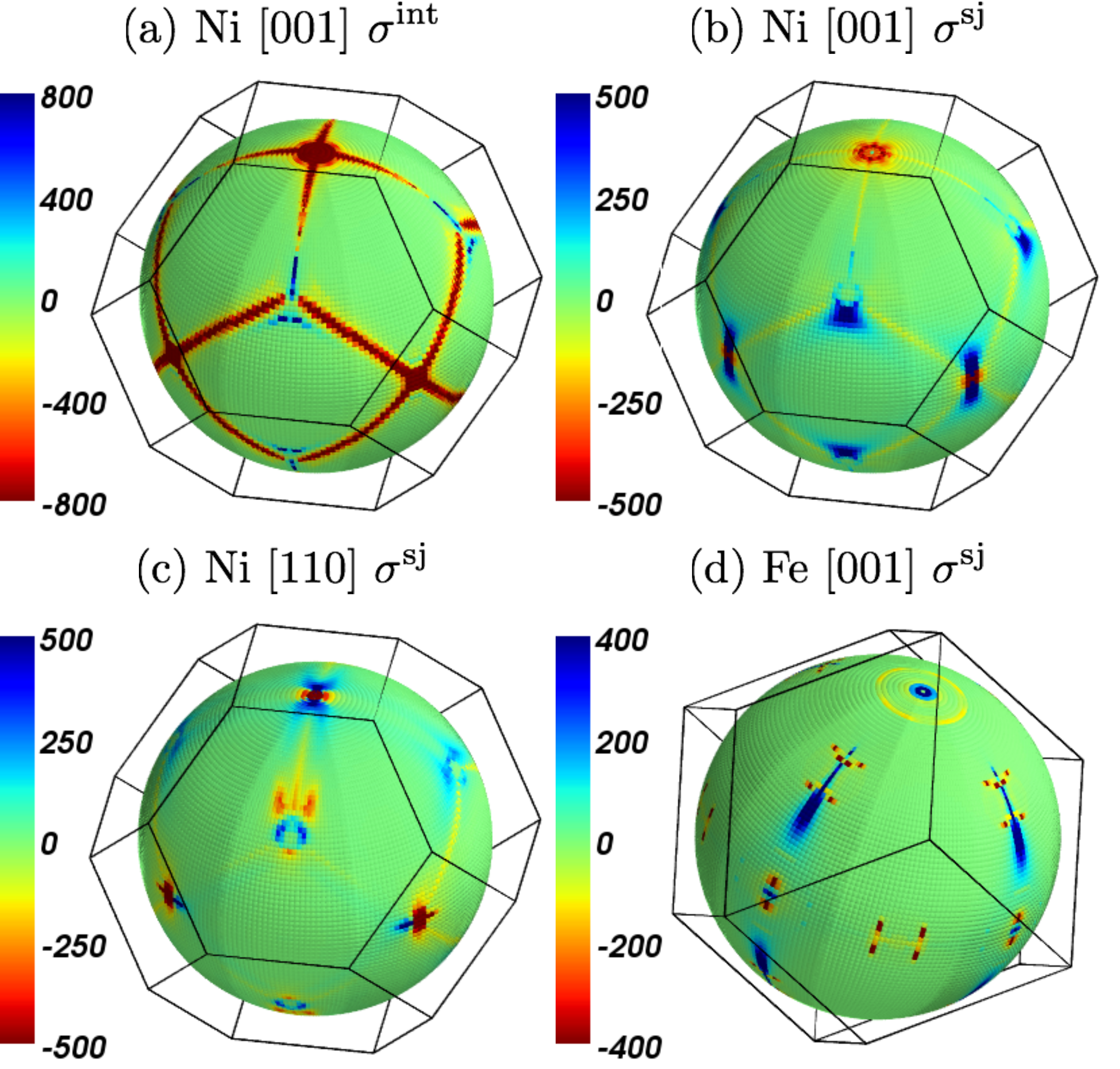}
\end{center}
\caption{\label{Side-Jump}
Angle-resolved side-jump conductivity in units of S/cm as a function of direction in the Brillouin
zone for fcc Ni for two different magnetization directions. Taken from~\cite{Weischenberg.2011.prl}.
}
\end{figure}

The phenomena considered in this work stem from the electronic structure of perfect idealized 
solids. In an experiment, especially at finite temperatures, one inevitably faces imperfections 
in the crystalline order due to impurities or disorder, phonons, magnons etc. For the Hall effects, disorder
in the system serves as a source of additional channels for the Hall signal due to so-called skew- and 
side-jump scattering~\cite{Berger:1970,Smit:1955,Smit:1958,Sinitsyn:2006}. In the perturbation
theory picture, any sort of effects due to impurity scattering should involve the matrix elements
of the scattering potential with the Bloch states of the perfect crystal. And while already the
Bloch wavefunctions in the solid, as we discussed, might display an anisotropy with respect to the SQA,
also the complicated structure of the impurity potential, especially if it is spin-polarized, should
exhibit strong anisotropy as well given an internal spin-orbit interaction and anisotropic crystal
field. Recently, assuming a disorder due to point-like delta-correlated defects which do not have
any internal structure of the potential, strong anisotropy of the side-jump contribution to the
anomalous Hall effect in ferromagnets has been demonstrated from first principles~\cite{Weischenberg.2011.prl} (c.f.~Fig.~\ref{Side-Jump}). In this case the calculated anisotropy is a consequence
of the anisotropic electronic structure of the perfect crystal, and the question of the anisotropy
of transverse transport due to microscopic details of the impurity potential, which can be treated
with high accuracy from {\it ab initio}, still remains open and serves as a fruitful subject for future
studies. 

Finally, we would like to remark, that following the same philosophy as outlined above,  
anisotropy of the transverse transport should be also large and experimentally observable for 
other effects driven by spin-orbit interaction, such as anomalous Nernst effect in 
ferromagnets~\cite{Miyasato:2007,Xiao:2006} and spin Nerst effect in paramagnets~\cite{Spin-Nernst,Chuu:2010}. On the other hand, in compounds which exhibit non-collinear magnetic order, the interplay
of magnetism and spin-orbit interaction becomes very complex, since  the non-collinearity of
the local spins can effectively play the role of the spin-flip part of the spin-orbit interaction, and the magnetic ground state 
itself can be very sensitive to the matrix elements of the spin-orbit interaction. In such a situation, strong
anisotropy of the transverse effects observed in this type of systems, such as magnon Hall effect~\cite{Onose:2010} and topological Hall effect~\cite{Wills:2012,Bruno:2004,Rosch:2009}, is guaranteed.
We are aware only of a single work in this direction~\cite{Onoda:2011}, while the phenomena mentioned
above still remain largely unexplored.

\section{Acknowledgements}

We acknowledge funding under 
DFG project MO 1731/3-1 and HGF-YIG program VH-NG-513.


\begin{thebibliography}{100}

\bibitem{Elliott.1954}
R.~J. Elliott, ``Theory of the effect of spin-orbit coupling on magnetic
  resonance in some semiconductors,'' {\em Phys. Rev.}, vol.~96, pp.~266--279,
  Oct 1954.

\bibitem{Zutic.2004.rmp}
I.~\ifmmode \check{Z}\else \v{Z}\fi{}uti\ifmmode~\acute{c}\else \'{c}\fi{},
  J.~Fabian, and S.~Das~Sarma, ``Spintronics: Fundamentals and
  {A}pplications,'' {\em Rev.\ Mod.\ Phys.}, vol.~76, Apr 2004.

\bibitem{mavropoulos:2009}
P.~Mavropoulos, ``Spin relaxation in nonmagnetic metals and semiconductors,''
  {\em 40. IFF Spring School book of lectures (Forschungszentrum J\"ulich)},
  2009.

\bibitem{Karplus:1954}
R.~Karplus and J.~M. Luttinger, ``Hall effect in ferromagnetics,'' {\em Phys.
  Rev.}, vol.~95, pp.~1154--1160, Sep 1954.

\bibitem{Dyakonov:1971}
M.~Dyakonov and V.~Perel, ``Current-induced spin orientation of electrons in
  semiconductors,'' {\em Physics Letters A}, vol.~35, no.~6, pp.~459 -- 460,
  1971.

\bibitem{Freeman:1993}
D.-S. Wang, R.~Wu, and A.~J. Freeman, ``First-principles theory of surface
  magnetocrystalline anisotropy and the diatomic-pair model,'' {\em Phys. Rev.
  B}, vol.~47, pp.~14932--14947, Jun 1993.

\bibitem{Laan:1998}
G.~van~der Laan, ``Microscopic origin of magnetocrystalline anisotropy in
  transition metal thin films,'' {\em J.\ Phys.: Condens.\ Matter}, vol.~10,
  p.~3239, Apr 1998.

\bibitem{Bruno:1989}
P.~Bruno, ``Tight-binding approach to the orbital magnetic moment and
  magnetocrystalline anisotropy of transition-metal monolayers,'' {\em Phys.
  Rev. B}, vol.~39, pp.~865--868, Jan 1989.

\bibitem{McGuire:1975}
T.~McGuire and R.~Potter, ``Anisotropic magnetoresistance in ferromagnetic 3d
  alloys,'' {\em Magnetics, IEEE Transactions on}, vol.~11, pp.~1018 -- 1038,
  jul 1975.

\bibitem{Bode:2002}
M.~Bode, S.~Heinze, A.~Kubetzka, O.~Pietzsch, X.~Nie, G.~Bihlmayer,
  S.~Bl\"ugel, and R.~Wiesendanger, ``Magnetization-direction-dependent local
  electronic structure probed by scanning tunneling spectroscopy,'' {\em Phys.
  Rev. Lett.}, vol.~89, p.~237205, Nov 2002.

\bibitem{Molenkamp:2004}
C.~Gould, C.~R\"uster, T.~Jungwirth, E.~Girgis, G.~M. Schott, R.~Giraud,
  K.~Brunner, G.~Schmidt, and L.~W. Molenkamp, ``Tunneling anisotropic
  magnetoresistance: A spin-valve-like tunnel magnetoresistance using a single
  magnetic layer,'' {\em Phys. Rev. Lett.}, vol.~93, p.~117203, Sep 2004.

\bibitem{BAMR:2005}
J.~Velev, R.~F. Sabirianov, S.~S. Jaswal, and E.~Y. Tsymbal, ``Ballistic
  anisotropic magnetoresistance,'' {\em Phys. Rev. Lett.}, vol.~94, p.~127203,
  Mar 2005.

\bibitem{Seemann:2011}
K.~M. Seemann, F.~Freimuth, H.~Zhang, S.~Bl\"ugel, Y.~Mokrousov, D.~E.
  B\"urgler, and C.~M. Schneider, ``Origin of the planar hall effect in
  nanocrystalline ${\mathrm{Co}}_{60}{\mathrm{Fe}}_{20}{\mathrm{B}}_{20}$,''
  {\em Phys. Rev. Lett.}, vol.~107, p.~086603, Aug 2011.

\bibitem{Shick:2008}
B.~G. Park, J.~Wunderlich, D.~A. Williams, S.~J. Joo, K.~Y. Jung, K.~H. Shin,
  K.~Olejnik, A.~B. Shick, and T.~Jungwirth, ``Tunneling anisotropic
  magnetoresistance in multilayer-(Co/Pt)/AlO$_x$/Pt structures,'' {\em Phys. Rev.
  Lett.}, vol.~100, p.~087204, Feb 2008.

\bibitem{AHE-RMP}
N.~Nagaosa, J.~Sinova, S.~Onoda, A.~H. MacDonald, and N.~P. Ong, ``Anomalous
  Hall effect,'' {\em Rev. Mod. Phys.}, vol.~82, pp.~1539--1592, May 2010.

\bibitem{Fivaz:1969}
R.~C. Fivaz, ``Transport theory for ferromagnets,'' {\em Phys. Rev.}, vol.~183,
  pp.~586--594, Jul 1969.

\bibitem{Roman.2009.prl}
E.~Roman, Y.~Mokrousov, and I.~Souza, ``Orientation dependence of the intrinsic
  anomalous Hall effect in hcp cobalt,'' {\em Phys. Rev. Lett.}, vol.~103,
  p.~097203, Aug 2009.

\bibitem{Weissman:1973}
A.~Hirsch and Y.~Weissman, ``Anisotropy of the Hall effect in iron,'' {\em
  Physics Letters A}, vol.~44, no.~4, pp.~239 -- 240, 1973.

\bibitem{Volkenshtein:1961}
N.~Volkenshtein, G.~Fedorov, and V.~Shirokovskii {\em Fiz. Metal. Metalloved.},
  vol.~11, p.~152, 1961.

\bibitem{Hiraoka:1968}
T.~Hiraoka {\em J. Sci. Hiroshima Univ., Ser. A-2}, vol.~32, p.~153, 1968.

\bibitem{Lee:1967}
R.~S. Lee and S.~Legvold, ``Hall effect of gadolinium, lutetium, and yttrium
  single crystals,'' {\em Phys. Rev.}, vol.~162, pp.~431--435, Oct 1967.

\bibitem{Ohgushi:2006}
K.~Ohgushi, S.~Miyasaka, and Y.~Tokura, ``Anisotropic anomalous Hall effect of
  topological origin in carrier-doped FeCr$_{2}$S$_{4}$,'' {\em Journal of the
  Physical Society of Japan}, vol.~75, no.~1, p.~013710, 2006.

\bibitem{Sales:2008}
B.~C. Sales, R.~Jin, and D.~Mandrus, ``Orientation dependence of the anomalous
  Hall resistivity in single crystals of
  ${\mathrm{Yb}}_{14}\mathrm{Mn}{\mathrm{Sb}}_{11}$,'' {\em Phys. Rev. B},
  vol.~77, p.~024409, Jan 2008.

\bibitem{Skokov:2008}
J.~Stankiewicz and K.~P. Skokov, ``Anomalous Hall effect in Y$_2$Fe$_{17-x}$Co$_x$
  single crystals,'' {\em Phys. Rev. B}, vol.~78, p.~214435, Dec 2008.

\bibitem{Stankiewicz:2011}
J.~Stankiewicz, D.~Karpenkov, and K.~P. Skokov, ``Fundamental magnetotransport
  anisotropy in ${R}_{2}$Fe${}_{17}$ single crystals,'' {\em Phys. Rev. B},
  vol.~83, p.~014419, Jan 2011.

\bibitem{Sih:2005}
V.~Sih, R.~C. Myers, Y.~K. Kato, W.~H. Lau, A.~C. Gossard, and D.~D. Awschalom,
  ``Spatial imaging of the spin Hall effect and current-induced polarization in
  two-dimensional electron gases,'' {\em Nat. Phys.}, vol.~1, no.~1,
  pp.~31--35, 2005.

\bibitem{Tombros.2008.prl}
N.~Tombros, S.~Tanabe, A.~Veligura, C.~Jozsa, M.~Popinciuc, H.~T. Jonkman, and
  B.~J. van Wees, ``Anisotropic spin relaxation in graphene,'' {\em Phys. Rev.
  Lett.}, vol.~101, p.~046601, Jul 2008.

\bibitem{Averkiev.2008}
N.~S. Averkiev and L.~E. Golub, ``Spin relaxation anisotropy: microscopic
  mechanisms for 2D systems,'' {\em Semiconductor Science and Technology},
  vol.~23, p.~114002, 2008.

\bibitem{Freimuth.2010.prl}
F.~Freimuth, S.~Bl\"ugel, and Y.~Mokrousov, ``Anisotropic spin Hall effect from
  first principles,'' {\em Phys.\ Rev.\ Lett.}, vol.~105, p.~246602, Dec 2010.

\bibitem{Zhang.2011.prl}
H.~Zhang, F.~Freimuth, S.~Bl\"ugel, Y.~Mokrousov, and I.~Souza, ``Role of
  spin-flip transitions in the anomalous Hall effect of FePt alloy,'' {\em
  Phys.\ Rev.\ Lett.}, vol.~106, p.~117202, Mar 2011.

\bibitem{Chudnovsky:2009}
E.~M. Chudnovsky, ``Intrinsic spin Hall effect in noncubic crystals,'' {\em
  Phys. Rev. B}, vol.~80, p.~153105, Oct 2009.

\bibitem{Zhang.prb.2011}
H.~Zhang, S.~Bl\"ugel, and Y.~Mokrousov, ``Anisotropic intrinsic anomalous Hall
  effect in ordered $3d$Pt alloys,'' {\em Phys. Rev. B}, vol.~84, p.~024401,
  Jul 2011.

\bibitem{Weischenberg.2011.prl}
J.~Weischenberg, F.~Freimuth, J.~Sinova, S.~Bl\"ugel, and Y.~Mokrousov,
  ``\textit{Ab~Initio} theory of the scattering-independent anomalous Hall
  effect,'' {\em Phys. Rev. Lett.}, vol.~107, p.~106601, Sep 2011.

\bibitem{Zimmermann:2012}
B.~Zimmermann, P.~Mavropoulos, S.~Heers, N.~H. Long, S.~Bl\"ugel, and
  Y.~Mokrousov, ``Anisotropy of spin relaxation in metals,'' {\em submitted},
  2012.

\bibitem{Fabian.1998.prl}
J.~Fabian and S.~Das~Sarma, ``Spin relaxation of conduction electrons in
  polyvalent metals: Theory and a realistic calculation,'' {\em Phys.\ Rev.\
  Lett.}, vol.~81, pp.~5624--5627, Dec 1998.

\bibitem{nagaosa:2006}
N.~Nagaosa, ``Anomalous Hall effect: A new perspective,'' {\em J. Phys. Soc.
  Jap.}, vol.~75, p.~042001, 2006.

\bibitem{sinitsyn:2008}
N.~A. Sinitsyn, ``Semiclassical theories of the anomalous Hall effect,'' {\em
  J.\ Phys.: Condens.\ Matter}, vol.~20, p.~023201, Jan 2008.

\bibitem{mokrousov:2009}
Y.~Mokrousov, ``Anomalous Hall effect,'' {\em 40. IFF Spring School book of
  lectures (Forschungszentrum J\"ulich)}, 2009.

\bibitem{Gradhand.2010.prl}
M.~Gradhand, D.~V. Fedorov, P.~Zahn, and I.~Mertig, ``Extrinsic spin Hall
  effect from first principles,'' {\em Phys. Rev. Lett.}, vol.~104, p.~186403,
  May 2010.

\bibitem{Lowitzer.2010.prl}
S.~Lowitzer, D.~K\"odderitzsch, and H.~Ebert, ``Coherent description of the
  intrinsic and extrinsic anomalous Hall effect in disordered alloys on an
  \textit{Ab~Initio} level,'' {\em Phys. Rev. Lett.}, vol.~105, p.~266604, Dec
  2010.

\bibitem{Yao:2004}
Y.~Yao, L.~Kleinman, A.~H. MacDonald, J.~Sinova, T.~Jungwirth, D.-S. Wang,
  E.~Wang, and Q.~Niu, ``First principles calculation of anomalous Hall
  conductivity in ferromagnetic bcc Fe,'' {\em Phys. Rev. Lett.}, vol.~92,
  p.~037204, Jan 2004.

\bibitem{Berry-RMP}
D.~Xiao, M.-C. Chang, and Q.~Niu, ``Berry phase effects on electronic
  properties,'' {\em Rev. Mod. Phys.}, vol.~82, pp.~1959--2007, Jul 2010.

\bibitem{Gradhand:2012}
M.~Gradhand, D.~Fedorov, F.~Pientka, P.~Zahn, I.~Mertig, and B.~Gy\"orffy,
  ``First-principle calculations of the Berry curvature of Bloch states for
  charge and spin transport of electrons,'' {\em J.\ Phys.: Condens.\ Matter},
  vol.~24, p.~213202, May 2012.

\bibitem{Fang:2003}
Z.~Fang, N.~Nagaosa, K.~S. Takahashi, A.~Asamitsu, R.~Mathieu, T.~Ogasawara,
  H.~Yamada, M.~Kawasaki, Y.~Tokura, and K.~Terakura, ``The anomalous Hall
  effect and magnetic monopoles in momentum space,'' {\em Science}, vol.~302,
  no.~5642, pp.~92--95, 2003.

\bibitem{Mathieu:2004}
R.~Mathieu, A.~Asamitsu, H.~Yamada, K.~S. Takahashi, M.~Kawasaki, Z.~Fang,
  N.~Nagaosa, and Y.~Tokura, ``Scaling of the anomalous Hall effect in
  Sr$_{1-x}$Ca$_x$RuO$_3$,'' {\em Phys. Rev. Lett.}, vol.~93, p.~016602, Jun 2004.

\bibitem{Wang:2006}
X.~Wang, J.~R. Yates, I.~Souza, and D.~Vanderbilt, ``\textit{Ab initio}
  calculation of the anomalous Hall conductivity by Wannier interpolation,''
  {\em Phys. Rev. B}, vol.~74, p.~195118, Nov 2006.

\bibitem{Wang:2007}
X.~Wang, D.~Vanderbilt, J.~R. Yates, and I.~Souza, ``Fermi-surface calculation
  of the anomalous Hall conductivity,'' {\em Phys. Rev. B}, vol.~76, p.~195109,
  Nov 2007.

\bibitem{Guo:2011}
H.-R. Fuh and G.-Y. Guo, ``Intrinsic anomalous Hall effect in nickel: A
  GGA+$U$ study,'' {\em Phys. Rev. B}, vol.~84, p.~144427, Oct 2011.

\bibitem{Mikitik:1999}
G.~P. Mikitik and Y.~V. Sharlai, ``Manifestation of Berry's phase in metal
  physics,'' {\em Phys. Rev. Lett.}, vol.~82, pp.~2147--2150, Mar 1999.

\bibitem{Thouless:1982}
D.~J. Thouless, M.~Kohmoto, M.~P. Nightingale, and M.~den Nijs, ``Quantized
  Hall conductance in a two-dimensional periodic potential,'' {\em Phys. Rev.
  Lett.}, vol.~49, pp.~405--408, Aug 1982.

\bibitem{Hasan-RMP}
M.~Z. Hasan and C.~L. Kane, ``\textit{Colloquium} : Topological insulators,''
  {\em Rev. Mod. Phys.}, vol.~82, pp.~3045--3067, Nov 2010.

\bibitem{Zhang-RMP}
X.-L. Qi and S.-C. Zhang, ``Topological insulators and superconductors,'' {\em
  Rev. Mod. Phys.}, vol.~83, pp.~1057--1110, Oct 2011.

\bibitem{Hirsch:1999}
J.~E. Hirsch, ``Spin Hall effect,'' {\em Phys. Rev. Lett.}, vol.~83,
  pp.~1834--1837, Aug 1999.

\bibitem{Kato:2004}
Y.~K. Kato, R.~C. Myers, A.~C. Gossard, and D.~D. Awschalom, ``Observation of
  the spin Hall effect in semiconductors,'' {\em Science}, vol.~306, no.~5703,
  pp.~1910--1913, 2004.

\bibitem{Uchida:2008}
K.~Uchida, S.~Takahashi, K.~Harii, W.~Ieda, J.and~Koshibae, K.~Ando,
  S.~Maekawa, and E.~Saitoh, ``Observation of spin Seebeck effect,'' {\em
  Nature}, vol.~455, no.~7214, pp.~778--781, 2008.

\bibitem{Buhrman:2012}
L.~Liu, C.-F. Pai, Y.~Li, H.~W. Tseng, D.~C. Ralph, and R.~A. Buhrman,
  ``Spin-torque switching with the giant spin Hall effect of tantalum,'' {\em
  Science}, vol.~336, no.~6081, pp.~555--558, 2012.

\bibitem{Murakami:2003}
S.~Murakami, N.~Nagaosa, and S.-C. Zhang, ``Dissipationless quantum spin
  current at room temperature,'' {\em Science}, vol.~301, no.~5638,
  pp.~1348--1351, 2003.

\bibitem{Murakami:2004}
S.~Murakami, N.~Nagaosa, and S.-C. Zhang, ``Spin-Hall insulator,'' {\em Phys.
  Rev. Lett.}, vol.~93, p.~156804, Oct 2004.

\bibitem{Bernevig:2006}
B.~A. Bernevig, T.~L. Hughes, and S.-C. Zhang, ``Quantum spin Hall effect and
  topological phase transition in HgTe quantum wells,'' {\em Science},
  vol.~314, no.~5806, pp.~1757--1761, 2006.

\bibitem{Konig:2007}
M.~K\"onig, S.~Wiedmann, C.~Br\"une, A.~Roth, H.~Buhmann, L.~W. Molenkamp,
  X.-L. Qi, and S.-C. Zhang, ``Quantum spin Hall insulator state in HgTe
  quantum wells,'' {\em Science}, vol.~318, no.~5851, pp.~766--770, 2007.

\bibitem{Gradhand.2011.01}
M.~Gradhand, D.~V. Fedorov, F.~Pientka, P.~Zahn, I.~Mertig, and B.~L.
  Gy\"orffy, ``Calculating the Berry curvature of Bloch electrons using the KKR
  method,'' {\em Phys.\ Rev.\ B}, vol.~84, p.~075113, Aug 2011.

\bibitem{Lowitzer:2011}
S.~Lowitzer, M.~Gradhand, D.~K\"odderitzsch, D.~V. Fedorov, I.~Mertig, and
  H.~Ebert, ``Extrinsic and intrinsic contributions to the spin Hall effect of
  alloys,'' {\em Phys. Rev. Lett.}, vol.~106, p.~056601, Feb 2011.

\bibitem{Sinova:2004}
J.~Sinova, D.~Culcer, Q.~Niu, N.~A. Sinitsyn, T.~Jungwirth, and A.~H.
  MacDonald, ``Universal intrinsic spin Hall effect,'' {\em Phys. Rev. Lett.},
  vol.~92, p.~126603, Mar 2004.

\bibitem{Guo:2008}
G.~Y. Guo, S.~Murakami, T.-W. Chen, and N.~Nagaosa, ``Intrinsic spin Hall
  effect in platinum: First-principles calculations,'' {\em Phys. Rev. Lett.},
  vol.~100, p.~096401, Mar 2008.

\bibitem{Shi:2006}
J.~Shi, P.~Zhang, D.~Xiao, and Q.~Niu, ``Proper definition of spin current in
  spin-orbit coupled systems,'' {\em Phys. Rev. Lett.}, vol.~96, p.~076604, Feb
  2006.

\bibitem{Vosko.1980.CanJPhys}
S.~H. Vosko, L.~Wilk, and M.~Nusair, ``Accurate spin-dependent electron liquid
  correlation energies for local spin density calculations: a critical
  analysis,'' {\em Can.~J.~Phys.}, vol.~58, no.~8, pp.~1200--1211, 1980.

\bibitem{KKR-code}
olymp.cup.uni-muenchen.de/ak/ebert/SPR-TB-KKR

\bibitem{Heers.PhD}
S.~Heers.
\newblock PhD thesis, RWTH Aachen, 2011.

\bibitem{fleur}
www.flapw.de

\bibitem{Souza:2002}
I.~Souza, N.~Marzari, and D.~Vanderbilt, ``Maximally localized Wannier
  functions for entangled energy bands,'' {\em Phys. Rev. B}, vol.~65,
  p.~035109, Dec 2001.

\bibitem{Freimuth:2008}
F.~Freimuth, Y.~Mokrousov, D.~Wortmann, S.~Heinze, and S.~Bl\"ugel, ``Maximally
  localized Wannier functions within the FLAPW formalism,'' {\em Phys. Rev. B},
  vol.~78, p.~035120, Jul 2008.

\bibitem{wannier90}
A.~A. Mostofi, J.~R. Yates, Y.-S. Lee, I.~Souza, D.~Vanderbilt, and N.~Marzari,
  ``wannier90: A tool for obtaining maximally-localised Wannier functions,''
  {\em Computer Physics Communications}, vol.~178, no.~9, pp.~685 -- 699, 2008.

\bibitem{Zener:1954}
C.~Zener, ``Classical theory of the temperature dependence of magnetic
  anisotropy energy,'' {\em Phys. Rev.}, vol.~96, pp.~1335--1337, Dec 1954.

\bibitem{Zhang:2012:Bi}
H.~{Zhang}, F.~{Freimuth}, G.~{Bihlmayer}, S.~{Bl{\"u}gel}, and Y.~{Mokrousov},
  ``{Topological phases of Bi(111) bilayer in an external exchange field},''
  {\em ArXiv e-prints, 1203.5025}, Mar. 2012.

\bibitem{Seemann:2010}
K.~M. Seemann, Y.~Mokrousov, A.~Aziz, J.~Miguel, F.~Kronast, W.~Kuch, M.~G.
  Blamire, A.~T. Hindmarch, B.~J. Hickey, I.~Souza, and C.~H. Marrows,
  ``Spin-orbit strength driven crossover between intrinsic and extrinsic
  mechanisms of the anomalous Hall effect in the epitaxial $L1_0$-ordered
  ferromagnets FePd and FePt,'' {\em Phys. Rev. Lett.}, vol.~104, p.~076402,
  Feb 2010.

\bibitem{Yau:1971}
K.~L. Yau and J.~T.~H. Chang, ``The planar Hall effect in thin foils of Ni-Fe
  alloy,'' {\em Journal of Physics F: Metal Physics}, vol.~1, no.~1, p.~38,
  1971.

\bibitem{Bowen:2005}
M.~Bowen, K.-J. Friedland, J.~Herfort, H.-P. Sch\"onherr, and K.~H. Ploog,
  ``Order-driven contribution to the planar Hall effect in
  ${\mathrm{Fe}}_{3}\mathrm{Si}$ thin films,'' {\em Phys. Rev. B}, vol.~71,
  p.~172401, May 2005.

\bibitem{Cooper:1965}
B.~R. Cooper, ``Theory of the interband ferromagnetic Kerr effect in nickel,''
  {\em Phys. Rev.}, vol.~139, pp.~A1504--A1514, Aug 1965.

\bibitem{Berger:1970}
L.~Berger, ``Side-jump mechanism for the Hall effect of ferromagnets,'' {\em
  Phys. Rev. B}, vol.~2, pp.~4559--4566, Dec 1970.

\bibitem{Smit:1955}
J.~Smit, ``The spontaneous Hall effect in ferromagnetics i,'' {\em Physica},
  vol.~21, pp.~877 -- 887, 1955.

\bibitem{Smit:1958}
J.~Smit, ``The spontaneous Hall effect in ferromagnetics ii,'' {\em Physica},
  vol.~24, pp.~39 -- 51, 1958.

\bibitem{Sinitsyn:2006}
N.~A. Sinitsyn, Q.~Niu, and A.~H. MacDonald, ``Coordinate shift in the
  semiclassical Boltzmann equation and the anomalous Hall effect,'' {\em Phys.
  Rev. B}, vol.~73, p.~075318, Feb 2006.

\bibitem{Miyasato:2007}
T.~Miyasato, N.~Abe, T.~Fujii, A.~Asamitsu, S.~Onoda, Y.~Onose, N.~Nagaosa, and
  Y.~Tokura, ``Crossover behavior of the anomalous Hall effect and anomalous
  Nernst effect in itinerant ferromagnets,'' {\em Phys. Rev. Lett.}, vol.~99,
  p.~086602, Aug 2007.

\bibitem{Xiao:2006}
D.~Xiao, Y.~Yao, Z.~Fang, and Q.~Niu, ``Berry-phase effect in anomalous
  thermoelectric transport,'' {\em Phys. Rev. Lett.}, vol.~97, p.~026603, Jul
  2006.

\bibitem{Spin-Nernst}
S.-G. Cheng, Y.~Xing, Q.-F. Sun, and X.~C. Xie, ``Spin Nernst effect and Nernst
  effect in two-dimensional electron systems,'' {\em Phys. Rev. B}, vol.~78,
  p.~045302, Jul 2008.

\bibitem{Chuu:2010}
C.-P. Chuu, M.-C. Chang, and Q.~Niu, ``Semiclassical dynamics and transport of
  the Dirac spin,'' {\em Solid State Communications}, vol.~150, pp.~533 -- 537,
  2010.

\bibitem{Onose:2010}
Y.~Onose, T.~Ideue, H.~Katsura, Y.~Shiomi, N.~Nagaosa, and Y.~Tokura,
  ``Observation of the magnon Hall effect,'' {\em Science}, vol.~329, no.~5989,
  pp.~297--299, 2010.

\bibitem{Wills:2012}
D.~Boldrin and A.~S. Wills, ``Anomalous Hall effect in geometrically frustrated
  magnets,'' {\em Advances in Condensed Matter Physics}, vol.~2012, 2012.

\bibitem{Bruno:2004}
P.~Bruno, V.~K. Dugaev, and M.~Taillefumier, ``Topological Hall effect and
  Berry phase in magnetic nanostructures,'' {\em Phys. Rev. Lett.}, vol.~93,
  p.~096806, Aug 2004.

\bibitem{Rosch:2009}
A.~Neubauer, C.~Pfleiderer, B.~Binz, A.~Rosch, R.~Ritz, P.~G. Niklowitz, and
  P.~B\"oni, ``Topological Hall effect in the $A$ phase of MnSi,'' {\em Phys.
  Rev. Lett.}, vol.~102, p.~186602, May 2009.

\bibitem{Onoda:2011}
L.~Balicas, S.~Nakatsuji, Y.~Machida, and S.~Onoda, ``Anisotropic hysteretic
  Hall effect and magnetic control of chiral domains in the chiral spin states
  of Pr$_2$Ir$_2$O$_7$,'' {\em Phys. Rev. Lett.}, vol.~106, p.~217204, May 2011.

\end{thebibliography}
\end{document}